\documentclass[footinbib,twocolumn,showpacs,amsmath,amstex,amssymb,mathfonts,superscriptaddress,prl]{revtex4-2}
\usepackage{graphicx}
\usepackage{color}
\usepackage{bm}
\usepackage[colorlinks,linkcolor=blue,anchorcolor=blue,citecolor=blue]{hyperref}
\usepackage{amsmath}
\usepackage{amssymb}
\usepackage{amsthm}
\usepackage{amsfonts}
\usepackage{verbatim}

\usepackage{bbm}

\newcommand{\fref}[1]{Fig. \ref{#1}}
\newcommand{\eref}[1]{Eq. \eqref{#1}}
\newcommand{\tref}[1]{Tab. \ref{#1}}

\begin{document}

\title{Emergent low-energy many-body Hilbert spaces in Chern bands}
\pacs{73.43.Lp, 71.10.Pm}

\author{Chen-Xin Jiang}
\affiliation{Department of Physics and Chongqing Key Laboratory for Strongly Coupled Physics, Chongqing University, Chongqing 401331, People's Republic of China}
\affiliation{Division of Physics and Applied Physics, Nanyang Technological University, Singapore 637371, Singapore}
\author{Zi-Xiang Hu\thanks{*Corresponding authors}}
\email{zxhu@cqu.edu.cn}
\affiliation{Department of Physics and Chongqing Key Laboratory for Strongly Coupled Physics, Chongqing University, Chongqing 401331, People's Republic of China}
\author{Bo Yang\thanks{*Corresponding authors}}
\email{yang.bo@ntu.edu.sg}
\affiliation{Division of Physics and Applied Physics, Nanyang Technological University, Singapore 637371, Singapore}

\date{\today}
\begin{abstract}
We show that in Landau levels and more generic Chern bands, the most relevant many-body subspace selected by realistic two-body interactions alone (e.g., in the limit of vanishing bandwidth and band gap) depends strongly on the particle density and can be constructed systematically. Using the well-defined hierarchical structure of conformal Hilbert spaces (CHS), each with a highest-density vacuum state, we compute, for both fermionic and bosonic systems, the many-body quantum metric that quantifies the interaction energy scales of these subspaces. For a large family of short-range, Coulomb-based interactions, particles preferentially occupy the many-body states in the smallest accommodating CHS. This allows us to describe the low-energy physics of the interacting system in terms of the appropriate anyonic degrees of freedom. Our results also explain why, in any Chern band (in lattice or continuum systems), band mixing tends to be suppressed even in the limit of strong interaction; this suppression is stronger when the band deviates less from the ideal trace condition and when the two-body interaction is shorter-ranged.
\end{abstract}

\maketitle

{\textit{Introduction--}} The central challenge in strongly correlated quantum materials is that
only the single-particle Hilbert space can be readily obtained,
whereas the associated many-body Hilbert space grows exponentially with system size and is analytically intractable.
Fractional topological phases in Chern bands are typical examples
where nontrivial physics emerges in the limit of ``infinite interaction strength'' (i.e., the flat-band limit) \cite{PhysRevLett.106.236802,PhysRevLett.106.236804,PhysRevLett.106.236803,PhysRevX.1.021014,Sheng2011,Cai2023,Zeng2023,Park2023,PhysRevX.13.031037,Lu2024,Bernevig2025Fractional}.
However, it should be emphasized that in such systems it is still the kinetic energy that selects the relevant sub-Hilbert space,
within which the interaction energy scale becomes dominant \cite{PARAMESWARAN2013816,BERGHOLTZ2013,PhysRevB.109.L121107}.
An intuitive, albeit somewhat crude, notion is to consider the band gap $\epsilon_g$,
the bandwidth $\epsilon_w$ of the band lying below this gap (corresponding to the sub-Hilbert space $\mathcal H_{\text{subgap}}$),
and the interaction energy scale $\epsilon_{\text{int}}$ within $\mathcal H_{\text{subgap}}$~\cite{footnote}.
For a strongly interacting system to be well defined, it is usually necessary that $\epsilon_g\gg\epsilon_{\text{int}}\gg\epsilon_w$ \cite{PARAMESWARAN2013816,BERGHOLTZ2013}.
As an illustration, for fractional topological phases, taking the limits $\epsilon_w\to 0$ and $\epsilon_g\to\infty$ is usually a good approximation,
and the resulting topological order arises from interactions confined to a single-band subspace \cite{PhysRevLett.106.236804,PhysRevX.1.021014}.

The opposite limit, $\epsilon_{\text{int}}\gg\epsilon_g,\epsilon_w$, is generally regarded as even more challenging, but it turns out that for topological systems the low-energy physics can be quite tractable in certain cases \cite{PhysRevLett.112.126806,PhysRevB.109.L121107,PhysRevB.112.075110}.
Let $\mathcal H_s$ be the physical single-particle Hilbert space, corresponding to either a single band or multiple bands of a condensed-matter system of fixed size, with dimension $n_o$ (i.e., the number of single-particle orbitals).
The corresponding many-body Hilbert space, or the Fock space,
can therefore be written as
\begin{eqnarray}\label{schs}
    \mathcal F_s=\bigoplus_{n_p=0}^{\infty}\mathcal H_{s,n_p},
\end{eqnarray}
where $n_p$ is the number of particles in $\mathcal H_s$,
such that $\mathcal H_{s,0}$ represents the vacuum and $\mathcal H_{s,1}=\mathcal H_s$.
The space $\mathcal H_{s,n_p}$ is spanned by product states containing $n_p$ particles,
and in the spinless fermionic case, $\mathcal H_{s,n_p>n_o}$ is empty.
We typically aim to understand the low-energy dynamics within $\mathcal F_s$,
focusing on the physically relevant subspace $\bar{\mathcal F}_{s} \subset \mathcal F_{s}$.
In the limit of large kinetic energy (i.e., $\epsilon_w\gg\epsilon_{\text{int}}$, the Fermi-liquid regime \cite{RevModPhys.66.129}), $\bar{\mathcal F}_{s}$ is spanned by the product states with the lowest kinetic energy. In the opposite limit of strong interaction, or vanishing kinetic energy within $\mathcal F_{s}$, a natural question is whether there are situations in which the correct low-energy subspace $\bar{\mathcal F}_{s}$ can be identified. Characterizing the $\bar{\mathcal F}_{s}$ preferred by the interaction is an important step towards understanding the low-energy physics of interacting particles within $\mathcal F_{s}$, and is also useful for numerical studies of larger system sizes.

    {\textit{Band-index ferromagnetism--}} We first present striking examples in which, even under strong interactions,
the ground state in topological bands is well described by product states.
This parallels quantum Hall ferromagnets \cite{PhysRevB.47.16419}, with the band index acting as an effective pseudospin for spinless fermions. Unlike in quantum Hall ferromagnets, however, this band pseudospin is generally not a good quantum number, since interaction-induced interband scattering can change the band occupation.
Let $\mathcal H_s$ denote the Hilbert space spanned by the $N$ lowest Landau levels (LLs), taken to be degenerate (i.e., with vanishing cyclotron energy),
so that all single-particle orbitals have the same energy.
One might expect interactions to mix these LLs substantially,
but in fact they do not.

\begin{figure}[!t]
    \centering
    \includegraphics[width=.45\textwidth]{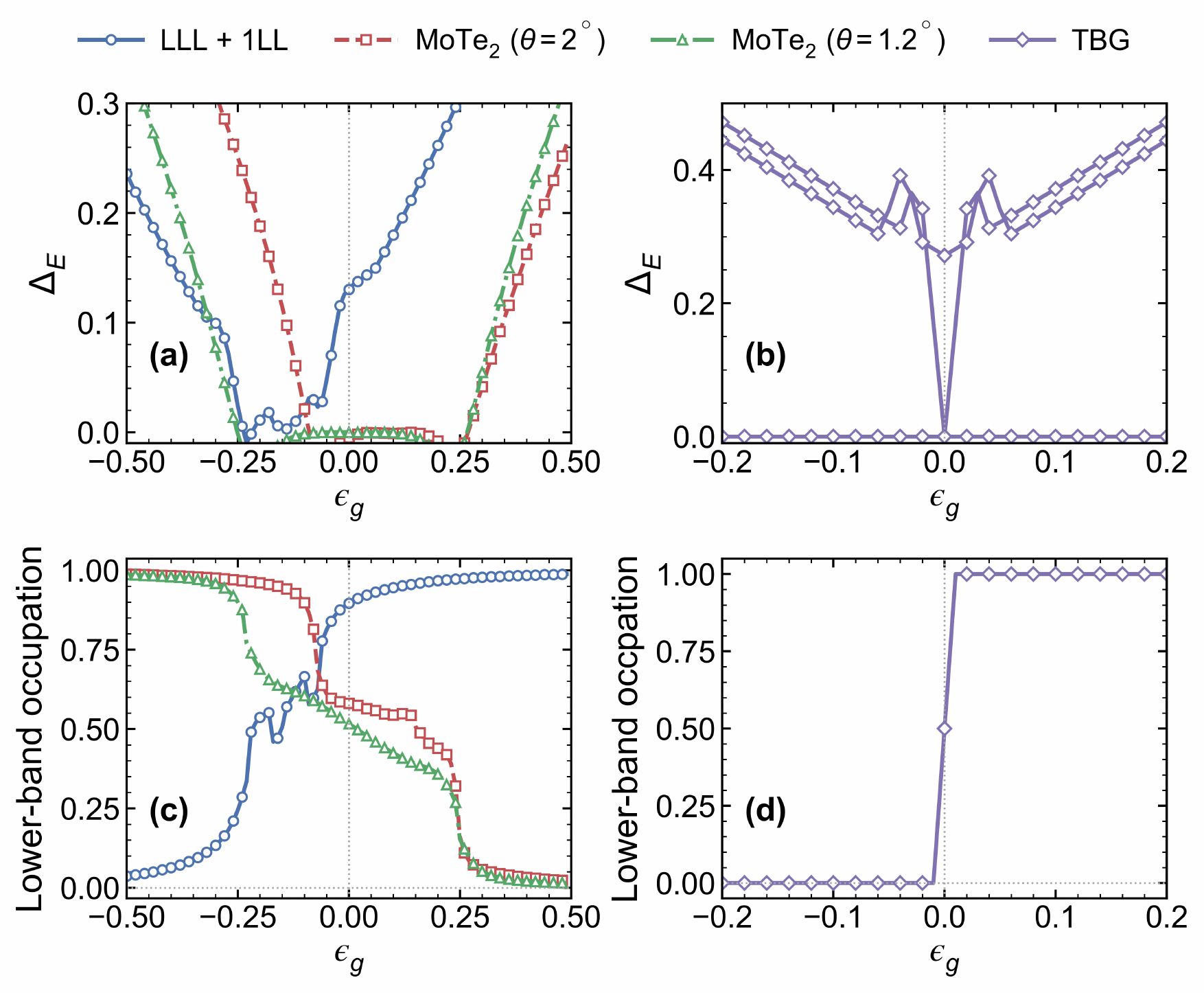}
    \caption{
Thermodynamic-limit results. (a) Energy gaps $\Delta_E$ between the ground and excited states as functions of $\epsilon_g$ for the LLL+1LL system and tMoTe$_2$ at twist angles $\vartheta=2^\circ$ and $1.2^\circ$.  (b) $\Delta_E$ as a function of $\epsilon_g$ for TBG at $\vartheta=1.132^\circ$, showing the three lowest excitation energies relative to the ground state. (c) Band occupation fraction of the LLL in the LLL+1LL system and of the band with smaller $\chi^{-1}$ in tMoTe$_2$. (d) Band occupation fraction of the $\mathcal C=-1$ TBG band. See Supplemental Material for all other parameters.
}
    \label{fig:1}
\end{figure}

As an example, consider $N=2$, i.e., the lowest LL (LLL) and the first excited LL (1LL), separated by a tunable band gap $\epsilon_g$ (the cyclotron energy), at filling factor $\nu=1$,
so that the electron number equals the number of orbitals in one LL.
The Hamiltonian takes the form
\begin{equation}
    \scalebox{0.93}{$
        \displaystyle
        \hat{H}=  \epsilon_g\sum_{\mathbf{k}}
        c^\dagger_{2\mathbf{k}} c_{2\mathbf{k}}
        +\sum_{\{l_i\mathbf{k}_i\}}V^{l_1l_2l_3l_4}_{\mathbf{k}_1\mathbf{k}_2\mathbf{k}_3\mathbf{k}_4}c^\dagger_{l_1\mathbf{k}_1}c^\dagger_{l_2\mathbf{k}_2}c_{l_4\mathbf{k}_4}c_{l_3\mathbf{k}_3},
    $}
\end{equation}
where $l_i=1,2$ index the two bands,
$\mathbf{k}$ is the crystal momentum,
$V^{l_1,l_2,l_3,l_4}_{\mathbf{k}_1,\mathbf{k}_2,\mathbf{k}_3,\mathbf{k}_4}$
represents the interaction matrix element restricted to the two-band subspace,
and $c^\dagger_{l,\mathbf{k}}$ and $c_{l,\mathbf{k}}$ denote the corresponding creation and annihilation operators.
In the limit $\epsilon_g\to\infty$,
all electrons are trivially confined to the LLL.
Counterintuitively, this picture largely persists as $\epsilon_g$ is reduced to zero.
As shown in \fref{fig:1}(a), the many-body gap at $\nu=1$ remains finite,
demonstrating that the ground state at $\epsilon_g=0$ can be adiabatically deformed into the ground state in the single-LL limit (i.e., $\epsilon_g\to\infty$).
Remarkably, even at $\epsilon_g=0$, most electrons remain in the LLL,
with the thermodynamic-limit occupation exceeding $85\%$.
Thus, interactions alone can stabilize the integer quantum Hall phase with two degenerate LLs,
where a robust band-polarized state with a many-body gap emerges even with vanishing $\epsilon_g$.

However, this behavior is not generic for Chern band systems.
For instance, in twisted MoTe$_2$ (tMoTe$_2$) \cite{PhysRevLett.122.086402,Cai2023,Zeng2023,Park2023,PhysRevX.13.031037},
tuning the gap between the two Chern bands leads to markedly different behavior.
As shown in \fref{fig:1}(a) and (c), the two bands become much more strongly mixed by the interaction when $\epsilon_g$ is small.
At twist angle $\vartheta=1.2^\circ$, the two band occupations are almost identical at $\epsilon_g=0$.
At $\vartheta=2^\circ$, band mixing is also strong and considerably exceeds that in the LL case.
Meanwhile, the many-body gap is strongly reduced and can collapse even before $\epsilon_g$ reaches zero.
Thus, in the small-gap regime tMoTe$_2$ lacks the robust band-polarized ground state found in the LL system.

For comparison, we also examine magic-angle twisted bilayer graphene (TBG).
Specifically, we focus on the chiral limit \cite{pnas.1108174108,PhysRevLett.122.106405} at a twist angle of $\vartheta=1.132^\circ$,
where the two flat bands are degenerate and have identical topological and geometric properties, except for their opposite Chern numbers.
The two bands reside in the same valley but are fully polarized on opposite sublattices,
which strongly suppresses interaction-induced scattering between them.
As $|\epsilon_g|$ approaches zero, the two lowest states become degenerate, with all particles mostly occupying one of the two bands in each state, while the gap to higher excitations remains finite.
This mirrors an Ising ferromagnet at zero Zeeman field,
with the band index effectively playing the role of the sublattice pseudospin.
The Chern numbers and quantum metric traces of all these systems are summarized in \tref{tab:1}.

\begin{table}[t]
    \caption{
        Chern numbers $\mathcal{C}$ and total quantum metric traces $\chi^{-1}$ of the two bands in \fref{fig:1}.
    }
    \begin{ruledtabular}
        \begin{tabular}{lcc}
            System                  & $(\mathcal{C}_1,\mathcal{C}_2)$ & $(\chi_1^{-1},\chi_2^{-1})$ \\
            LLL+1LL                 & $(1,1)$                         & $(1,3)$                     \\
            tMoTe$_2$ ($1.2^\circ$) & $(1,-1)$                        & $(1.30,1.14)$               \\
            tMoTe$_2$ ($2^\circ$)   & $(-1,-1)$                       & $(3.74,1.15)$               \\
            TBG ($1.132^\circ$)     & $(-1,1)$                        & $(1,1)$                     \\
        \end{tabular}
    \end{ruledtabular}
    \label{tab:1}
\end{table}

{\textit{Interaction energy scale of conformal Hilbert spaces--}} The numerical results above call for careful interpretation: the tendency of electrons to occupy one band rather than another reflects the lower interaction energy associated with that band. Previous works have shown that the quantum (Fubini--Study) metric of Bloch states is closely linked to their occupation: as electrons are added to a band, they preferentially fill Bloch states with a smaller quantum metric \cite{PhysRevResearch.5.L012015,PhysRevResearch.6.L032063}. More specifically, for a completely occupied band and to leading order in the momentum transfer of the interaction, the Fock contribution to the energy is proportional to the Brillouin-zone (BZ) integral of the \emph{trace of the quantum metric} \cite{PhysRevResearch.6.L032063}, $\chi^{-1}=(2\pi)^{-1}\int_{\rm BZ} d^2{\bf k}\ {\rm Tr} \mathcal G_{\bf k}$. In particular, this explains why the completely filled LLL has the lowest Fock energy among all product states with LL mixing. It serves as a guiding principle for understanding why interactions favor certain single-particle orbitals when the kinetic energy is neglected. Thus, at the single-particle level, the BZ quantum metric provides an intuitive explanation of the numerical results in \fref{fig:1} and \tref{tab:1}.

\begin{figure}[!t]
    \centering
    \includegraphics[width=.45\textwidth]{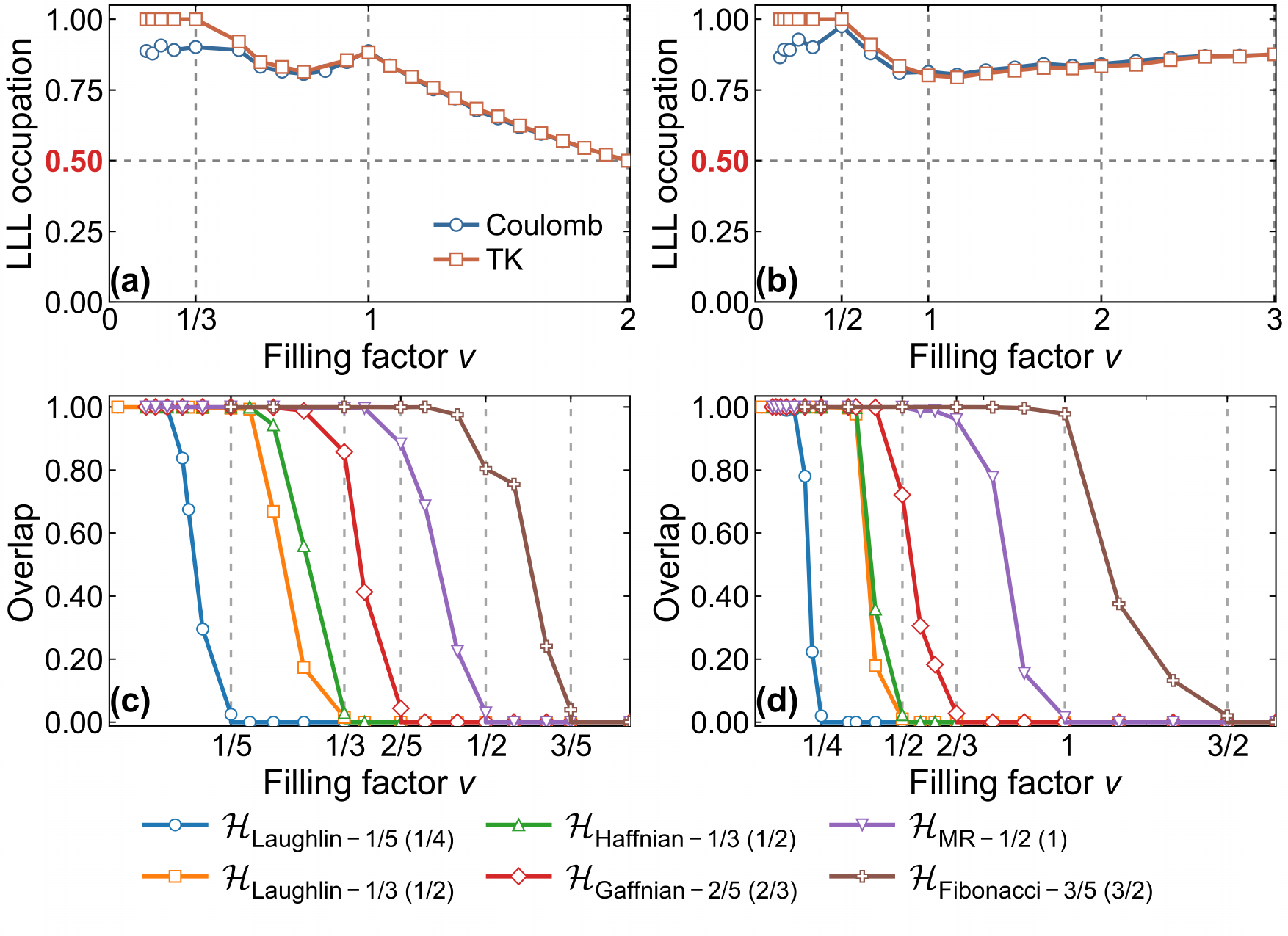}
    \caption{
(a) and (b) LLL occupation fraction in the LLL+1LL system for fermions and bosons, respectively, with Coulomb and Trugman--Kivelson (TK) interactions in the strong-interaction limit, $\epsilon_{\rm int}/\epsilon_g\rightarrow\infty$. (c) and (d) Total overlap of the lowest 200 eigenstates of the Coulomb interaction projected into the LLL with representative CHS \cite{PhysRevB.83.241302,PhysRevB.75.075317,MooreRead1991,PhysRevB.59.8084} as a function of filling factor for fermions and bosons, respectively. Here, MR denotes
the Moore--Read state. Different curves correspond to the indicated CHS.
}
    \label{fig:2}
\end{figure}

It is important to emphasize, however, that the quantum metric is a single-particle quantity and therefore cannot fully capture interaction-driven, inherently many-body physics. What we need to understand is which part of the \emph{many-body Hilbert space} $\mathcal F_{s}$ is preferentially occupied as particles are added. The quantum metric of Bloch states also depends on the chosen basis, which is unsatisfactory, particularly for flat or degenerate bands, where individual single-particle wavefunctions of the same energy are not physically measurable. Moreover, the Fock-energy argument does not apply to bosons, since their Fock term has the opposite sign and bosons never fully occupy a band. This raises a nontrivial question: when bosons are added to a small set of degenerate LLs, do they preferentially occupy specific LLs, and how does the bosonic $\bar{\mathcal F}_s$ differ from its fermionic counterpart? The answer can be seen in \fref{fig:2}(b), and we explain it in detail later.

To address these issues, we focus on LLs as the simplest case and define an interaction energy scale for a specific class of $\bar{\mathcal F}_s$. In doing so, we show that the interaction alone naturally favors these particular Hilbert spaces. These special $\bar{\mathcal F}_s$ are the conformal Hilbert spaces (CHS), which can be constructed exactly in LLs \cite{PhysRevB.100.241302,S0217979222300031,PhysRevB.105.035144,Wang2023NatCommun} and, more generally, in ideal flat bands (IFB) \cite{PhysRevB.90.165139,PhysRevLett.114.236802,PhysRevLett.127.246403,PhysRevResearch.2.023237}. Although a CHS can be defined more generally (e.g., via local exclusion conditions \cite{PhysRevB.100.241302}), for simplicity it can be understood as the space spanned by the zero modes of a frustration-free operator (e.g., a model interaction Hamiltonian\cite{PhysRevLett.51.605,PhysRevB.75.075318,PhysRevB.75.195306,PhysRevLett.125.176402}), physically representing the ground state and quasihole states of an integer or fractional quantum (anomalous) Hall phase. Each CHS contains a highest-density ground state, with the maximal particle number $n_p$ allowed for a given number of orbitals $n_o$, which we take as the \emph{vacuum} of the CHS. Lower-density states within the same CHS are obtained by inserting flux (i.e., adding orbitals) or removing particles from this vacuum, thereby creating quasiholes that may obey anyonic statistics and act as the ``elementary particles'' of that CHS. The defining property of a CHS is the bulk-edge correspondence: the entanglement spectrum of the vacuum is in one-to-one correspondence with all the states in the same CHS \cite{PhysRevLett.101.010504,PhysRevLett.106.100405}.

\begin{figure}[!t]
    \centering
    \includegraphics[width=.45\textwidth]{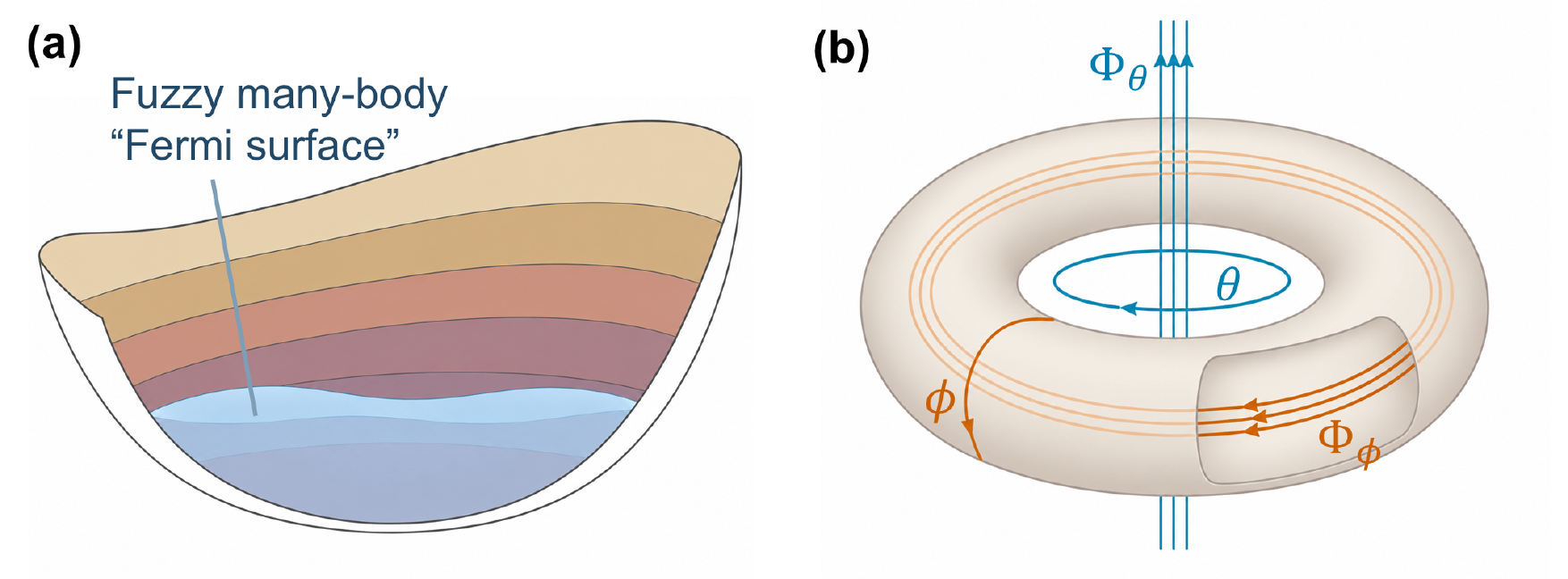}
    \caption{
Schematic illustrations of the CHS hierarchy and flux insertion on a torus.
(a) Schematic illustration of a representative CHS hierarchy for either
fermionic or bosonic systems, with nested regions indicating subspace
inclusion. The blue region represents how the many-body Hilbert space fills up when particles are added, mimicking a fuzzy many-body ``Fermi
surface'' that is well defined for both interacting fermions and bosons.
(b) Fluxes through the two noncontractible cycles of the torus define the twist angles $\theta$ and $\phi$.
}
    \label{fig:3}
\end{figure}

The CHS vacuum is a well-defined, unique many-body ground state ${|\psi_g(\theta,\phi)\rangle}$ at filling $\nu = n_p/n_o$ (up to topological degeneracies on manifolds of genus $g>0$ \cite{PhysRevB.41.9377}). On the torus geometry, these states are parametrized by the two fluxes $0 \le \theta,\phi \le 2\pi$ threading the handles of the torus \cite{PhysRevB.31.3372} [see \fref{fig:3}(b)]. A many-body quantum geometric tensor can therefore be defined on the $(\theta,\phi)$ plane \cite{PhysRevB.62.1666,PhysRevResearch.1.032019,PhysRevX.14.011052,PhysRevResearch.7.023158}, and the associated quantum metric is characterized by the dimensionless trace
\begin{eqnarray}
    \chi^{-1}_{\rm mb}=\frac{1}{2\pi}\int_0^{2\pi}d\theta d\phi\ \text{Tr}\ \mathcal G\left(\theta,\phi\right).
\end{eqnarray}
It defines the intrinsic CHS length scale $\ell_{\mathcal F}\sim \chi_{\rm mb}\ell_B$, where $\ell_B$ is the magnetic length. More generally, $\ell_{\mathcal F}$ is the product of the smallest physical length scale (e.g., $\ell_B$ or the lattice constant, which is independent of the subspace) and a subspace-dependent factor $\chi$. This in turn sets the characteristic interaction energy scale, which for the Coulomb interaction is proportional to $1/\ell_{\mathcal F}$. For $\mathcal F_s$ constructed from the single-particle Hilbert space $\mathcal H_s$ of Bloch states, this description is fully equivalent to using the quantum metric of the Bloch states in the BZ. If $\mathcal H_s=\mathcal H_{LLL}$ consists only of the LLL, the fermionic vacuum in $\mathcal F_{LLL}$ is the state with the LLL completely filled (the highest-density state). If instead $\mathcal H_s=\mathcal H_{LLL+1LL}$ includes the two lowest LLs, the fermionic vacuum in $\mathcal F_{LLL+1LL}$ is the state in which both levels are completely filled. Since here $\ell_{\mathcal F_{LLL+1LL}}<\ell_{\mathcal F_{LLL}}$, electrons (or fermions in general) tend to fill the LLL first, where the interaction energy scale is smaller, before occupying the 1LL significantly.

For fermions, it is easy to show that both $\mathcal F_{LLL}$ and $\mathcal F_{LLL+1LL}$ (or, more generally, the Fock space of the lowest $N$ LLs) are CHS. This, however, does not hold for bosons. Although every single-particle orbital in $\mathcal H_s$ can still be filled using a permanent (instead of a Slater determinant), this configuration is not the highest-density state, since more bosons can always be added. Hence, no highest-density state exists at the single-particle level, and the lowest $N$ LLs do not form a CHS. In contrast to fermions, bosonic CHS can therefore only be constructed from many-body states [i.e., there are no bosonic CHS of the form of \eref{schs}], and it is from these that the bosonic interaction energy scale is determined. As we will see, this explains why bosons remain in the LLL even at very high filling factors.

The Fock space in \eref{schs} is spanned by product states and thus describes single-particle physics. For both fermionic and bosonic systems, \emph{many-body} CHS can be introduced as subspaces of \eref{schs}. Since interactions drive particles preferentially into the LLL, we now restrict \eref{schs} to $\mathcal F_s=\mathcal F_{LLL}$. This is particularly relevant for realistic systems, where the LLL forms a flat band with minimal kinetic energy. A prototypical many-body CHS is the zero-energy subspace of the Trugman--Kivelson interaction~\cite{PhysRevB.31.5280}
\begin{eqnarray}\label{tk}
    V\left(\mathbf r\right)=\nabla^n\delta^{\left(2\right)}\left(\mathbf r\right),
\end{eqnarray}
where $\mathbf r$ is the relative coordinate of the two particles, and the order of the derivative is $n=0$ for bosons and $n=2$ for fermions. Projected into the LLL, this interaction generates the Haldane pseudopotentials \cite{PhysRevLett.51.605} that serve as the model Hamiltonian for the Laughlin state at $\nu=1/2$ for bosons and $\nu=1/3$ for fermions \cite{PhysRevLett.50.1395}. The null space of \eref{tk} (i.e., the set of zero-energy states) is thus spanned by the Laughlin ground state and its quasihole states, forming a proper subspace of \eref{schs}. The vacuum of this many-body CHS is the Laughlin ground state, for which we can evaluate the quantum geometric tensor in the $(\theta,\phi)$ plane. As shown below, the resulting many-body quantum metric fixes the relevant interaction energy scale of the many-body CHS, and thereby indicates which many-body Hilbert space is energetically favored by the two-body interaction.

    {\textit{Hierarchy of many-body CHS--}} For both fermions and bosons, many-body CHS arise as subspaces of \eref{schs}, even though \eref{schs} itself defines a CHS only in the fermionic case. The hierarchical structure of different CHS is purely algebraic and has previously been used as a tool to understand the structure of, and the relationships between, different types of anyons \cite{PhysRevB.100.241302,S0217979222300031,PhysRevB.105.035144,Wang2023NatCommun}. Each CHS has a vacuum, namely its highest-density state, i.e., the state at the largest filling $\nu$. There is a notable difference between fermions and bosons: for fermions, all many-body CHS within the LLL have vacua with $\nu\le 1$, while there is no such restriction for bosons [see \fref{fig:3}(a)]. For each CHS, the many-body quantum metric of the vacuum can be readily computed and used as an estimate of the interaction energy scale within that CHS. This allows us to predict which CHS will be preferentially occupied as particles are added to the LLL.

\begin{figure}[!t]
    \centering
    \includegraphics[width=.45\textwidth]{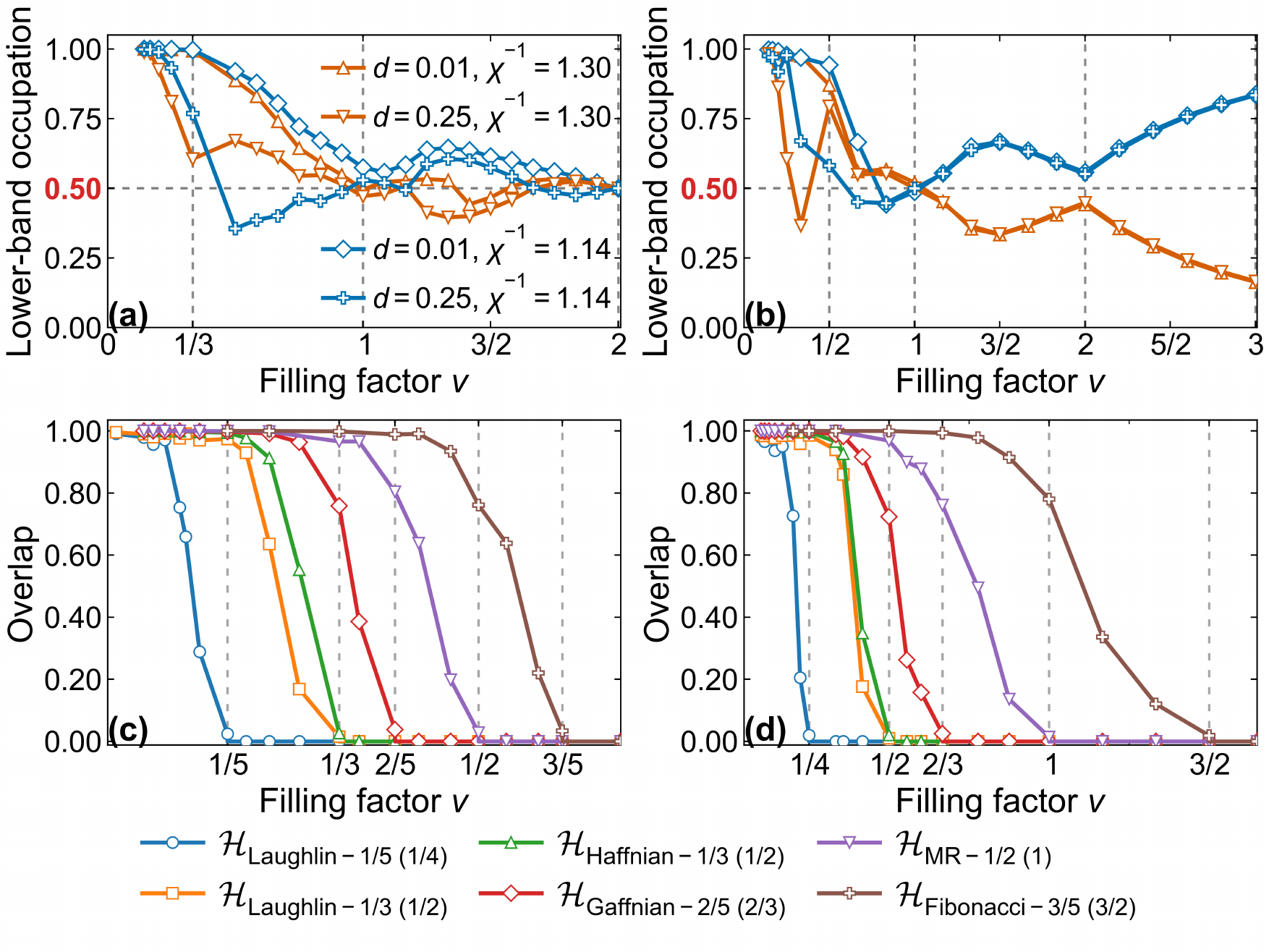}
    \caption{
(a) and (b) Occupation fraction of the lower-energy band in tMoTe$_2$ at twist angle $\vartheta=1.2^\circ$ for fermions and bosons, respectively, with screened Coulomb interactions in the strong-interaction limit, $\epsilon_{\rm int}/|\epsilon_g|\rightarrow\infty$. Here $d$ denotes the screening parameter (see Supplemental Material). The legends indicate the $\chi^{-1}$ value of the lower-energy band. (c) and (d) Total overlap of the lowest 200 eigenstates of the Coulomb interaction in tMoTe$_2$ ($\vartheta=1.2^\circ$, $\chi^{-1}=1.14$ band) with representative CHS as a function of filling factor for fermions and bosons, respectively. Different curves correspond to the indicated CHS.
}
    \label{fig:4}
\end{figure}

We evaluate the many-body quantum metric $\chi_{\rm mb}^{-1,(\alpha)}$ for the vacuum of each CHS $\mathcal H_\alpha$. Remarkably, the corresponding many-body quantum geometric tensor satisfies the ideal flat-band condition \cite{PhysRevB.90.165139,PhysRevLett.127.246403}: the trace of the many-body quantum metric exactly equals the many-body Berry curvature, or equivalently the Hall conductivity of the vacuum. Consequently, the integrated many-body quantum metric is directly determined by the vacuum filling factor, $\chi_{\rm mb}^{-1,(\alpha)}=\nu_{\rm vac}^{(\alpha)}$. A larger quantum metric corresponds to a shorter intrinsic length scale $\ell_{\mathcal F}$ and, consequently, to a larger characteristic Coulomb interaction scale $
    E_{\rm int}^{(\alpha)}
    \sim e^2/(\epsilon\ell_{\mathcal F}^{(\alpha)})$.
At a fixed filling factor $\nu$, the low-energy interaction physics is therefore governed primarily by many-body states in CHS whose vacuum filling satisfies $\nu_{\rm vac}^{(\alpha)} > \nu$. To test this, we compute, for various filling factors, the total overlap between the 200 lowest Coulomb eigenstates and the different CHS [\fref{fig:2}(c,d)]. In general, the Coulomb eigenstates at $\nu$ lie predominantly within CHS with $\nu_{\rm vac}^{(\alpha)}>\nu$, consistent with the interaction-energy hierarchy inferred from the many-body quantum metric. As $\nu$ approaches $\nu_{\rm vac}^{(\alpha)}$, the overlap decreases, reflecting the rapid reduction in the dimension of the corresponding CHS. The same general picture holds for both fermions and bosons: as particles are added to the LLL, physical two-body interactions fill up the CHS sequentially, in order of increasing $\nu_{\rm vac}^{(\alpha)}$.

    {\textit{Implications and discussion--}} It is worth emphasizing that, while many-body CHS can be realized as the zero-energy states of artificial few-body model Hamiltonians \cite{PhysRevB.75.075318,PhysRevB.75.195306,PhysRevLett.125.176402,S0217979222300031,PhysRevB.105.035144}, they can also be constructed algebraically even when no parent model
Hamiltonian is known \cite{PhysRevB.100.241302,Yang_2021}. It is thus more appropriate to regard the CHS as an intricate algebraic structure of Chern bands that requires no prior information about the dynamics. With realistic two-body interactions, however, this algebraic hierarchy of CHS also gives rise to an emergent hierarchy of \emph{interaction energy scales}. The immediate implication is that for particle filling $\nu<\nu_{\bar{\mathcal F}_s}$, where $\nu_{\bar{\mathcal F}_s}$ is the filling factor of the vacuum of the CHS $\bar{\mathcal F}_s$, the low-energy physics (both gapped and gapless) is well captured within $\bar{\mathcal F}_s$. Since the interaction energy scale of $\bar{\mathcal F}_s$ from the many-body quantum metric decreases with $\nu_{\bar{\mathcal F}_s}$, one can always choose the $\bar{\mathcal F}_s$ with the smallest $\nu_{\bar{\mathcal F}_s}>\nu$ that still captures the low-energy physics at filling factor $\nu$. It is worth noting that the dimension of the CHS also decreases with $\nu_{\bar{\mathcal F}_s}$.

This can already be exploited numerically: for example, instead of performing exact diagonalization in the full Hilbert space (i.e., the LLL), one can obtain the low-energy physics by diagonalizing within a much smaller $\bar{\mathcal F}_s$ \cite{PhysRevB.100.241302,Yang_2021}. More importantly, since the elementary anyonic excitations within $\bar{\mathcal F}_s$ (denoted $\bar{\mathcal F}_s$-anyons) are generally well known, the low-energy physics at $\nu$ can be understood in terms of quantum fluids of $\bar{\mathcal F}_s$-anyons \cite{PhysRevB.103.115102,S0217979222300031}. The many-body quantum metric and the interaction energy scales of the CHS thus explain the numerical observation that, with realistic interactions, the gapped excitations of the Laughlin phase live almost entirely within the Moore--Read CHS \cite{trung2026nonabelianbraidingabelianfractional}. This is the crucial ingredient that could enable non-Abelian braiding in Abelian topological phases \cite{trung2026nonabelianbraidingabelianfractional}.

The CHS formalism is also very useful for predicting band mixing by strong interactions in the presence of a finite band gap. For LLs, the band mixing is always suppressed, since both the kinetic energy and the interaction favor lower LLs. In more generic Chern bands, however, the band favored by the kinetic energy need not coincide with the band favored by the interaction, as can occur, for example, in valence-band moir\'e systems [e.g., \fref{fig:4}(a,b)]. We therefore focus on this latter situation. The analytic tools developed here generalize readily, since CHS can be constructed exactly in any Chern band (see Supplemental Material). Although CHS are exact null spaces of pseudopotential Hamiltonians only in IFB \cite{PhysRevLett.127.246403,PhysRevResearch.2.023237,MeraOzawa2021Kahler}, both realistic interactions and deviations from the IFB condition can be treated as perturbations, and the preference for different CHS within a single Chern band remains a general feature (see \fref{fig:4}). In particular, even when the interaction dominates and the two bands have similar quantum metric traces for the fully filled band (i.e., similar single-particle quantum metrics), strong band mixing does not always occur: it depends crucially on the particle density, or filling factor. At low filling factors, band mixing is strongly suppressed owing to the smaller many-body quantum metric trace of the relevant many-body CHS. In an IFB with a model pseudopotential two-body interaction, all particles remain exactly within the CHS of a single band (as null states), with no band mixing at all. Deviations from the IFB condition or from the pseudopotential interaction (e.g., a longer-range screened Coulomb interaction) act as perturbations; the stronger the perturbation, the more band mixing emerges. We expect this to be a generic feature of Chern bands; further examples are given in the Supplemental Material.

{\sl Acknowledgements.} C.-X. Jiang thanks Bo Peng and Yuzhu Wang for fruitful discussions. This work is supported by Singapore Ministry of Education (MOE) Academic Research Fund Tier 3 Grant (No. MOE-MOET32023-0003) ``Quantum Geometric Advantage'', and Singapore Ministry of Education (MOE) Academic Research Fund Tier 2 Grant (No. MOE-T2EP50124-0017).  Z.-X. Hu is supported by the National Natural Science Foundation of China Grant (No.12474140 and No.12547101) and the Fundamental Research Funds for the Central Universities Grant (No. 2025CDJ-IAISYB-029). C.-X. Jiang acknowledges the support of the China Scholarship Council Grant (No. 202406050101).
\bibliography{ref.bib}

\end{document}


\title{Supplementary Information for ``Emergent low-energy many-body Hilbert spaces in Chern bands''}
 
\author{Chen-Xin Jiang}
\affiliation{Department of Physics and Chongqing Key Laboratory for Strongly Coupled Physics, Chongqing University, Chongqing 401331, People's Republic of China}
\affiliation{Division of Physics and Applied Physics, Nanyang Technological University, Singapore 637371, Singapore}
\author{Zi-Xiang Hu\thanks{*Corresponding authors}}
\email{zxhu@cqu.edu.cn}
\affiliation{Department of Physics and Chongqing Key Laboratory for Strongly Coupled Physics, Chongqing University, Chongqing 401331, People's Republic of China}
\author{Bo Yang\thanks{*Corresponding authors}}
\email{yang.bo@ntu.edu.sg}
\affiliation{Division of Physics and Applied Physics, Nanyang Technological University, Singapore 637371, Singapore}

\date{\today}

\maketitle

This Supplementary Material provides details regarding the theoretical models, numerical methods, and additional results complementing the main text. We first introduce the single-particle continuum Hamiltonians for twisted bilayer graphene (TBG) and twisted bilayer MoTe$_2$ (tMoTe$_2$), along with the parent Hamiltonians that define conformal Hilbert spaces (CHS) and the screened Coulomb interaction potentials adopted in our calculations. We then present a rigorous derivation of the many-body quantum geometric tensor (QGT) on a torus under twisted boundary conditions, including explicit evaluations of the many-body quantum metric and many-body Chern number. We further elaborate on the mapping protocol that transfers CHS from the lowest Landau level (LLL) to generic Chern bands.
We provide extensive numerical evidence verifying the hierarchical structure of CHS by quantifying wave-function overlaps between low-energy many-body eigenstates and canonical CHS basis states. Our analysis covers LLL, first Landau level (1LL), TBG, and tMoTe$_2$ systems for both fermionic and bosonic configurations under diverse interaction strengths and screening conditions. Finite-size scaling analyses are performed to characterize band-mixing behaviors in LLL–1LL, TBG, and tMoTe$_2$ systems via band occupation statistics and excitation gap scaling. Finally, we investigate interaction-driven band mixing in the strong-coupling limit $\epsilon_{\rm int}/|\epsilon_g|\rightarrow\infty$ at low fractional fillings for interactions with varying spatial ranges.

\section{Continuum models for moir\'e systems}
\paragraph{Twisted bilayer graphene.}We describe TBG using the Bistritzer--MacDonald continuum model \cite{pnas.1108174108}. Since we focus on single-component  CHS and interaction-driven band mixing, we neglect the spin degree of freedom and consider a single valley. Specifically, we take the graphene valley centered at
$\mathbf K=(4\pi/3a,0)$, where $a\simeq2.46\,\text{\AA}$ is the graphene lattice constant. We introduce the four-component annihilation-operator spinor
$\Psi(\mathbf{k})=[\psi_b(\mathbf k),\psi_t(\mathbf k)]^{T}$, with
$\psi_{\lambda}=(\psi_{A_\lambda},\psi_{B_\lambda})^{T}$,
where \(\lambda=b,t\) labels the bottom and top layers, respectively, and \(A,B\) denote the two sublattices, the single-particle Hamiltonian is given by \cite{PhysRevLett.124.106803}
\begin{equation}
\begin{aligned}
H_{\mathrm{TBG}}
=&\sum_{\mathbf{k}}\psi^\dagger_t(\mathbf{k})h_{-\frac{\vartheta}{2}}(\mathbf{k}-\mathbf{K}^t)\psi_t(\mathbf{k})+\sum_{\mathbf{k}}\psi^\dagger_b(\mathbf{k})h_{\frac{\vartheta}{2}}(\mathbf{k}-\mathbf{K}^b)\psi_b(\mathbf{k})\\
&+\sum_{\mathbf{k}}\sum_{j=0}^2\left[\psi^\dagger_t(\mathbf{k}-\mathbf{g}_0+\mathbf{g}_j)T_j\psi_b(\mathbf{k})+{\rm h.c.}\right]+M\sum_{\mathbf{k},\lambda}\psi_\lambda^\dagger(\mathbf{k})\sigma_z\psi_\lambda(\mathbf{k}),
\end{aligned}
\label{eq:TBG}
\end{equation}
where $\mathbf{K}^t=\frac{4\pi}{3a}\left(\cos\frac{\vartheta}{2},\sin\frac{\vartheta}{2}\right)$, $\mathbf{K}^b=\frac{4\pi}{3a}\left(\cos\frac{\vartheta}{2},-\sin\frac{\vartheta}{2}\right)$, and the intralayer Dirac Hamiltonian is $h_{\vartheta}(\mathbf{k})=\hbar v_F\mathbf{k}\cdot\boldsymbol{\sigma}_{\vartheta}$, with $\boldsymbol{\sigma}_{\vartheta}=e^{-{\rm i}\vartheta\sigma_z/2}
\boldsymbol{\sigma}
e^{{\rm i}\vartheta\sigma_z/2}$ and \(\boldsymbol{\sigma}=(\sigma_x,\sigma_y)\) acts on the sublattice degree of freedom. The Fermi velocity is given by
$\hbar v_F=\frac{\sqrt{3}}{2}a t_0$, where $t_0=2.62\,\mathrm{eV}$ is the nearest-neighbor hopping amplitude. The last term describes a sublattice-staggered potential induced by aligned
hexagonal boron nitride (hBN) \cite{Long2022}. The interlayer moir\'e hopping takes the form \cite{PhysRevLett.122.106405}
\begin{equation}
T_j
=
w_{AA}\sigma_0
-w_{AB}\cos(j\phi)\sigma_x
+w_{AB}\sin(j\phi)\sigma_y,
\qquad
\phi=\frac{2\pi}{3},
\label{eq:TBG_Tj}
\end{equation}
where $w_{AA}$ and $w_{AB}$ denote the tunneling amplitudes in the AA and AB/BA stacking regions, respectively. We work in the chiral limit by setting $w_{AA}=0$,
while retaining \(w_{AB}=110\,\mathrm{meV}\). The three interlayer momentum-transfer vectors are
\begin{equation}
\begin{aligned}
\mathbf g_0
&=\frac{8\pi}{3a}\sin\frac{\vartheta}{2}(0,-1),\\
\mathbf g_1
&=\frac{8\pi}{3a}\sin\frac{\vartheta}{2}
\left(\frac{\sqrt{3}}{2},\frac{1}{2}\right),\\
\mathbf g_2
&=\frac{8\pi}{3a}\sin\frac{\vartheta}{2}
\left(-\frac{\sqrt{3}}{2},\frac{1}{2}\right).
\end{aligned}
\end{equation}
The reciprocal vectors of the moiré superlattice can be chosen as
\begin{equation}
\mathbf G_{1,2}
=
\frac{8\pi}{\sqrt{3}a}\sin\left(\frac{\vartheta}{2}\right)
\left(\frac{1}{2},\pm\frac{\sqrt{3}}{2}\right).
\end{equation}

\begin{figure}[!ht]
    \centering
    \includegraphics[width=.8\textwidth]{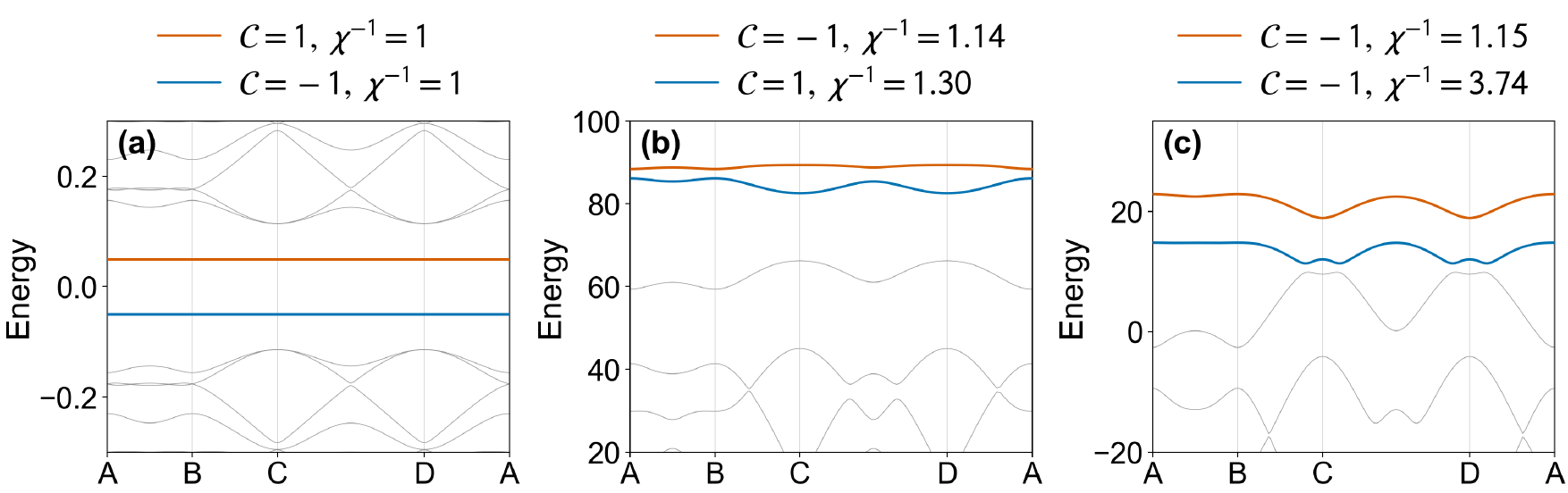}
    \caption{
 band structures along the momentum-space path
$A\rightarrow B\rightarrow C\rightarrow D\rightarrow A$.
Here,
$A=(\mathbf{b}_1-\mathbf{b}_2)/3$,
$B=\mathbf{0}$,
$C=(\mathbf{b}_2-\mathbf{b}_1)/3$, and
$D=(2\mathbf{b}_1+\mathbf{b}_2)/3$, where $\mathbf{b}_1$ and
$\mathbf{b}_2$ are the reciprocal vectors of the
moir\'e Brillouin zone.
(a) TBG at $\vartheta=1.132^\circ$, with the two flat bands
highlighted.
(b) tMoTe$_2$ at
$\vartheta=1.2^\circ$.
(c) tMoTe$_2$ at
$\vartheta=2^\circ$.
In (b) and (c), the two bands considered in our calculations are
highlighted. The upper and lower bands are shown in orange and blue,
respectively, and their Chern number $\mathcal C$ and total trace of quantum metric $\chi^{-1}$ are specified in the corresponding panels. The
remaining bands are shown in gray.
}
    \label{fig:S101}
\end{figure}

We set $M=50\,\mathrm{meV}$ and focus on the first magic angle, $\vartheta=1.132^\circ$, where the two low-energy bands become exactly flat in the ideal chiral limit and carry opposite Chern numbers $C=\pm1$ (see \fref{fig:S101}(a)). For numerical calculations, we truncate the plane-wave basis to
$\mathbf{k}+m\mathbf{G}_1+n\mathbf{G}_2$, with $m,n=-d,\ldots,d$.

\paragraph{Twisted MoTe$_2$.}
For tMoTe$_2$, we adopt the continuum model of Ref.~\cite{PhysRevLett.122.086402} describing the moir\'e valence bands. Owing to the strong spin-orbit coupling and spin-valley locking, we restrict ourselves to a single valley and its corresponding spin sector. In the layer basis
$\Psi(\mathbf{k})=[\psi_b(\mathbf{k}),\psi_t(\mathbf{k})]^T$,
the single-particle Hamiltonian is written as
\begin{equation}
\begin{aligned}
H_{\mathrm{tMoTe_2}}
=&
-\frac{\hbar^2}{2m^*}
\sum_{\mathbf{k}}
\left[
\psi_t^\dagger(\mathbf{k})
\left|\mathbf{k}-\mathbf K^{\prime t}\right|^2
\psi_t(\mathbf{k})
+
\psi_b^\dagger(\mathbf{k})
\left|\mathbf{k}-\mathbf{K}^{\prime b}\right|^2
\psi_b(\mathbf{k})
\right]
\\
&+
\mathcal{V}\sum_{\mathbf{k}}\sum_{j=1,3,5}
\left[
e^{{\rm i}\varphi}\psi_t^\dagger(\mathbf{k}+\mathbf{G}^\prime_j)
\psi_t(\mathbf{k})
+
e^{-{\rm i}\varphi}\psi_b^\dagger(\mathbf{k}+\mathbf{G}^\prime_j)
\psi_b(\mathbf{k})
+\mathrm{h.c.}
\right]
\\
&+
\mathcal{W}\sum_{\mathbf{k}}\sum_{j=0}^{2}
\left[
\psi_t^\dagger(\mathbf{k}+\mathbf{g}^\prime_j)
\psi_b(\mathbf{k})
+\mathrm{h.c.}
\right].
\end{aligned}
\label{eq:tMoTe2}
\end{equation}
where $\mathbf{K}^{\prime t}=\frac{4\pi}{3a^\prime}\left(\cos\frac{\vartheta}{2},\sin\frac{\vartheta}{2}\right)$, $\mathbf{K}^{\prime b}=\frac{4\pi}{3a^\prime}\left(\cos\frac{\vartheta}{2},-\sin\frac{\vartheta}{2}\right)$, with the monolayer lattice constant $a^\prime=3.52\ {\rm \AA}$. The moir\'e reciprocal vectors take the form
\begin{equation}
    \mathbf{G}^{\prime}_j=\frac{8\pi\sin\frac{\vartheta}{2}}{\sqrt{3}a^\prime}(\cos\frac{\pi(j-1)}{3},\sin\frac{\pi(j-1)}{3}).
\end{equation}
The three interlayer momentum-transfer vectors are
\begin{equation}
\begin{aligned}
\mathbf g^\prime_0
&=\frac{8\pi}{3a^\prime}\sin\frac{\vartheta}{2}(0,-1),\\
\mathbf g^\prime_1
&=\frac{8\pi}{3a^\prime}\sin\frac{\vartheta}{2}
\left(\frac{\sqrt{3}}{2},\frac{1}{2}\right),\\
\mathbf g^\prime_2
&=\frac{8\pi}{3a^\prime}\sin\frac{\vartheta}{2}
\left(-\frac{\sqrt{3}}{2},\frac{1}{2}\right).
\end{aligned}
\end{equation}
Here $m^*$ is the effective mass of electron, $\mathcal W$ is the interlayer
tunneling amplitude, $\mathcal V$ and $\varphi$ characterize the amplitude and shape of the intralayer moir\'e potential, respectively. We use the parameter set $(\mathcal V,\varphi,\mathcal W,m^*)=(8\,\mathrm{meV},-89.6^\circ,-8.5\,\mathrm{meV},0.62m_0)$ where $m_0$ is the mass of electron, and consider twist angles
$\vartheta=1.2^\circ$ and $2^\circ$ (see \fref{fig:S101}(b,c)). For numerical calculations, we truncate the plane-wave basis to
$\mathbf{k}+m\mathbf{G}^\prime_2+n\mathbf{G}^\prime_0$, with $m,n=-d,\ldots,d$.

\section{Two-body interactions in multiband systems}

The two-body interaction can be written as
\begin{equation}
\hat{H}_{\rm int}=\frac{1}{2}\sum_{\{\mathbf{k}_i\}}V^{l_1,l_2,l_3,l_4}_{\mathbf{k}_1,\mathbf{k}_2,\mathbf{k}_3,\mathbf{k}_4}c^\dagger_{l_1,\mathbf{k}_1}c^\dagger_{l_2,\mathbf{k}_2}c_{l_4,\mathbf{k}_4}c_{l_3,\mathbf{k}_3},
\end{equation}
where $l_i$ denotes the band index. Using a gauge-invariant Bloch-state formulation, $V^{l_1,l_2,l_3,l_4}_{\mathbf{k}_1,\mathbf{k}_2,\mathbf{k}_3,\mathbf{k}_4}$ for landau levels (LLs) can be written as \cite{PhysRevLett.127.246403}
\begin{equation}
    V_{\mathbf{k}_1,\mathbf{k}_2,\mathbf{k}_3,\mathbf{k}_4}^{l_1,l_2,l_3,l_4}=\sum_{\mathbf{b}}U(\mathbf{k}_1-\mathbf{k}_3-\mathbf{b})f_{\mathbf{b}}^{l_1,\mathbf{k}_1,l_3,\mathbf{k}_3}f_{-\mathbf{b}+\delta\mathbf{b}}^{l_2,\mathbf{k}_2,l_4,\mathbf{k}_4},
\end{equation}
where
$\mathbf b=m\mathbf b_1+n\mathbf b_2$,
with $\mathbf b_1$ and $\mathbf b_2$ the primitive reciprocal-lattice vectors,
$\delta\mathbf b=\mathbf k_1+\mathbf k_2-\mathbf k_3-\mathbf k_4$, $U(\mathbf q)$ is the Fourier transform of the interaction
$U(\mathbf r_1-\mathbf r_2)$, and $f^{l_1,\mathbf{k}_1,l_2,\mathbf{k}_2}_{\mathbf{q}}$ is the form factor 
\begin{equation}
    f^{l_1,\mathbf{k}_1,l_2,\mathbf{k}_2}_{\mathbf{q}}=F_{l_1,l_2}(\mathbf{q})\langle\mathbf{k}_1|e^{{\rm i}\mathbf{q}\cdot\mathbf{\bar{R}}}|\mathbf{k}_2\rangle
    =F_{l_1,l_2}(\mathbf{q})\eta_{\mathbf{b}} e^{\frac{\rm i}{2}\mathbf{k}_1\times\mathbf{k}_2}e^{\frac{\rm i}{2}(\mathbf{k}_1+\mathbf{k}_2)\times\mathbf{b}}e^{-\frac{1}{4}|\mathbf{q}|^2}\delta_{\mathbf{q},\mathbf{k}_1-\mathbf{k}_2-\mathbf{b}}.
\end{equation}
where $\mathbf{\bar{R}}$ is the guiding center and $\eta_{\mathbf{b}}=1$ if $\mathbf{b}/2$ is a reciprocal lattice vector and $-1$ otherwise. $F_{l_1,l_2}(\mathbf{q})$ is the LL form factor, which can be expressed as \cite{an2025fractional}
\begin{equation}
    \begin{aligned}
        F_{l_1,l_2}(\mathbf{q})=  \sqrt{\frac{\min(l_1,l_2)!}{\max(l_1,l_2)!}}\left[\frac{{\rm sgn}(l_1-l_2)q_y-iq_x}{\sqrt{2}}\right]^{|l_1-l_2|} L^{|l_1-l_2|}_{\min(l_1,l_2)}(q^2/2),
    \end{aligned}
\end{equation}
where ${\rm sgn}(x)$ is the sign function, $L_n^m (x)$ is the Laguerre polynomial.
The Coulomb interaction can be written as
\begin{equation}
    U_{\rm coulomb}(\mathbf{q})=\frac{e^2}{4\pi\varepsilon S}\frac{2\pi}{q},
    \label{eq:3}
\end{equation}
where $\varepsilon$ is the permittivity of the material and $S$ is the real-space area of the 2D system. For Landau levels, one has $S = 2\pi N_{\phi} \ell_B^2$, where $N_{\phi}$ is the number of flux quanta and $\ell_B$ is the magnetic length. To facilitate comparison across different models, we set $\ell_B=1$ and rescale the unit-cell area of all models to $2\pi N_{\phi}$. To account for Coulomb screening arising from realistic dielectric
environments and gate geometries, we consider two commonly used forms
of screened Coulomb interaction: the dual-gate-screened Coulomb
interaction and the Yukawa interaction
\cite{PhysRevResearch.3.L032070,PhysRevLett.124.106803},
\begin{equation}
    U_{\mathrm{gate}}(q)
    =\frac{e^2}{2\epsilon S q}\tanh(qd),
    \qquad
    U_{\mathrm{Y}}(q)
    =\frac{e^2}
    {2\epsilon S\sqrt{q^2+\kappa^2}},
    \label{eq:screen}
\end{equation}
where $d$ is the the separation between the electrode and
Moir\'e superlattice, while $\kappa$ measures the screening strength. 
For other continuum models, such as TBG or tMoTe$_2$, the interaction can be written as
\begin{equation}
    V_{\mathbf{k}_1,\mathbf{k}_2,\mathbf{k}_3,\mathbf{k}_4}^{l_1,l_2,l_3,l_4}=\sum_{\mathbf{q}}U(\mathbf{q})\mathcal{F}_{\mathbf{q}}^{l_1,\mathbf{k}_1,l_3,\mathbf{k}_3}\mathcal{F}_{-\mathbf{q}}^{l_2,\mathbf{k}_2,l_4,\mathbf{k}_4}.
\end{equation}
where the form factor can be expressed as\cite{PhysRevLett.124.106803}
\begin{equation}
        \mathcal{F}^{l_1,\mathbf{k}_1,l_2,\mathbf{k}_2}_{\mathbf{q}}=\sum_{\alpha}\sum_{\{m_i,n_i\}=-d}^{d}\delta_{\mathbf{k}_1-\mathbf{k}_2+\mathbf{G}_1-\mathbf{G}_2,\mathbf{q}} u^*_{l_1,\alpha,m_1,n_1}(\mathbf{k}_1)u_{l_2,\alpha,m_2,n_2}(\mathbf{k}_2),
\end{equation}
where $\alpha$ is the sublattice index or layer index, $u_{l,\alpha,m,n}(\mathbf{k})$ denotes the $\alpha$ component of the Bloch eigenvector of the $l$-th band obtained from \eref{eq:TBG} or \eref{eq:tMoTe2}, and $d$ is the cutoff for the number of reciprocal lattice vectors. By
introducing $\mathbf b=\mathbf G_1-\mathbf G_2$, the form factor can be rewritten in a form
analogous to that of the LL $\mathcal{F}^{l_1,\mathbf{k}_1,l_2,\mathbf{k}_2}_{\mathbf{b}}$.

\section{Single-band parent Hamiltonians}

The representative CHS considered in this work can be generated as the
null subspaces (i.e., the sets of zero-energy states) of model
pseudopotential Hamiltonians. We denote by
$\hat{V}_{m,F}^{(n)}$ and $\hat{V}_{m,B}^{(n)}$ the fermionic and bosonic
$n$-body pseudopotentials, respectively, in the relative-angular-momentum
channel $m$. For two-body interactions, the corresponding momentum-space
pseudopotential on the torus depends on a single transferred momentum
$\mathbf q$, whereas higher-body pseudopotentials generally depend on
multiple independent momenta. The model pseudopotentials used in this work
are summarized in \tref{tab:pseudopotentials}.

For the Laughlin CHS, the fermionic vacua at
$\nu_{\rm vac}=1/3$ and $1/5$ are generated by
$\hat{V}_{1,F}^{(2)}$ and
$\hat{V}_{1,F}^{(2)}+\hat V_{3,F}^{(2)}$, respectively, while the bosonic vacua at
$\nu_{\rm vac}=1/2$ and $1/4$ are generated by
$\hat V_{0,B}^{(2)}$ and
$\hat V_{0,B}^{(2)}+\hat V_{2,B}^{(2)}$, respectively
\cite{PhysRevLett.51.605,PhysRevB.31.5280}.
For the Moore--Read CHS, the bosonic and fermionic states are generated by
$\hat V_{0,B}^{(3)}$ and $\hat V_{3,F}^{(3)}$, respectively
\cite{PhysRevB.54.16864}.
The Gaffnian CHS are generated by
$\hat V_{0,B}^{(3)}+\hat V_{2,B}^{(3)}$ for bosons and
$\hat V_{3,F}^{(3)}+\hat V_{5,F}^{(3)}$ for fermions
\cite{PhysRevB.75.075317}.
For the Haffnian CHS, the corresponding parent Hamiltonians are
$\hat V_{0,B}^{(3)}+\hat V_{2,B}^{(3)}+\hat V_{3,B}^{(3)}$ for bosons and
$\hat V_{3,F}^{(3)}+\hat V_{5,F}^{(3)}+\hat V_{6,F}^{(3)}$ for fermions
\cite{PhysRevB.83.241302}.
Finally, the Fibonacci ($\mathbb Z_3$ Read--Rezayi) CHS are generated by
the shortest-range four-body pseudopotentials,
$\hat V_{0,B}^{(4)}$ for bosons and $\hat V_{6,F}^{(4)}$ for fermions
\cite{PhysRevB.59.8084}.

We next construct the three- and four-body interaction matrix elements projected onto the LLL. The three-body matrix element can be written as \cite{1zg9-qbd6}
\begin{equation}
V_{\mathbf k_1\mathbf k_2\mathbf k_3
\mathbf k_4\mathbf k_5\mathbf k_6}
=
\frac{1}{S^2}
\sum_{\mathbf b,\mathbf b^\prime}
V(\mathbf{k}_1-\mathbf{k}_4-\mathbf{b},\mathbf k_2-\mathbf k_5-\mathbf b^\prime)\,
f^{\mathbf k_1,\mathbf k_4}_{\mathbf b}
f^{\mathbf k_2,\mathbf k_5}_{\mathbf b^\prime}
f^{\mathbf k_3,\mathbf k_6}_
{\delta \mathbf{b}^\prime-\mathbf{b}-\mathbf{b}^\prime},
\label{eq:three_body_matrix}
\end{equation}
where $\delta\mathbf{b}^\prime=\mathbf{k}_1+\mathbf{k}_2+\mathbf{k}_3-\mathbf{k}_4-\mathbf{k}_5-\mathbf{k}_6$ and $f^{0,\mathbf{k}_1,0,\mathbf{k}_2}_\mathbf{b}\equiv f^{\mathbf{k}_1,\mathbf{k}_2}_\mathbf{b}$. 
Similarly, the four-body matrix element is
\begin{equation}
\begin{aligned}
V_{\mathbf k_1\mathbf k_2\mathbf k_3\mathbf k_4
\mathbf k_5\mathbf k_6\mathbf k_7\mathbf k_8}
=
\frac{1}{S^3}
\sum_{\mathbf b,\mathbf b^\prime,\mathbf{b}^{\prime\prime}}
V(\mathbf{k}_1-\mathbf{k}_5-\mathbf{b},\mathbf k_2-\mathbf k_6-\mathbf b^\prime,\mathbf k_3-\mathbf k_7-\mathbf b^{\prime\prime})\,
f^{\mathbf k_1,\mathbf k_5}_{\mathbf b}
f^{\mathbf k_2,\mathbf k_6}_{\mathbf b^\prime}
f^{\mathbf k_3,\mathbf k_7}_{\mathbf{b}^{\prime\prime}}
f^{\mathbf k_4,\mathbf k_8}_
{\delta \mathbf{b}^{\prime\prime}-\mathbf{b}-\mathbf{b}^\prime-\mathbf{b}^{\prime\prime}},
\end{aligned}
\label{eq:four_body_matrix}
\end{equation}
where $\delta \mathbf{b}^{\prime\prime}=\mathbf{k}_1+\mathbf{k}_2+\mathbf{k}_3+\mathbf{k}_4-\mathbf{k}_5-\mathbf{k}_6-\mathbf{k}_7-\mathbf{k}_8$. The corresponding three- and four-body interaction Hamiltonians can be written as

\begin{equation}
\hat{H}_{\rm 3-body}=\frac{1}{6}\sum_{\{\mathbf{k}_i\}}V_{\mathbf{k}_1,\mathbf{k}_2,\mathbf{k}_3,\mathbf{k}_4,\mathbf{k}_5,\mathbf{k}_6}c^\dagger_{\mathbf{k}_1}c^\dagger_{\mathbf{k}_2}c^\dagger_{\mathbf{k}_3}c_{\mathbf{k}_6}c_{\mathbf{k}_5}c_{\mathbf{k}_4},
\end{equation}
\begin{equation}
\hat{H}_{\rm 4-body}=\frac{1}{24}\sum_{\{\mathbf{k}_i\}}V_{\mathbf{k}_1,\mathbf{k}_2,\mathbf{k}_3,\mathbf{k}_4,\mathbf{k}_5,\mathbf{k}_6,\mathbf{k}_7,\mathbf{k}_8}c^\dagger_{\mathbf{k}_1}c^\dagger_{\mathbf{k}_2}c^\dagger_{\mathbf{k}_3}c^\dagger_{\mathbf{k}_4}c_{\mathbf{k}_8}c_{\mathbf{k}_7}c_{\mathbf{k}_6}c_{\mathbf{k}_5},
\end{equation}
where $c_{0,\mathbf{k}}\equiv c_\mathbf{k}$.

\begin{table}[t]
\centering
\caption{
Momentum-space forms of the pseudopotentials used in this work.
Overall normalization factors are omitted. Here $L_m$ denotes the
Laguerre polynomial, $x_1=\frac14|\mathbf q_1-\mathbf q_2|^2$ and
$x_2=\frac34|\mathbf q_1+\mathbf q_2|^2$.
}
\label{tab:pseudopotentials}
\begin{tabular}{c c}
\hline
Pseudopotential & Momentum-space form \\
\hline

$V_{m,F/B}^{(2)}(\mathbf{q})$
&
$L_m(|\mathbf{q}|^2)$
\\[4pt]

$V_{0,B}^{(3)}(\mathbf{q}_1,\mathbf{q}_2)$
&
$1$
\\[4pt]

$V_{2,B}^{(3)}(\mathbf{q}_1,\mathbf{q}_2)$
&
$L_2(x_1)+L_1(x_1)L_1(x_2)+L_2(x_2)$
\\[4pt]

$V_{3,B}^{(3)}(\mathbf{q}_1,\mathbf{q}_2)$
&
$L_3(x_1)+L_2(x_1)L_1(x_2)+L_1(x_1)L_2(x_2)+L_3(x_2)$
\\[4pt]

$V_{3,F}^{(3)}(\mathbf{q}_1,\mathbf{q}_2)$
&
$|\mathbf q_1|^2|\mathbf q_2|^2|\mathbf q_1-\mathbf q_2|^2$
\\[4pt]

$V_{5,F}^{(3)}(\mathbf{q}_1,\mathbf{q}_2)$
&
$|\mathbf q_1|^2|\mathbf q_2|^2|\mathbf q_1-\mathbf q_2|^2
\left[L_2(x_1)+L_1(x_1)L_1(x_2)+L_2(x_2)\right]$
\\[4pt]

$V_{6,F}^{(3)}(\mathbf{q}_1,\mathbf{q}_2)$
&
$|\mathbf q_1|^2|\mathbf q_2|^2|\mathbf q_1-\mathbf q_2|^2
\left[L_3(x_1)+L_2(x_1)L_1(x_2)+L_1(x_1)L_2(x_2)+L_3(x_2)\right]$
\\[4pt]

$V_{0,B}^{(4)}(\mathbf{q}_1,\mathbf{q}_2,\mathbf{q}_3)$
&
$1$
\\[4pt]

$V_{6,F}^{(4)}(\mathbf{q}_1,\mathbf{q}_2,\mathbf{q}_3)$
&
$|\mathbf q_1|^2|\mathbf q_2|^2|\mathbf q_3|^2
|\mathbf q_1-\mathbf q_2|^2
|\mathbf q_1-\mathbf q_3|^2
|\mathbf q_2-\mathbf q_3|^2$
\\

\hline
\end{tabular}
\end{table}

\section{Many-body quantum geometric tensor}

In this section, we provide a detailed evaluation of the many-body QGT on a torus using twisted boundary conditions. For an $n_p$-particle wave function
$\Psi(\mathbf r_1,\ldots,\mathbf r_{n_p})$, we introduce two twist angles
$(\theta,\phi)\in[0,2\pi)^2$ along the two directions of the torus \cite{PhysRevB.31.3372},
\begin{equation}
\begin{aligned}
    \Psi(\mathbf r_1,\ldots,\mathbf r_i+\mathbf L_1,\ldots,\mathbf r_{n_p})
=&
e^{{\rm i}\theta}
\Psi(\mathbf r_1,\ldots,\mathbf r_i,\ldots,\mathbf r_{n_p}),
\\
\Psi(\mathbf r_1,\ldots,\mathbf r_i+\mathbf L_2,\ldots,\mathbf r_{n_p})
=&
e^{{\rm i}\phi}
\Psi(\mathbf r_1,\ldots,\mathbf r_i,\ldots,\mathbf r_{n_p}),
\end{aligned}
\end{equation}
where $\mathbf L_1$ and $\mathbf L_2$ are the two period vectors defining the torus.
In the Bloch-state representation, the twisted boundary conditions are
equivalent to shifting every single-particle crystal momentum according to
\begin{equation}
\mathbf k_\mathbf{m}
\rightarrow
\mathbf k_\mathbf{m}(\theta,\phi)=\mathbf k_\mathbf{m}
+
\frac{\theta}{2\pi N_1}\mathbf b_1
+
\frac{\phi}{2\pi N_2}\mathbf b_2,
\end{equation}
where $\mathbf m=(m_1,m_2)$ denotes the discrete momentum index, 
$N_1$ and $N_2$ denote the number of unit cells along the two torus directions. The discrete crystal momenta are
\begin{equation}
    \mathbf{k}_\mathbf{m}=\frac{m_1}{N_1}\mathbf{b}_1+\frac{m_2}{N_2}\mathbf{b}_2,\qquad
m_i=0,\ldots,N_i-1.
\label{eq:momentum1}
\end{equation}
Under the twisted boundary conditions, the single-particle Bloch states become $
|\psi_{l,\mathbf m}(\theta,\phi)\rangle$, and form an
orthonormal basis at each twist point.
We define the single-particle overlap between two twist points
$\boldsymbol{\Omega}=(\theta,\phi)$ and
$\boldsymbol{\Omega}'=(\theta',\phi')$ as
\begin{equation}
\mathcal S_{\alpha,\alpha^\prime}
(\boldsymbol{\Omega},\boldsymbol{\Omega}^\prime)
=
\left\langle
\psi_{\alpha}(\boldsymbol{\Omega})
\middle|
\psi_{\alpha^\prime}(\boldsymbol{\Omega}^\prime)
\right\rangle,
\end{equation}
where $\alpha=(l,\mathbf{m})$.
A subtle point is that Bloch states at different twist angles generally carry different crystal momenta. A direct spatial overlap of the full Bloch wave functions would therefore include the overlap of plane-wave factors with different momenta and can vanish trivially. To compare states at different twist angles consistently, we first remove the twist-dependent plane-wave phase so that the resulting states obey the same periodic boundary conditions. For Bloch states, this amounts to evaluating the overlap between their cell-periodic parts. Writing the Bloch state at twist $\boldsymbol{\Omega}$ as
\begin{equation}
\psi_{l,\mathbf m}(\boldsymbol{\Omega};\mathbf r)
=
e^{{\rm i}\mathbf k_{\mathbf m}(\boldsymbol{\Omega})\cdot\mathbf r}
u_{l,\mathbf m}(\boldsymbol{\Omega};\mathbf r),
\end{equation}
the single-particle overlap between two twist points
$\boldsymbol{\Omega}$ and $\boldsymbol{\Omega}'$ is evaluated from the
cell-periodic parts. Therefore, it takes the form
\begin{equation}
\mathcal S_{\alpha,\alpha'}
(\boldsymbol{\Omega},\boldsymbol{\Omega}')
=
\delta_{\mathbf m,\mathbf m'}
\left\langle
u_{l,\mathbf m}(\boldsymbol{\Omega})
\middle|
u_{l',\mathbf m'}(\boldsymbol{\Omega}')
\right\rangle=
\delta_{\mathbf m,\mathbf m'}
f_{\mathbf 0}^{\,l,\mathbf{k}_\mathbf{m}(\mathbf \Omega),
l',\mathbf{k}_\mathbf{m^\prime}(\mathbf \Omega^\prime)}.
\end{equation}

We now turn to the many-body overlap. For a fermionic many-body basis state specified by a set of occupied single-particle orbitals
$I=\{\alpha_1,\ldots,\alpha_{n_p}\}$, the corresponding Fock state at twist
$\boldsymbol{\Omega}$ is denoted by
\begin{equation}
|I;\boldsymbol{\Omega}\rangle_F
=
C_{\alpha_1}^\dagger(\boldsymbol{\Omega})
C_{\alpha_2}^\dagger(\boldsymbol{\Omega})
\cdots
C_{\alpha_{n_p}}^\dagger(\boldsymbol{\Omega})
|0\rangle,
\end{equation}
where \(C_\alpha^\dagger(\boldsymbol{\Omega})\) is the fermionic creation operator associated with the single-particle orbital \(\alpha\) at twist \(\boldsymbol{\Omega}\). 
The overlap between two such basis states at
$\boldsymbol{\Omega}$ and $\boldsymbol{\Omega}'$ is given by the determinant of the single-particle overlap matrix,
\begin{equation}
{}_F\langle I;\boldsymbol{\Omega}
|
J;\boldsymbol{\Omega}'\rangle_F
=
\det
\begin{pmatrix}
\mathcal S_{\alpha_1,\beta_1}(\boldsymbol{\Omega},\boldsymbol{\Omega}') &
\cdots &
\mathcal S_{\alpha_1,\beta_{n_p}}(\boldsymbol{\Omega},\boldsymbol{\Omega}') \\
\vdots & \ddots & \vdots \\
\mathcal S_{\alpha_{n_p},\beta_1}(\boldsymbol{\Omega},\boldsymbol{\Omega}') &
\cdots &
\mathcal S_{\alpha_{n_p},\beta_{n_p}}(\boldsymbol{\Omega},\boldsymbol{\Omega}')
\end{pmatrix},
\end{equation}
where
$I=\{\alpha_1,\ldots,\alpha_{n_p}\}$ and
$J=\{\beta_1,\ldots,\beta_{n_p}\}$.
Similarly, for bosons, we denote the corresponding Fock state by
\begin{equation}
|I;\boldsymbol{\Omega}\rangle_B
=
\frac{1}{\sqrt{\prod_\alpha n_\alpha!}}
\prod_\alpha
\left[
B_\alpha^\dagger(\boldsymbol{\Omega})
\right]^{n_\alpha}
|0\rangle,
\end{equation}
where $B_\alpha^\dagger(\boldsymbol{\Omega})$ is the bosonic creation operator
for the single-particle orbital $\alpha$ at twist $\boldsymbol{\Omega}$, and $n_\alpha$ is its occupation number.
The overlap between two bosonic Fock states is given by the permanent of the
single-particle overlap matrix,
\begin{equation}
{}_B\langle I;\boldsymbol{\Omega}
|
J;\boldsymbol{\Omega}^\prime\rangle_B
=\frac{1}{{
\sqrt{
\left(\prod_\alpha n_\alpha^{(I)}!\right)
\left(\prod_\beta n_\beta^{(J)}!\right)
}
}}
\operatorname{perm}
\begin{pmatrix}
\mathcal S_{\alpha_1,\beta_1}(\boldsymbol{\Omega},\boldsymbol{\Omega}^\prime) &
\cdots &
\mathcal S_{\alpha_1,\beta_{n_p}}(\boldsymbol{\Omega},\boldsymbol{\Omega}^\prime) \\
\vdots & \ddots & \vdots \\
\mathcal S_{\alpha_{n_p},\beta_1}(\boldsymbol{\Omega},\boldsymbol{\Omega}^\prime) &
\cdots &
\mathcal S_{\alpha_{n_p},\beta_{n_p}}(\boldsymbol{\Omega},\boldsymbol{\Omega}^\prime)
\end{pmatrix}.
\end{equation}
At each twist point $\boldsymbol{\Omega}$, we consider a $q$-fold degenerate
many-body manifold spanned by
$\{|\Psi_i(\boldsymbol{\Omega})\rangle\}_{i=1}^{q}$.
The overlap between the degenerate manifolds at
$\boldsymbol{\Omega}$ and $\boldsymbol{\Omega}'$ is described by the
$q\times q$ matrix
\begin{equation}
\mathcal M
(\boldsymbol{\Omega},\boldsymbol{\Omega}')
=
\begin{pmatrix}
\langle\Psi_1(\boldsymbol{\Omega})|\Psi_1(\boldsymbol{\Omega}')\rangle
&
\cdots
&
\langle\Psi_1(\boldsymbol{\Omega})|\Psi_q(\boldsymbol{\Omega}')\rangle
\\
\vdots
&
\ddots
&
\vdots
\\
\langle\Psi_q(\boldsymbol{\Omega})|\Psi_1(\boldsymbol{\Omega}')\rangle
&
\cdots
&
\langle\Psi_q(\boldsymbol{\Omega})|\Psi_q(\boldsymbol{\Omega}')\rangle
\end{pmatrix},
\end{equation}
where each many-body state is expanded in the corresponding fermionic or bosonic Fock basis as
\begin{equation}
|\Psi_i^{(\eta)}(\boldsymbol{\Omega})\rangle
=
\sum_I
C_{I,\eta}^{(i)}(\boldsymbol{\Omega})
|I;\boldsymbol{\Omega}\rangle_{\eta},
\qquad
i=1,\ldots,q,
\qquad
\eta=F,B.
\end{equation}
where $\eta=F$ and $B$ denote fermionic and bosonic cases, respectively.

The many-body QGT of the $q$-fold degenerate manifold can be defined in terms of the projector
\begin{equation}
P(\boldsymbol{\Omega})
=
\sum_{i=1}^{q}
|\Psi_i(\boldsymbol{\Omega})\rangle
\langle\Psi_i(\boldsymbol{\Omega})|.
\label{eq:proj}
\end{equation}
Writing
$\partial_\mu\equiv\partial/\partial\Omega_\mu$,
with $\Omega_\mu\in\{\theta,\phi\}$, the many-body QGT is \cite{Ma2010NonAbelianQGT}
\begin{equation}
\mathcal Q_{\mu\nu}
=
\sum_{i=1}^{q}
\left\langle
\partial_\mu\Psi_i
\right|
\left[1-P(\boldsymbol{\Omega})\right]
\left|
\partial_\nu\Psi_i
\right\rangle
=
\mathcal G_{\mu\nu}
+\frac{{\rm i}}{2}\mathcal B_{\mu\nu},
\label{eq:qgt}
\end{equation}
where
$\mathcal G_{\mu\nu}=\mathrm{Re}\,\mathcal Q_{\mu\nu}$
is the many-body quantum metric and
$\mathcal B_{\mu\nu}=2\,\mathrm{Im}\,\mathcal Q_{\mu\nu}$
is the Berry curvature, with $\mathrm{Re}$ and $\mathrm{Im}$ denoting the real and imaginary parts, respectively.
On the $(\theta,\phi)$ parameter space, both quantities can be
extracted directly from the overlap matrix
$\mathcal M(\boldsymbol{\Omega},\boldsymbol{\Omega}')$.

We first consider the quantum metric. The projector defined in \eref{eq:proj}
satisfies $P^2=P$, $(1-P)^2=1-P$ and $\mathrm{Tr}\,P=q$. A gauge-invariant distance between
the many-body subspaces at two twist points
$\boldsymbol{\Omega}$ and $\boldsymbol{\Omega}'$ can then be defined from
the Hilbert--Schmidt distance between their projectors \cite{qdv4-79lc},
\begin{equation}
D^2(\boldsymbol{\Omega},\boldsymbol{\Omega}')
=
\frac{1}{2}
\mathrm{Tr}
\left[
P(\boldsymbol{\Omega})-P(\boldsymbol{\Omega}')
\right]^2 .
\label{eq:projector_distance}
\end{equation}
Using the projector properties, this reduces to
\begin{equation}
\begin{aligned}
D^2(\boldsymbol{\Omega},\boldsymbol{\Omega}')
=&
q-
\mathrm{Tr}
\left[
P(\boldsymbol{\Omega})
P(\boldsymbol{\Omega}')
\right]\\
=&q-\sum_{i,j=1}^{q}
\left|
\langle
\Psi_i(\boldsymbol{\Omega})
|
\Psi_j(\boldsymbol{\Omega}')
\rangle
\right|^2\\
=&q-
\mathrm{Tr}
\left[
\mathcal M(\boldsymbol{\Omega},\boldsymbol{\Omega}')
\mathcal M^\dagger(\boldsymbol{\Omega},\boldsymbol{\Omega}')
\right].
\end{aligned}
\end{equation}
Differentiating the projector identity with respect to $\Omega_\mu$, we obtain
\begin{equation}
(\partial_\mu P)P+P(\partial_\mu P)=\partial_\mu P.
\end{equation}
Left-multiplying this equation by $P$ gives
\begin{equation}
P(\partial_\mu P)P=0.
\end{equation}
Right-multiplying this relation by $(1-P)$ then yields
\begin{equation}
(1-P)(\partial_\mu P)(1-P)=0.
\end{equation}
Therefore, $\partial_\mu P$ can be written as
\begin{equation}
\partial_\mu P
=
P(\partial_\mu P)(1-P)
+
(1-P)(\partial_\mu P)P .
\label{eq:dP_decomposition}
\end{equation}
It then follows that
\begin{equation}
\mathrm{Tr}
\left[
(\partial_\mu P)(\partial_\nu P)
\right]
=
\mathrm{Tr}
\left[
P(\partial_\mu P)(1-P)(\partial_\nu P)P
\right]+
\mathrm{Tr}
\left[
(1-P)(\partial_\mu P)P(\partial_\nu P)(1-P)
\right].
\end{equation}
Noting that
\begin{equation}
\begin{aligned}
    \mathrm{Tr}\left[
P(\partial_\mu P)(1-P)(\partial_\nu P)P
\right]^\dagger
=&
\mathrm{Tr}\left[P(\partial_\nu P)(1-P)(\partial_\mu P)P\right]\\
=&\mathrm{Tr}\left[(1-P)(\partial_\mu P)P(\partial_\nu P)\right]\\
=&\mathrm{Tr}\left[(1-P)(\partial_\mu P)P(\partial_\nu P)-(1-P)(\partial_\mu P)P(\partial_\nu P)P\right]\\
=&\mathrm{Tr}\left[(1-P)(\partial_\mu P)P(\partial_\nu P)(1-P)\right].
\end{aligned}
\end{equation}
Therefore,
\begin{equation}
\begin{aligned}
\mathrm{Tr}
\left[
(\partial_\mu P)(\partial_\nu P)
\right]
&=
2\,\mathrm{Re}\,
\mathrm{Tr}
\left[
P(\partial_\mu P)(1-P)(\partial_\nu P)P
\right]
\\
&=
2\,\mathrm{Re}
\sum_{i=1}^{q}
\langle\Psi_i|
(\partial_\mu P)(1-P)(\partial_\nu P)
|\Psi_i\rangle
\\
&=
2\,\mathrm{Re}
\sum_{i=1}^{q}
\langle\partial_\mu\Psi_i|
(1-P)
|\partial_\nu\Psi_i\rangle
\\
&=
2\,\mathrm{Re}\,\mathcal Q_{\mu\nu}\\
&=2 \mathcal G_{\mu\nu},
\end{aligned}
\end{equation}
where we have used the relation obtained by differentiating
$P|\Psi_i\rangle=|\Psi_i\rangle$,
\begin{equation}
(\partial_\mu P)|\Psi_i\rangle
=
(1-P)|\partial_\mu\Psi_i\rangle.
\end{equation}
For two neighboring twist points,
$\boldsymbol{\Omega}'=\boldsymbol{\Omega}+d\boldsymbol{\Omega}$, with
$d\boldsymbol{\Omega}=(d\theta,d\phi)$, the projector can be expanded as
\begin{equation}
P(\boldsymbol{\Omega}+d\boldsymbol{\Omega})
=
P(\boldsymbol{\Omega})
+
\partial_\mu P\,d\Omega_\mu
+
O(d\Omega^2),
\end{equation}
where repeated indices $\mu,\nu=\theta,\phi$ are summed over.
Substituting this expansion into ~\eref{eq:projector_distance} gives
\begin{equation}
\begin{aligned}
    D^2
=&
\frac{1}{2}
\mathrm{Tr}
\left[
\partial_\mu P\,\partial_\nu P
\right]
d\Omega_\mu d\Omega_\nu
+
O(d\Omega^3)\\
=&\mathcal G_{\mu\nu}
d\Omega_\mu d\Omega_\nu
+
O(d\Omega^3).
\end{aligned}
\end{equation}
Explicitly, on the $(\theta,\phi)$ parameter space,
\begin{equation}
\mathcal G
=
\begin{pmatrix}
\mathcal G_{\theta\theta} & \mathcal G_{\theta\phi}\\
\mathcal G_{\theta\phi} & \mathcal G_{\phi\phi}
\end{pmatrix},
\end{equation}
and therefore
\begin{equation}
D^2
=
\mathcal G_{\theta\theta}(d\theta)^2
+
2\mathcal G_{\theta\phi}d\theta\,d\phi
+
\mathcal G_{\phi\phi}(d\phi)^2
+
O(d\Omega^3).
\end{equation}
The individual components of the metric can therefore be extracted by
choosing different infinitesimal displacements in the $(\theta,\phi)$ space.
\begin{equation}
\mathcal G_{\theta\theta}
=
\frac{
D^2\left(
\boldsymbol{\Omega},
\boldsymbol{\Omega}+d\theta\,\hat{\boldsymbol{\theta}}
\right)
}{
(d\theta)^2
},
\qquad
\mathcal G_{\phi\phi}
=
\frac{
D^2\left(
\boldsymbol{\Omega},
\boldsymbol{\Omega}+d\phi\,\hat{\boldsymbol{\phi}}
\right)
}{
(d\phi)^2
},
\end{equation}
\begin{equation}
\mathcal G_{\theta\phi}
=
\frac{
D^2\!\left(
\boldsymbol{\Omega},
\boldsymbol{\Omega}
+d\theta\,\hat{\boldsymbol{\theta}}
+d\phi\,\hat{\boldsymbol{\phi}}
\right)
-
D^2\!\left(
\boldsymbol{\Omega},
\boldsymbol{\Omega}
+d\theta\,\hat{\boldsymbol{\theta}}
\right)
-
D^2\!\left(
\boldsymbol{\Omega},
\boldsymbol{\Omega}
+d\phi\,\hat{\boldsymbol{\phi}}
\right)
}{
2\,d\theta\,d\phi
}.
\end{equation}
We note that the two twist directions need not be orthogonal. Denoting the
angle between $\mathbf L_1$ and $\mathbf L_2$ by $\gamma$, the corresponding
reciprocal-space twist directions are characterized by the Gram matrix
\begin{equation}
\Gamma=
\begin{pmatrix}
1 & -\cos\gamma\\
-\cos\gamma & 1
\end{pmatrix}.
\end{equation}
Accordingly,
\begin{equation}
\mathrm{Tr}\,\mathcal G
\rightarrow
\mathrm{Tr}\left(\Gamma^{-1}\mathcal G\right),
\qquad
\det\mathcal G
\rightarrow
\frac{\det\mathcal G}{\det\Gamma}.
\end{equation}
Finally, the dimensionless total trace of the many-body quantum metric is defined as
\begin{equation}
\chi_{\rm mb}^{-1}
=
\frac{1}{2\pi}
\int d^2\boldsymbol{\Omega}\,
\mathrm{Tr}
\left[
\Gamma^{-1}\mathcal G(\boldsymbol{\Omega})
\right].
\end{equation}

We next turn to the Berry curvature. From the imaginary part of the
many-body QGT, the Berry curvature is 
\begin{equation}
\begin{aligned}
\mathcal B_{\mu\nu}
&=2\,\mathrm{Im}\,\mathcal Q_{\mu\nu}\\
&=
-{\rm i}
\left(
\mathcal Q_{\mu\nu}
-
\mathcal Q_{\nu\mu}
\right)\\
&=
-{\rm i}\left[
\sum_{i=1}^{q}
\langle
\partial_\nu\Psi_i
|
(1-P)
|
\partial_\mu\Psi_i
\rangle-\sum_{i=1}^{q}
\langle
\partial_\mu\Psi_i
|
(1-P)
|
\partial_\nu\Psi_i
\rangle\right]\\
&=-{\rm i}\sum_{i=1}^{q}\left[\langle \Psi_i|P(\partial_u P)(\partial_v P)|\Psi_i\rangle-\langle \Psi_i|P(\partial_v P)(\partial_u P)|\Psi_i\rangle\right]\\
&=-{\rm i}\,\mathrm{Tr}\left(
P
\left[
\partial_\mu P,
\partial_\nu P
\right]
\right)
.
\end{aligned}
\end{equation}
For two neighboring points in the $(\theta,\phi)$ parameter space along the
$\Omega_\mu$ direction, the overlap matrix is
\begin{equation}
\mathcal M
\left(
\boldsymbol{\Omega},
\boldsymbol{\Omega}+d\Omega_\mu\hat{\boldsymbol{\mu}}
\right)
=
\begin{pmatrix}
\langle\Psi_1(\boldsymbol{\Omega})|
\Psi_1(\boldsymbol{\Omega}+d\Omega_\mu\hat{\boldsymbol{\mu}})\rangle
&
\cdots
&
\langle\Psi_1(\boldsymbol{\Omega})|
\Psi_q(\boldsymbol{\Omega}+d\Omega_\mu\hat{\boldsymbol{\mu}})\rangle
\\
\vdots
&
\ddots
&
\vdots
\\
\langle\Psi_q(\boldsymbol{\Omega})|
\Psi_1(\boldsymbol{\Omega}+d\Omega_\mu\hat{\boldsymbol{\mu}})\rangle
&
\cdots
&
\langle\Psi_q(\boldsymbol{\Omega})|
\Psi_q(\boldsymbol{\Omega}+d\Omega_\mu\hat{\boldsymbol{\mu}})\rangle
\end{pmatrix}.
\label{eq:overlap1}
\end{equation}
Expanding
$|\Psi_j(\boldsymbol{\Omega}+d\Omega_\mu\hat{\boldsymbol{\mu}})\rangle$
to first order gives
\begin{equation}
|\Psi_j(\boldsymbol{\Omega}+d\Omega_\mu\hat{\boldsymbol{\mu}})\rangle
=
|\Psi_j(\boldsymbol{\Omega})\rangle
+
|\partial_\mu\Psi_j(\boldsymbol{\Omega})\rangle
d\Omega_\mu
+
O(d\Omega_\mu^2).
\end{equation}
Therefore, the overlap matrix (\eref{eq:overlap1}) can be written as
\begin{equation}
\mathcal M
=
I
+
\begin{pmatrix}
\langle\Psi_1(\boldsymbol{\Omega})|\partial_\mu\Psi_1(\boldsymbol{\Omega})\rangle &
\cdots &
\langle\Psi_1(\boldsymbol{\Omega})|\partial_\mu\Psi_q(\boldsymbol{\Omega})\rangle
\\
\vdots & \ddots & \vdots
\\
\langle\Psi_q(\boldsymbol{\Omega})|\partial_\mu\Psi_1(\boldsymbol{\Omega})\rangle &
\cdots &
\langle\Psi_q(\boldsymbol{\Omega})|\partial_\mu\Psi_q(\boldsymbol{\Omega})\rangle
\end{pmatrix}
d\Omega_\mu
+
O(d\Omega_\mu^2),
\end{equation}
where $I$ is identity matrix. The non-Abelian Berry connection can be written as \cite{WilczekZee1984}
\begin{equation}
\mathcal A_\mu
=
-{\rm i}
\begin{pmatrix}
\langle\Psi_1(\boldsymbol{\Omega})|\partial_\mu\Psi_1(\boldsymbol{\Omega})\rangle &
\cdots &
\langle\Psi_1(\boldsymbol{\Omega})|\partial_\mu\Psi_q(\boldsymbol{\Omega})\rangle
\\
\vdots & \ddots & \vdots
\\
\langle\Psi_q(\boldsymbol{\Omega})|\partial_\mu\Psi_1(\boldsymbol{\Omega})\rangle &
\cdots &
\langle\Psi_q(\boldsymbol{\Omega})|\partial_\mu\Psi_q(\boldsymbol{\Omega})\rangle
\end{pmatrix}.
\end{equation}
Accordingly, the overlap matrix takes the form
\begin{equation}
\mathcal M
=
I
+
{\rm i}\,\mathcal A_\mu d\Omega_\mu
+
O(d\Omega_\mu^2).
\end{equation}
Taking the determinant,
\begin{equation}
\begin{aligned}
    \det\mathcal M
=&
1
+
{\rm i}\,\mathrm{Tr}\,\mathcal A_\mu\,d\Omega_\mu
+
O(d\Omega_\mu^2)\\
=&e^{
{\rm i}\,\mathrm{Tr} \mathcal A_\mu d\Omega_\mu}+
O(d\Omega_\mu^2).
\end{aligned}
\end{equation}
The phase of the overlap matrix can therefore be extracted as
\begin{equation}
U_\mu(\boldsymbol{\Omega})
=
\frac{
\det\mathcal M
\left(
\boldsymbol{\Omega},
\boldsymbol{\Omega}
+\delta\Omega_\mu\hat{\boldsymbol{\mu}}
\right)
}{
\left|
\det\mathcal M
\left(
\boldsymbol{\Omega},
\boldsymbol{\Omega}
+\delta\Omega_\mu\hat{\boldsymbol{\mu}}
\right)
\right|
}.
\end{equation}
For an elementary plaquette in the $(\theta,\phi)$ parameter space, the
phase accumulated along its boundary is \cite{FukuiHatsugaiSuzuki2005,Hatsugai2005}
\begin{equation}
W_{\theta\phi}(\boldsymbol{\Omega})
=
U_\theta(\boldsymbol{\Omega})\,
U_\phi(\boldsymbol{\Omega}
+\delta\theta\,\hat{\boldsymbol{\theta}})
U_\theta^{-1}(\boldsymbol{\Omega}
+\delta\phi\,\hat{\boldsymbol{\phi}})
\,U_\phi^{-1}(\boldsymbol{\Omega}).
\end{equation}
For sufficiently small $\delta\theta$ and $\delta\phi$,
\begin{equation}
U_\mu(\boldsymbol{\Omega})
=
e^{{\rm i}\,\mathrm{Tr}\mathcal A_\mu(\boldsymbol{\Omega})
\delta\Omega_\mu}
+
O(\delta\Omega_\mu^2).
\end{equation}
Therefore,
\begin{equation}
\begin{aligned}
W_{\theta\phi}(\boldsymbol{\Omega})
\simeq &
\exp\Big\{
{\rm i}\big[\mathrm{Tr}\mathcal A_\theta(\theta,\phi)\,\delta\theta
+
\mathrm{Tr}\mathcal A_\phi(\theta+\delta\theta,\phi)\,\delta\phi-
\mathrm{Tr}\mathcal A_\theta(\theta,\phi+\delta\phi)\,\delta\theta
-
\mathrm{Tr}\mathcal A_\phi(\theta,\phi)\,\delta\phi
\big]
\Big\}\\
=&\exp\left\{
{\rm i}
\left[
\partial_\theta\mathrm{Tr}\mathcal A_\phi
-
\partial_\phi\mathrm{Tr}\mathcal A_\theta
\right]
\delta\theta\,\delta\phi
\right\}
+
O(\delta\Omega^3),
\end{aligned}
\end{equation}
where the Berry connections have been expanded to first order as
\begin{equation}
\begin{aligned}
\mathrm{Tr}\mathcal A_\phi(\theta+\delta\theta,\phi)
&=
\mathrm{Tr}\mathcal A_\phi(\theta,\phi)
+
\partial_\theta
\mathrm{Tr}\mathcal A_\phi\,
\delta\theta
+
O(\delta\theta^2),
\\
\mathrm{Tr}\mathcal A_\theta(\theta,\phi+\delta\phi)
&=
\mathrm{Tr}\mathcal A_\theta(\theta,\phi)
+
\partial_\phi
\mathrm{Tr}\mathcal A_\theta\,
\delta\phi
+
O(\delta\phi^2).
\end{aligned}
\end{equation}
It is worth noting that
\begin{equation}
\mathrm{Tr}\,\mathcal A_\mu
=
-{\rm i}\sum_{i=1}^{q}
\langle\Psi_i|\partial_\mu\Psi_i\rangle.
\end{equation}
Therefore,
\begin{equation}
\begin{aligned}
\partial_\theta\mathrm{Tr}\mathcal A_\phi
-
\partial_\phi\mathrm{Tr}\mathcal A_\theta
=&
-{\rm i}\sum_{i=1}^{q}
\Big[
\langle\partial_\theta\Psi_i|\partial_\phi\Psi_i\rangle
-
\langle\partial_\phi\Psi_i|\partial_\theta\Psi_i\rangle
\Big]\\
=&-{\rm i}\sum_{i=1}^{q}
\Big[
\langle\partial_\theta\Psi_i|
(1-P)
|\partial_\phi\Psi_i\rangle-
\langle\partial_\phi\Psi_i|
(1-P)
|\partial_\theta\Psi_i\rangle
\Big]\\
=&-{\rm i}\sum_{i=1}^{q}
\Big[
\langle\Psi_i|
P(\partial_\theta P)(\partial_\phi P)
|\Psi_i\rangle-
\langle\Psi_i|
P(\partial_\phi P)(\partial_\theta P)
|\Psi_i\rangle
\Big]\\
=&-{\rm i}\sum_{i=1}^{q}
\Big[
\langle\Psi_i|
P[\partial_\theta P,\partial_\phi P]
|\Psi_i\rangle
\Big]\\
=&-{\rm i}\,\mathrm{Tr}\left(
P
\left[
\partial_\theta P,
\partial_\phi P
\right]
\right)\\
=&\mathcal B_{\theta\phi},
\end{aligned}
\end{equation}
where we have used the following relations,
\begin{equation}
\sum_{i=1}^{q}
\Big[
\langle\partial_\theta\Psi_i|P|\partial_\phi\Psi_i\rangle
-
\langle\partial_\phi\Psi_i|P|\partial_\theta\Psi_i\rangle
\Big]
=\sum_{i,j=1}^{q}
\Big[
\langle\Psi_i|\partial_\theta\Psi_j\rangle
\langle\Psi_j|\partial_\phi\Psi_i\rangle
-
\langle\Psi_i|\partial_\phi\Psi_j\rangle
\langle\Psi_j|\partial_\theta\Psi_i\rangle
\Big]=0,
\end{equation}
together with
$\langle\partial_\mu\Psi_i|\Psi_j\rangle
=
-\langle\Psi_i|\partial_\mu\Psi_j\rangle$.
Accordingly,
\begin{equation}
W_{\theta\phi}(\boldsymbol{\Omega})
=
e^{{\rm i}\,\mathcal B_{\theta\phi}(\boldsymbol{\Omega})
\delta\theta\,\delta\phi}
+
O(\delta\Omega^3),
\end{equation}
and the Berry curvature can be extracted as
\begin{equation}
\mathcal B_{\theta\phi}(\boldsymbol{\Omega})
=
\lim_{\delta\theta,\delta\phi\rightarrow 0}
\frac{
\arg W_{\theta\phi}(\boldsymbol{\Omega})
}{
\delta\theta\,\delta\phi
}.
\end{equation}
Finally, the many-body Chern number is defined as
\begin{equation}
\mathcal C
=
\frac{1}{2\pi}
\int d^2\boldsymbol{\Omega}\,
\mathcal B_{\theta\phi}(\boldsymbol{\Omega}).
\end{equation}
It is worth noting that the same construction applies when
$|\Psi\rangle$ is either a single-particle Bloch state, with
$\mu,\nu$ labeling the crystal-momentum directions, or the vacuum sector
of Eq.~(1) of the main text.
The single-particle quantum metric and Berry curvature can therefore be
viewed as special cases of their many-body counterparts. For all CHS vacua considered here, we find
\begin{equation}
    \chi_{\rm mb}^{-1}
    =
    \mathcal C
    =
    \nu_{\rm vac},
\end{equation}
where $\nu_{\rm vac}$ denotes the vacuum filling. From a geometric perspective, the CHS also has the important property of being "ideal". For a two-dimensional parameter space, the
quantum metric and Berry curvature satisfy
\begin{equation}
    \frac{1}{2}\mathrm{Tr}\,\mathcal G
    \geq
    \sqrt{\det\mathcal G}
    \geq
    \frac{1}{2}\left|\mathcal B_{\mu\nu}\right|.
    \label{kahler}
\end{equation}
If the second inequality in \eref{kahler} is saturated, the
corresponding state manifold has the special geometric property of forming
a holomorphic line bundle over a K\"ahler manifold \cite{MeraOzawa2021Kahler,MeraOzawa2021Engineering}. If both inequalities in \eref{kahler} are saturated, Eq.~(1) of the main text is ``ideal'' and inherits a number of special geometric properties,
including those associated with ideal flat bands and saturation of the
trace condition \cite{Roy2014BandGeometry,Claassen2015PositionMomentum,
Lee2017BandEngineering}. Remarkably, the CHS vacua
considered here satisfy this ideal condition.
Thus, the ideal quantum geometry familiar from single-particle flat bands
extends naturally to the many-body CHS.

\section{Mapping conformal Hilbert spaces from the lowest Landau level to Chern bands}

In this section, we describe how the CHS constructed in the LLL are
mapped onto a generic Chern band. Alternatively, the corresponding parent
Hamiltonians can also be projected directly into the Chern band and used
to construct the CHS. For a nonideal Chern band, however, the projected
parent Hamiltonian generally no longer possesses exact zero-energy states.
This issue is particularly severe for gapless states, such as the
Gaffnian and Haffnian, for which the corresponding CHS can be difficult to
separate unambiguously from the rest of the spectrum. We therefore adopt
a direct projection approach instead. The basic idea is to identify the single-particle orbitals of
the LLL with those of the target Chern band and use this correspondence to
transfer the corresponding many-body Fock states. 
This provides a
way to compare the low-energy subspaces of interacting Chern bands with
the known CHS of Landau levels.
Let $\psi_{\mathbf m}^{\rm LLL}(\mathbf{r})$ denote a single-particle state in
the LLL and $\psi_{\mathbf m}^{\rm CB}(\mathbf{r})$ the corresponding state in
the target Chern band, where $\mathbf m$ labels the discrete momentum index
on the torus (\eref{eq:momentum1}).
For a single-particle orbital $\mathbf m$, the real-space density is
\begin{equation}
\rho_{\mathbf m}^{\rm LLL}(\mathbf r)
=
\left|
\psi_{\mathbf m}^{\rm LLL}(\mathbf r)
\right|^2 ,
\end{equation}
and it can be expanded in reciprocal-lattice
vectors as
\begin{equation}
\rho_{\mathbf m}^{\rm LLL}(\mathbf r)
=
\sum_{\mathbf b}
\rho_{\mathbf m,\mathbf b}^{\rm LLL}
e^{{\rm i}\mathbf b\cdot\mathbf r}.
\end{equation}
where 
\begin{equation}
\begin{aligned}
\rho_{\mathbf m,\mathbf b}^{\rm LLL}
=&
\frac{1}{S}
\int_S d^2r\,
\left[
\psi_{\mathbf m}^{\rm LLL}(\mathbf r)
\right]^*
e^{-{\rm i}\mathbf b\cdot\mathbf r}
\psi_{\mathbf m}^{\rm LLL}(\mathbf r)\\
=&\frac{1}{S}
f_{\mathbf b}^{\mathbf k_{\mathbf m},
\mathbf k_{\mathbf m}}.
\end{aligned}
\end{equation}
Therefore,
\begin{equation}
\rho_{\mathbf m}^{\rm LLL}(\mathbf r)
=
\frac{1}{S}
\sum_{\mathbf b}
f_{\mathbf b}^{\mathbf k_{\mathbf m},\mathbf k_{\mathbf m}}
e^{{\rm i}\mathbf b\cdot\mathbf r}.
\label{eq:rho_LLL_ff}
\end{equation}
Similarly, for the target Chern band,
\begin{equation}
\rho_{\mathbf m}^{\rm CB}(\mathbf r)
=
\frac{1}{S}
\sum_{\mathbf b}
\mathcal F_{\mathbf b}^{\mathbf k_{\mathbf m},\mathbf k_{\mathbf m}}
e^{{\rm i}\mathbf b\cdot\mathbf r}.
\end{equation}
For a given band $l$, we suppress the band index for notational simplicity
and use the shorthand
$
\mathcal F_{\mathbf b}^{\,l,\mathbf k_{\mathbf m},
l,\mathbf k_{\mathbf m}}
\equiv
\mathcal F_{\mathbf b}^{\mathbf k_{\mathbf m},\mathbf k_{\mathbf m}}$.
We construct an approximate mapping between the LLL orbitals and those of
the target Chern band in real space as
\begin{equation}
\psi_{\mathbf m}^{\rm CB}(\mathbf r)
\simeq
\mathcal N_{\mathbf m}
e^{{\rm i}\zeta_{\mathbf m}}
\mathcal R(\mathbf r)
\psi_{\mathbf m}^{\rm LLL}(\mathbf r),
\label{eq:approx_mapping}
\end{equation}
where $\mathcal R(\mathbf r)$ describes an orbital-independent spatial
deformation, while $\mathcal N_{\mathbf m}$ and $\zeta_{\mathbf m}$ denote
the orbital-dependent normalization factor and the relative phase, respectively,
with $\zeta_{\mathbf m}$ accounting for the gauge mismatch, including the
arbitrary phase generated in the single-particle diagonalization.
For an ideal $\mathcal C=1$ Chern band with the appropriate holomorphic structure,
the above factorization can become exact \cite{PhysRevLett.127.246403}. For a generic Chern band,
however, such a factorization is not guaranteed, and
\eref{eq:approx_mapping} should be regarded as an approximate mapping. In the following, all equations are understood within the approximate mapping introduced above, and equality signs are used for notational simplicity.

We first extract the spatial deformation $\mathcal R(\mathbf r)$ and the
orbital-dependent coefficient $\mathcal N_{\mathbf m}$ from the
single-particle densities. For an orbital $\mathbf m$ the densities are related by
\begin{equation}
\rho_{\mathbf m}^{\rm CB}(\mathbf r)
=
|\mathcal N_{\mathbf m}|^2
|\mathcal R(\mathbf r)|^2
\rho_{\mathbf m}^{\rm LLL}(\mathbf r).
\label{eq:density_mapping}
\end{equation}
To determine the $\mathcal{R}(r)$, we take the
orbital with $\mathbf m=\mathbf{0}$ as the reference. Since an overall constant
factor can be absorbed into $\mathcal R(\mathbf r)$, we choose
$\mathcal N_{\mathbf 0}=1$. The deformation profile is then determined by
\begin{equation}
|\mathcal R(\mathbf r)|^2
=
\frac{
\rho_{\mathbf 0}^{\rm CB}(\mathbf r)
}{
\rho_{\mathbf 0}^{\rm LLL}(\mathbf r)
}.
\label{eq:R_density}
\end{equation}
For the remaining orbitals, the coefficients $\mathcal N_{\mathbf m}$ are
determined by normalization. Imposing
\begin{equation}
\int_{S} d^2r\,
\rho_{\mathbf m}^{\rm CB}(\mathbf r)
=
1,
\end{equation}
therefore,
\begin{equation}
|\mathcal N_{\mathbf m}|^2
\int_{S} d^2r\,
|\mathcal R(\mathbf r)|^2
\rho_{\mathbf m}^{\rm LLL}(\mathbf r)
=
1.
\end{equation}
Hence,
\begin{equation}
\begin{aligned}
\mathcal N_{\mathbf m}
=&
\left[
\int_{S} d^2r\,
|\mathcal R(\mathbf r)|^2
\rho_{\mathbf m}^{\rm LLL}(\mathbf r)
\right]^{-1/2}\\
=&
\left[
\int_{S} d^2r\,
\rho_{\mathbf m}^{\rm LLL}(\mathbf r)
\frac{
\rho_{\mathbf 0}^{\rm CB}(\mathbf r)
}{
\rho_{\mathbf 0}^{\rm LLL}(\mathbf r)
}
\right]^{-1/2}.
\end{aligned}
\label{eq:Nm_density}
\end{equation}
The density determines $\mathcal{N}_m$ but contains no
information about the relative phase $\zeta_{\mathbf m}$. We therefore
determine $\zeta_{\mathbf m}$ from the form factors. We again choose $\mathbf m=0$ as the
reference orbital. From \eref{eq:approx_mapping}, 
the form factor connecting the reference orbital $\mathbf 0$ and an
arbitrary orbital $\mathbf m$ is
\begin{equation}
\begin{aligned}
\mathcal F_{\mathbf b}^{\mathbf 0,\mathbf k_\mathbf{m}}
=&
\int_S d^2r\,
\left[\psi_{\mathbf 0}^{\rm CB}(\mathbf r)\right]^*
e^{{\rm i}(\mathbf 0
-\mathbf k_{\mathbf m}-\mathbf b)\cdot\mathbf r}
\psi_{\mathbf m}^{\rm CB}(\mathbf r)\\
=&\mathcal N_{\mathbf 0}\mathcal N_{\mathbf m}
e^{{\rm i}(\zeta_{\mathbf m}-\zeta_{\mathbf 0})}\int_S d^2r\,
|\mathcal R(\mathbf r)|^2
\left[\psi_{\mathbf 0}^{\rm LLL}(\mathbf r)\right]^*
e^{{\rm i}(\mathbf 0
-\mathbf k_{\mathbf m}-\mathbf b)\cdot\mathbf r}
\psi_{\mathbf m}^{\rm LLL}(\mathbf r).
\end{aligned}
\label{eq:cb_ff_phase}
\end{equation}
we define
$W(\mathbf r)
=
|\mathcal R(\mathbf r)|^2$, and
 it can be expanded in
reciprocal-lattice vectors as
\begin{equation}
W(\mathbf r)
=
\sum_{\mathbf G}
W_{\mathbf G}
e^{{\rm i}\mathbf G\cdot\mathbf r},
\label{eq:W_fourier}
\end{equation}
with
\begin{equation}
W_{\mathbf G}
=
\frac{1}{S}
\int_S d^2r\,
W(\mathbf r)
e^{-{\rm i}\mathbf G\cdot\mathbf r}.
\label{eq:W_fourier_coeff}
\end{equation}
Using \eref{eq:W_fourier}, the spatial integral in
\eref{eq:cb_ff_phase} becomes
\begin{equation}
\begin{aligned}
\int_S d^2r\,
W(\mathbf r)
\left[\psi_{\mathbf 0}^{\rm LLL}(\mathbf r)\right]^*
e^{{\rm i}(\mathbf 0
-\mathbf k_{\mathbf m}-\mathbf b)\cdot\mathbf r}
\psi_{\mathbf m}^{\rm LLL}(\mathbf r)
=&
\sum_{\mathbf G}
W_{\mathbf G}
\int_S d^2r\,
\left[\psi_{\mathbf 0}^{\rm LLL}(\mathbf r)\right]^*
e^{{\rm i}(\mathbf 0
-\mathbf k_{\mathbf m}-\mathbf b+\mathbf G)\cdot\mathbf r}
\psi_{\mathbf m}^{\rm LLL}(\mathbf r)\\
=&\sum_\mathbf{G}W_\mathbf{G}f^{\mathbf 0,\mathbf k_\mathbf{m}}_{\mathbf b-\mathbf G}.
\end{aligned}
\end{equation}
Introducing
$\mathbf b'=\mathbf b-\mathbf G$, we obtain
\begin{equation}
\sum_{\mathbf G}
W_{\mathbf G}
f^{\mathbf 0,\mathbf k_\mathbf{m}}_{\mathbf b-\mathbf G}
=
\sum_{\mathbf b^\prime}
W_{\mathbf b-\mathbf b^\prime}
f^{\mathbf 0,\mathbf k_\mathbf{m}}_{\mathbf b^\prime}.
\end{equation}
Therefore,
\begin{equation}
\mathcal F_{\mathbf b}^{\mathbf 0,\mathbf k_\mathbf{m}}
=
\mathcal N_{\mathbf 0}\mathcal N_{\mathbf m}
e^{{\rm i}(\zeta_{\mathbf m}-\zeta_{\mathbf 0})}
\sum_{\mathbf b'}
\mathcal K_{\mathbf b\mathbf b^\prime}
f^{\mathbf 0,\mathbf k_\mathbf{m}}_{\mathbf b^\prime},
\end{equation}
where
\begin{equation}
\mathcal K_{\mathbf b\mathbf b'}
=
W_{\mathbf b'-\mathbf b}.
\label{eq:K_definition}
\end{equation}
Since $W(\mathbf r)=|\mathcal R(\mathbf r)|^2$ is real, its Fourier
components satisfy
$
W_{\mathbf G}^*
=
W_{-\mathbf G}
$. Therefore, 
\begin{equation}
\mathcal K_{\mathbf b\mathbf b^\prime}^*
=
W_{\mathbf b^\prime-\mathbf b}^*
=
W_{\mathbf b-\mathbf b^\prime}
=
\mathcal K_{\mathbf b^\prime\mathbf b},
\end{equation}
so that
$
\mathcal K^\dagger=\mathcal K
$.
Moreover,  for an arbitrary complex vector
$\mathbf v=\{v_{\mathbf b}\}$,
\begin{equation}
\begin{aligned}
\mathbf v^\dagger \mathcal K \mathbf v
&=
\sum_{\mathbf b,\mathbf b^\prime}
v_{\mathbf b}^*
\mathcal K_{\mathbf b\mathbf b^\prime}
v_{\mathbf b^\prime}
\\
&=
\sum_{\mathbf b,\mathbf b^\prime}
v_{\mathbf b}^*
W_{\mathbf b^\prime-\mathbf b}
v_{\mathbf b^\prime}\\
&=
\frac{1}{S}
\int_S d^2r\,
W(\mathbf r)
\sum_{\mathbf b,\mathbf b^\prime}
v_{\mathbf b}^*
v_{\mathbf b^\prime}
e^{{\rm i}\mathbf b\cdot\mathbf r}
e^{-{\rm i}\mathbf b^\prime\cdot\mathbf r}
\\
&=
\frac{1}{S}
\int_S d^2r\,
W(\mathbf r)
\left|
\sum_{\mathbf b}
v_{\mathbf b}
e^{-{\rm i}\mathbf b\cdot\mathbf r}
\right|^2
\geq 0.
\end{aligned}
\end{equation}
Hence, $\mathcal K$ is positive semidefinite.
We define the overlap between the LLL and Chern-band form factors as
\begin{equation}
\begin{aligned}
\mathcal P_{\mathbf m} =
\sum_{\mathbf b}
\left[
f_{\mathbf b}^{\mathbf 0,\mathbf k_\mathbf{m}}
\right]^*
\mathcal F_{\mathbf b}^{\mathbf 0,\mathbf k_\mathbf{m}}
=\mathcal N_{\mathbf 0}\mathcal N_{\mathbf m}
e^{{\rm i}(\zeta_{\mathbf m}-\zeta_{\mathbf 0})}
\sum_{\mathbf b,\mathbf b^\prime}
\left[
f_{\mathbf b}^{\mathbf 0,\mathbf k_\mathbf{m}}
\right]^*
\mathcal K_{\mathbf b\mathbf b^\prime}
f_{\mathbf b^\prime}^{\mathbf 0,\mathbf k_\mathbf{m}}.
\end{aligned}
\label{eq:phase_overlap}
\end{equation}
We note that, due to the positive-semidefinite property of $\mathcal K$,
\begin{equation}
\sum_{\mathbf b,\mathbf b^\prime}
\left[
f_{\mathbf b}^{\mathbf 0,\mathbf k_\mathbf{m}}
\right]^*
\mathcal K_{\mathbf b\mathbf b^\prime}
f_{\mathbf b^\prime}^{\mathbf 0,\mathbf k_\mathbf{m}}
\geq 0.
\end{equation}
Thus, all factors in the second line of
\eref{eq:phase_overlap} are real and non-negative, except for the
$e^{{\rm i}(\zeta_{\mathbf m}-\zeta_{\mathbf 0})}$.
It follows that,
\begin{equation}
\arg \mathcal P_{\mathbf m}
=
\zeta_{\mathbf m}-\zeta_{\mathbf 0}.
\end{equation}
Since $\zeta_{\mathbf 0}$ is common to all orbitals, it represents an
overall phase and can therefore be neglected.
For a Chern band with $\mathcal C=-1$, the corresponding LLL
has the opposite chirality and is related to the $\mathcal C=1$ LLL form factors
by complex conjugation. Accordingly, 
\begin{equation}
\mathcal P_{\mathbf{m}}^{(-)}
=
\sum_{\mathbf b}
f_{\mathbf b}^{\mathbf 0,\mathbf{k}_\mathbf{m}}
\mathcal F_{\mathbf b}^{\mathbf 0,\mathbf{k}_\mathbf{m}},
\end{equation}
instead of \eref{eq:phase_overlap}. The relative phase is then
extracted in the same way from
$\arg \mathcal P_{\mathbf m}^{(-)}$.

\section{Hierarchy of conformal Hilbert spaces in interacting Chern bands}
In this section, we examine how the hierarchical structure of the CHS
is manifested in the many-body spectra of different Chern bands for both
fermionic and bosonic systems. We consider several representative
single-particle bands, including the LLL, the 1LL, and generic Chern bands,
together with different interaction potentials. For each system, we
diagonalize the interacting Hamiltonian and retain the lowest 200
many-body eigenstates. Their span defines a low-energy subspace, whose
overlap with the corresponding CHS is then evaluated.

For a CHS $\mathcal H_s$, we denote its projector by
\begin{equation}
P_s
=
\sum_{a\in\mathcal H_s}
|\Phi_a\rangle
\langle\Phi_a|,
\end{equation}
where $\{|\Phi_a\rangle\}$ forms an orthonormal basis of the CHS.
Similarly, the projector onto the subspace spanned by the lowest 200
many-body eigenstates is
\begin{equation}
P_{200}
=
\sum_{n=1}^{200}
|\Psi_n\rangle
\langle\Psi_n|.
\end{equation}
The overlap between the low-energy subspace and the CHS is defined as
\begin{equation}
\mathcal O_s^{(200)}
=
\frac{1}{\mathcal D_s}
\Tr\left(P_{200}P_s\right)
=
\frac{1}{\mathcal D_s}
\sum_{n=1}^{200}
\sum_{a\in\mathcal H_s}
\left|
\langle\Phi_a|\Psi_n\rangle
\right|^2 .
\label{eq:chs_overlap_spectrum}
\end{equation}
where $\mathcal D_s$ denotes the dimension of $\mathcal H_s$.

\begin{figure}[!ht]
    \centering
    \includegraphics[width=.7\textwidth]{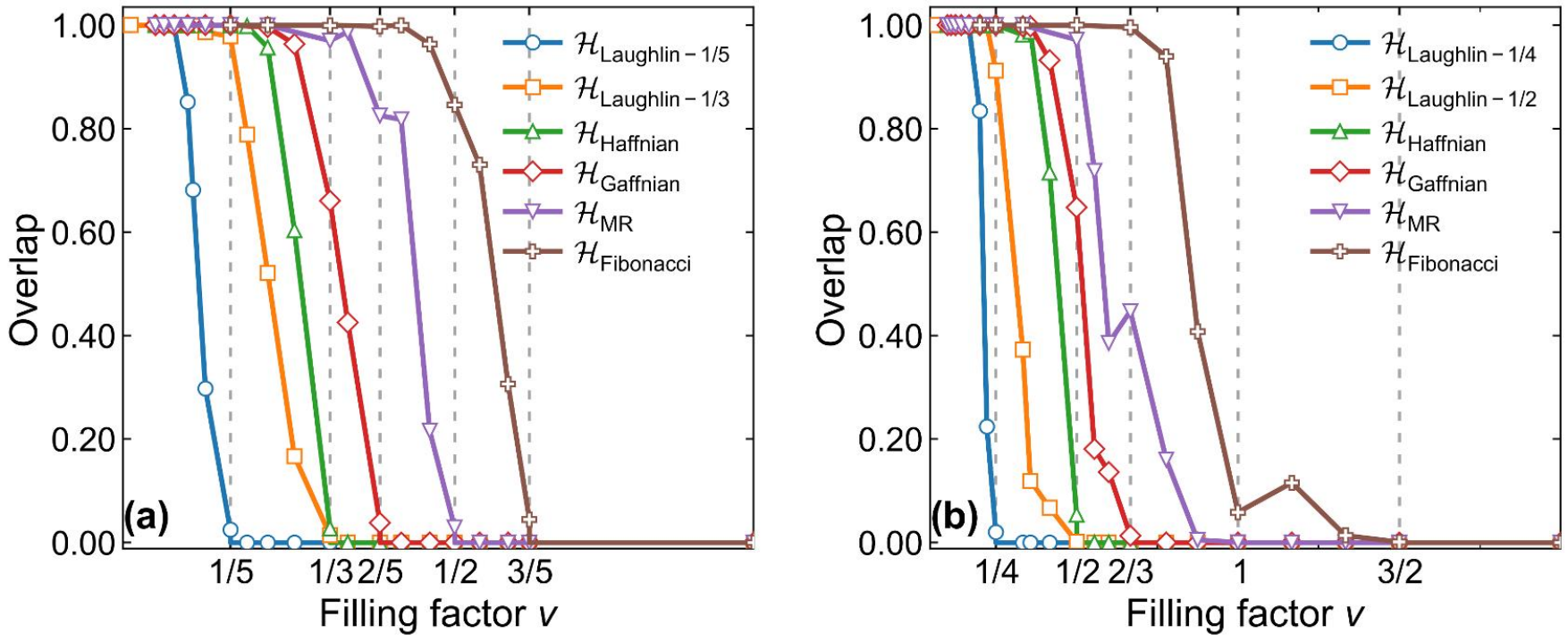}
    \caption{(a) and (b) Total overlap of the lowest 200 eigenstates of the Coulomb interaction projected into the 1LL with representative CHS as a function of filling factor for fermions and bosons, respectively. Different curves correspond to the indicated CHS.
}
    \label{fig:S601}
\end{figure}

\begin{figure}[!ht]
    \centering
    \includegraphics[width=.7\textwidth]{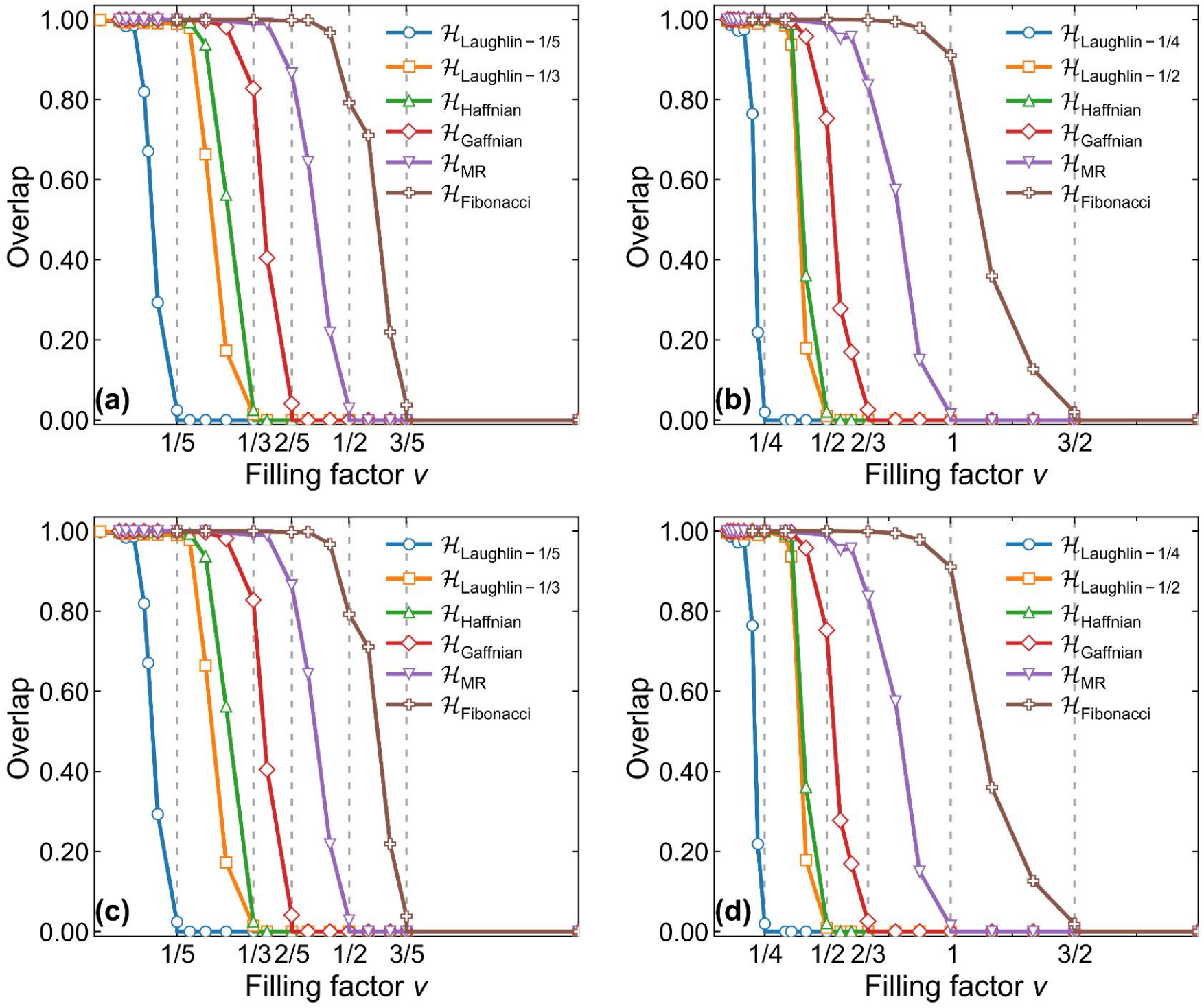}
    \caption{(a) and (b) Total overlap of the lowest 200 eigenstates of the Coulomb
interaction projected into the $\mathcal C=1$ band of TBG with
representative CHS as a function of filling factor for fermions and
bosons, respectively. (c) and (d) Corresponding results for the
$\mathcal C=-1$ band. Different curves correspond to the indicated CHS.
}
    \label{fig:S602}
\end{figure}

\begin{figure}[!ht]
    \centering
    \includegraphics[width=.7\textwidth]{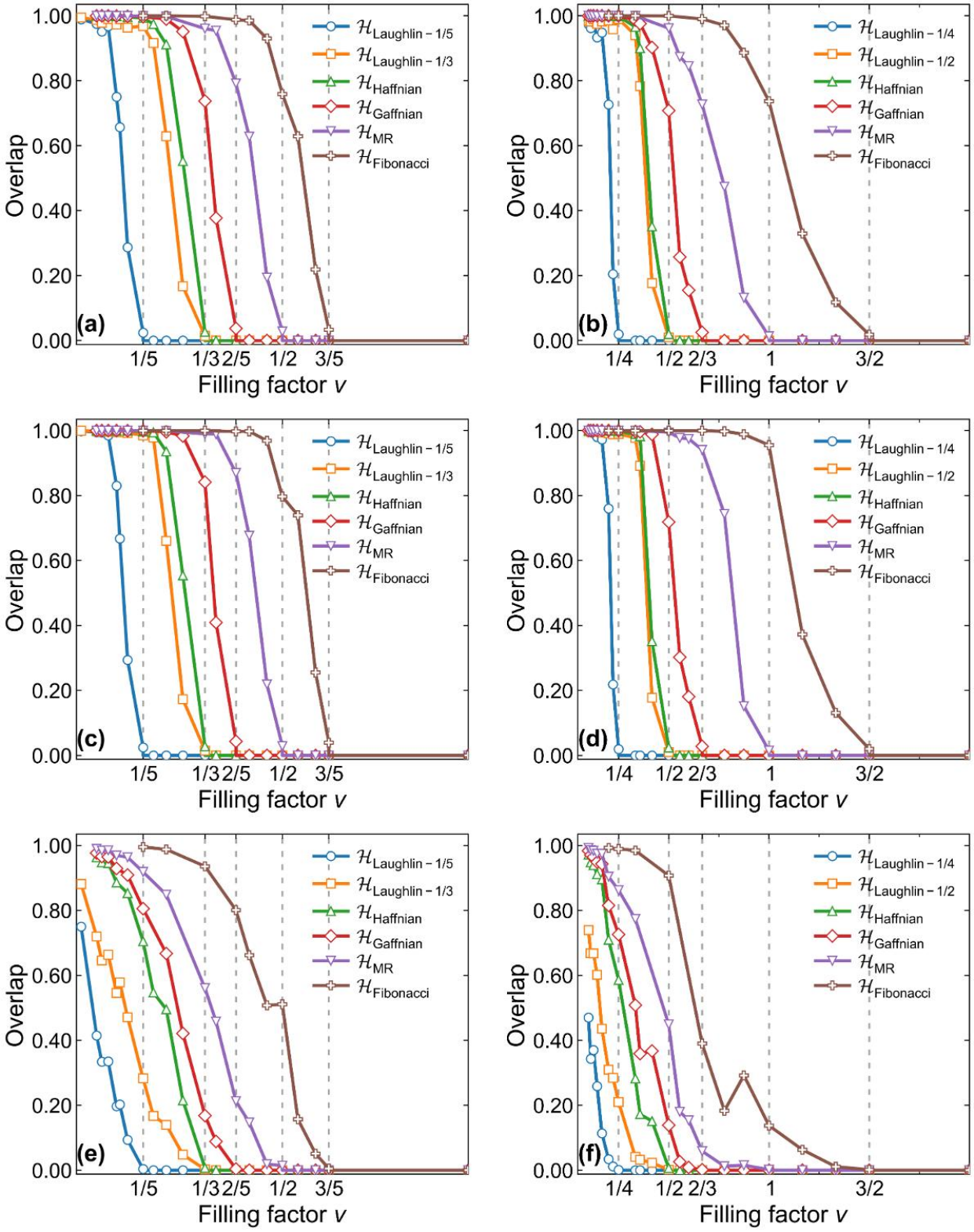}
    \caption{(a) and (b) Total overlap of the lowest 200 eigenstates of the Coulomb
interaction in tMoTe$_2$ at $\vartheta=1.2^\circ$, projected into the
$\chi^{-1}=1.30$ band, with representative CHS as a function of filling
factor for fermions and bosons, respectively. (c) and (d) Corresponding
results at $\vartheta=2^\circ$ for the $\chi^{-1}=1.15$ band, and
(e) and (f) for the $\chi^{-1}=3.74$ band. Different curves correspond
to the indicated CHS.
}
    \label{fig:S603}
\end{figure}

\begin{figure}[!ht]
    \centering
    \includegraphics[width=.9\textwidth]{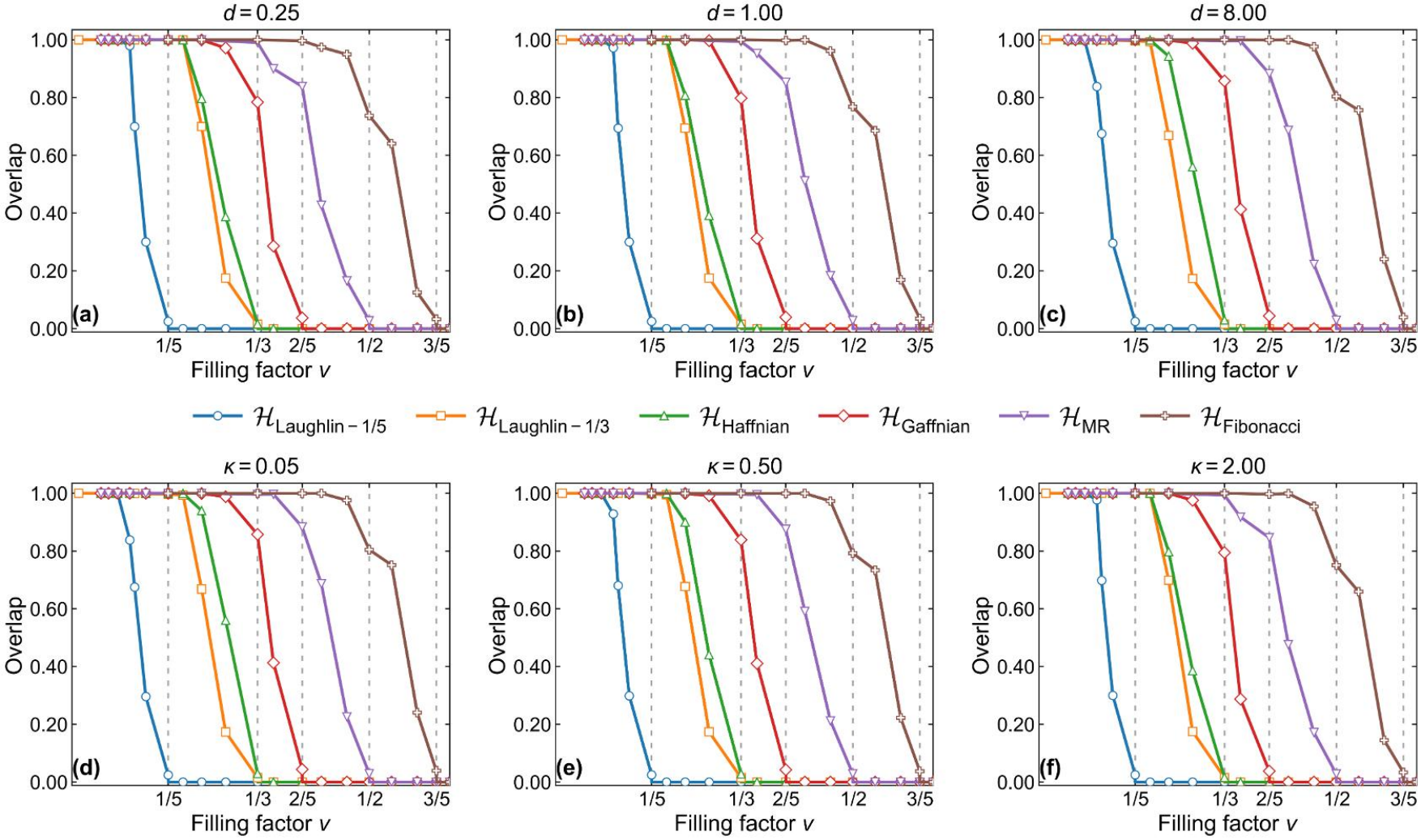}
    \caption{
Total overlap of the lowest 200 eigenstates of the screened Coulomb
interaction (\eref{eq:screen}), projected into the LLL, with
representative CHS as a function of filling factor for fermions.
(a)--(c) Results for $d=0.25$, $1$, and $8$, respectively.
(d)--(f) Results for $\kappa=0.05$, $0.5$, and $2$, respectively.
Different curves correspond to the indicated CHS.
}
    \label{fig:S604}
\end{figure}

\begin{figure}[!ht]
    \centering
    \includegraphics[width=.9\textwidth]{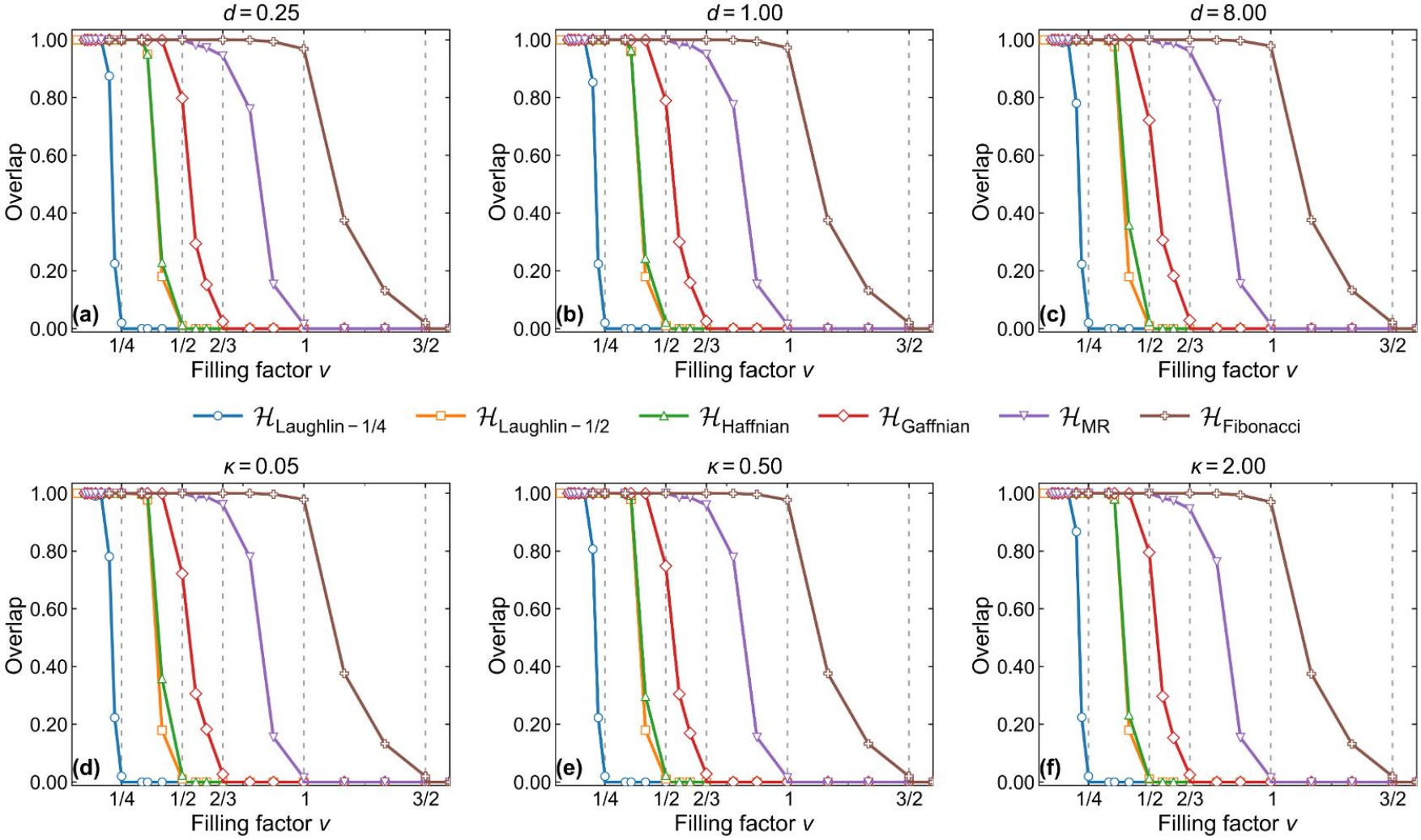}
    \caption{
Total overlap of the lowest 200 eigenstates of the screened Coulomb
interaction (\eref{eq:screen}), projected into the LLL, with
representative CHS as a function of filling factor for bosons.
(a)--(c) Results for $d=0.25$, $1$, and $8$, respectively.
(d)--(f) Results for $\kappa=0.05$, $0.5$, and $2$, respectively.
Different curves correspond to the indicated CHS.
}
    \label{fig:S605}
\end{figure}

\begin{figure}[!ht]
    \centering
    \includegraphics[width=.9\textwidth]{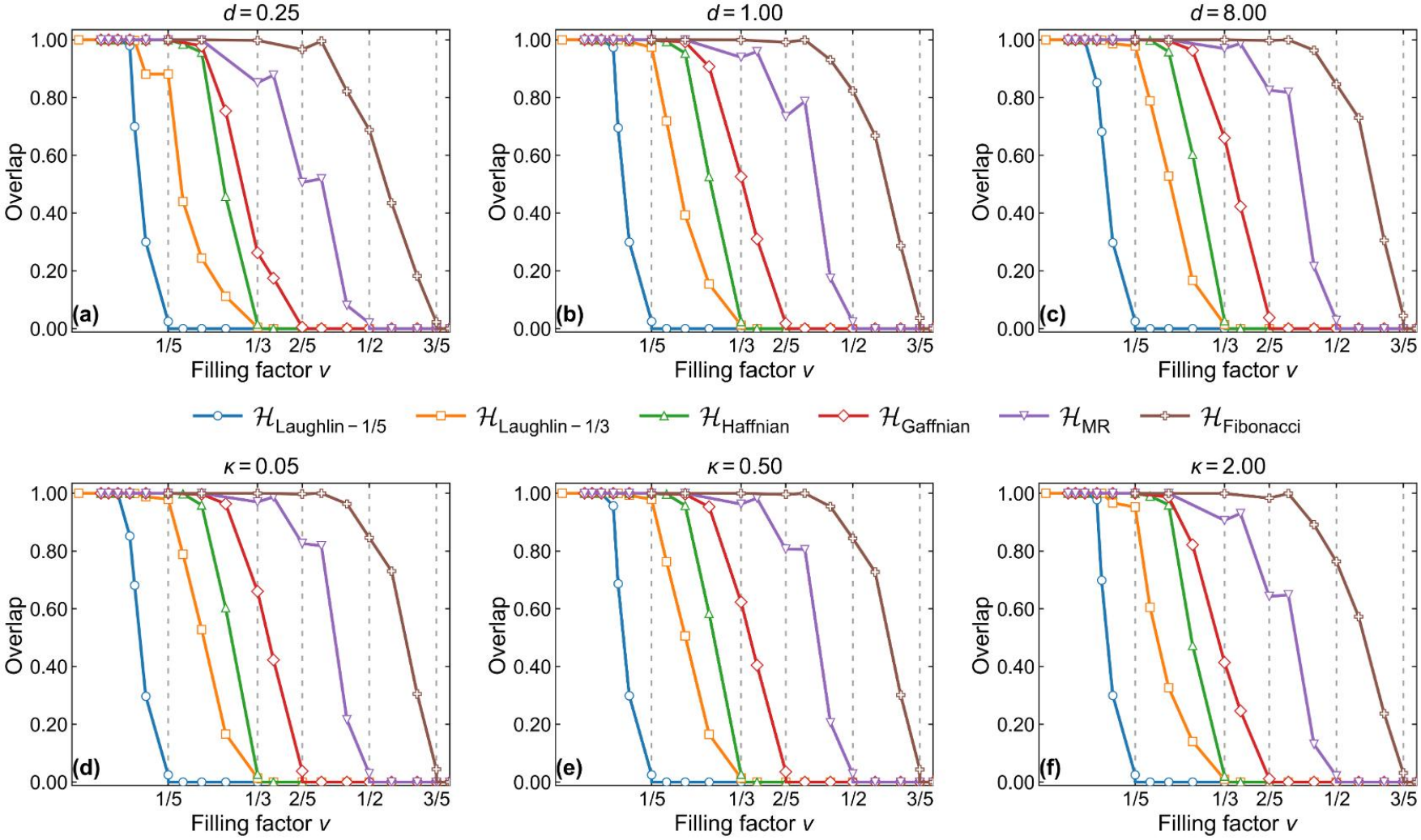}
    \caption{
Total overlap of the lowest 200 eigenstates of the screened Coulomb
interaction (\eref{eq:screen}), projected into the 1LL, with
representative CHS as a function of filling factor for fermions.
(a)--(c) Results for $d=0.25$, $1$, and $8$, respectively.
(d)--(f) Results for $\kappa=0.05$, $0.5$, and $2$, respectively.
Different curves correspond to the indicated CHS.
}
    \label{fig:S606}
\end{figure}

\begin{figure}[!ht]
    \centering
    \includegraphics[width=.9\textwidth]{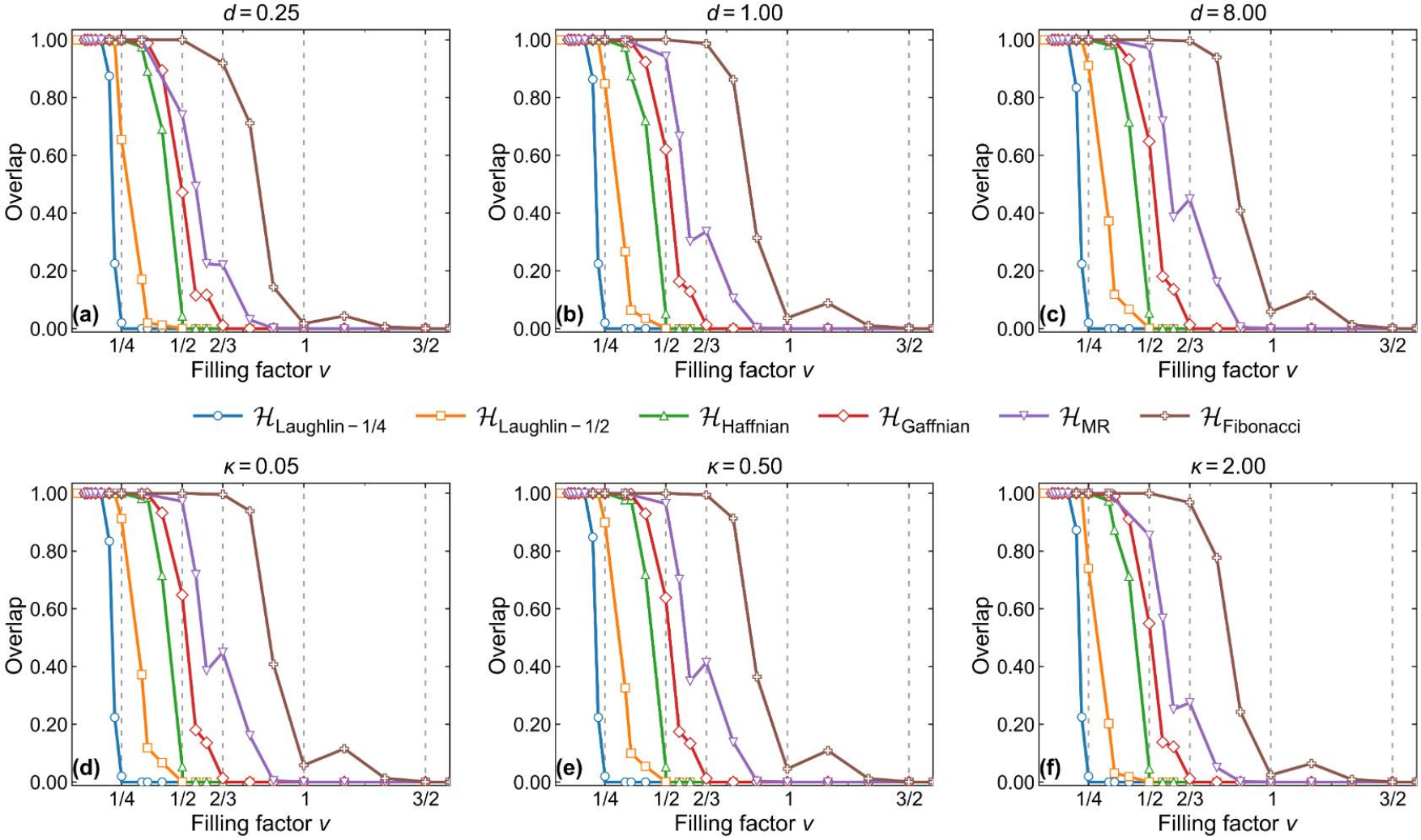}
    \caption{
Total overlap of the lowest 200 eigenstates of the screened Coulomb
interaction (\eref{eq:screen}), projected into the 1LL, with
representative CHS as a function of filling factor for bosons.
(a)--(c) Results for $d=0.25$, $1$, and $8$, respectively.
(d)--(f) Results for $\kappa=0.05$, $0.5$, and $2$, respectively.
Different curves correspond to the indicated CHS.
}
    \label{fig:S607}
\end{figure}

\begin{figure}[!ht]
    \centering
    \includegraphics[width=.9\textwidth]{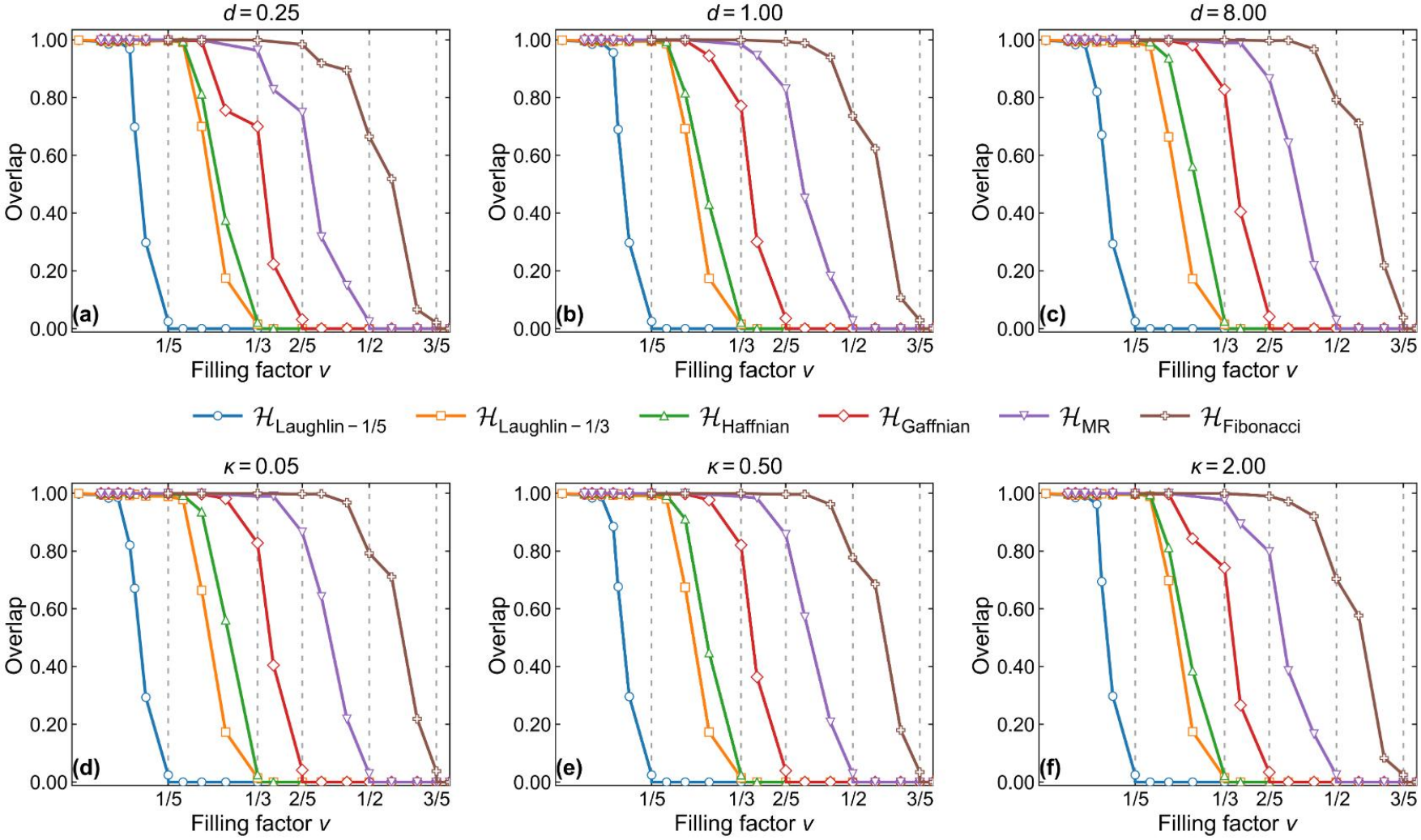}
    \caption{
Total overlap of the lowest 200 eigenstates of the screened Coulomb
interaction (\eref{eq:screen}), projected into the $\mathcal{C}=1$
band of TBG, with representative CHS as a function of filling factor
for fermions.
(a)--(c) Results for $d=0.25$, $1$, and $8$, respectively.
(d)--(f) Results for $\kappa=0.05$, $0.5$, and $2$, respectively.
Different curves correspond to the indicated CHS.
}
    \label{fig:S608}
\end{figure}

\begin{figure}[!ht]
    \centering
    \includegraphics[width=.9\textwidth]{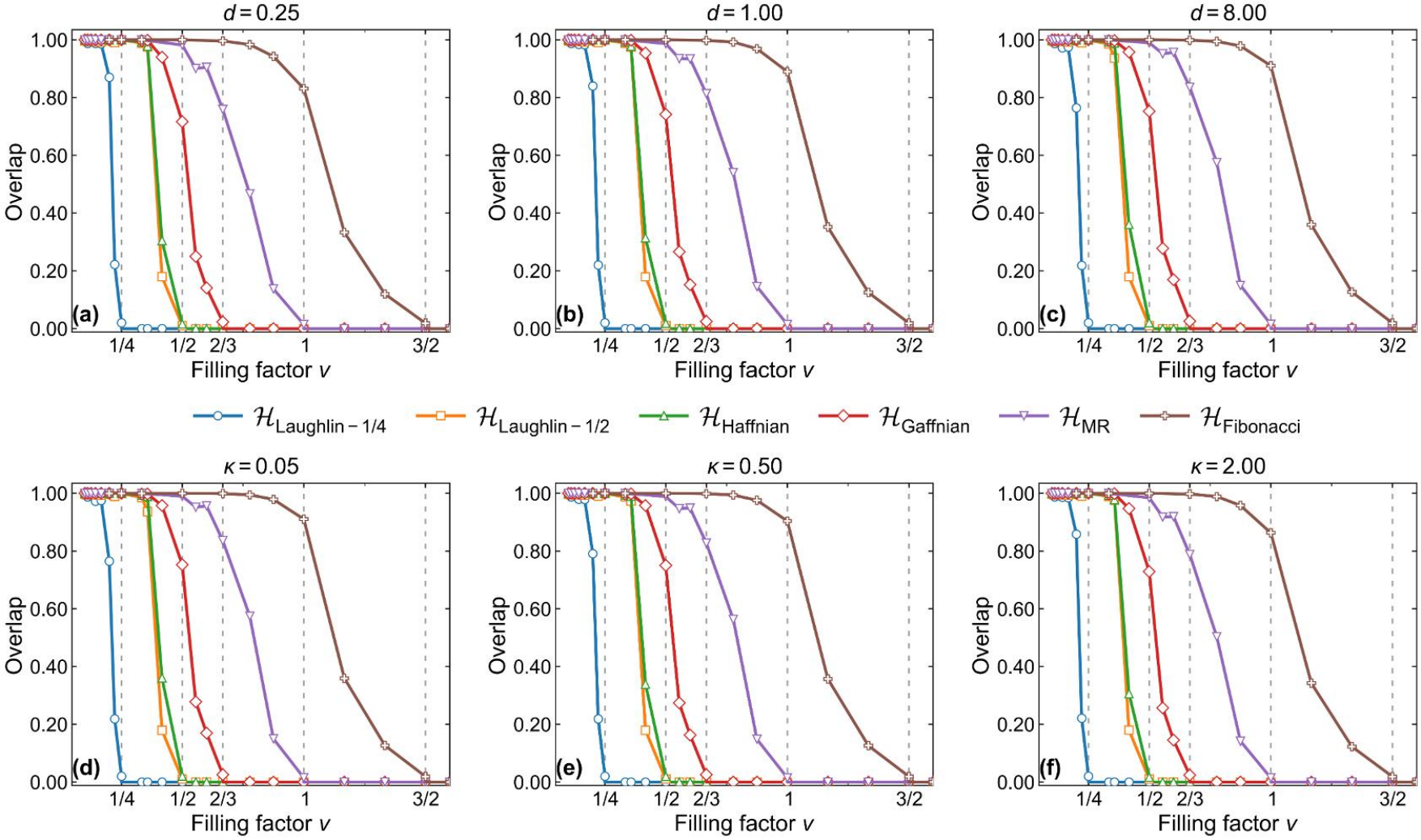}
    \caption{
Total overlap of the lowest 200 eigenstates of the screened Coulomb
interaction (\eref{eq:screen}), projected into the $\mathcal{C}=1$
band of TBG, with representative CHS as a function of filling factor
for bosons.
(a)--(c) Results for $d=0.25$, $1$, and $8$, respectively.
(d)--(f) Results for $\kappa=0.05$, $0.5$, and $2$, respectively.
Different curves correspond to the indicated CHS.
}
    \label{fig:S609}
\end{figure}

\begin{figure}[!ht]
    \centering
    \includegraphics[width=.9\textwidth]{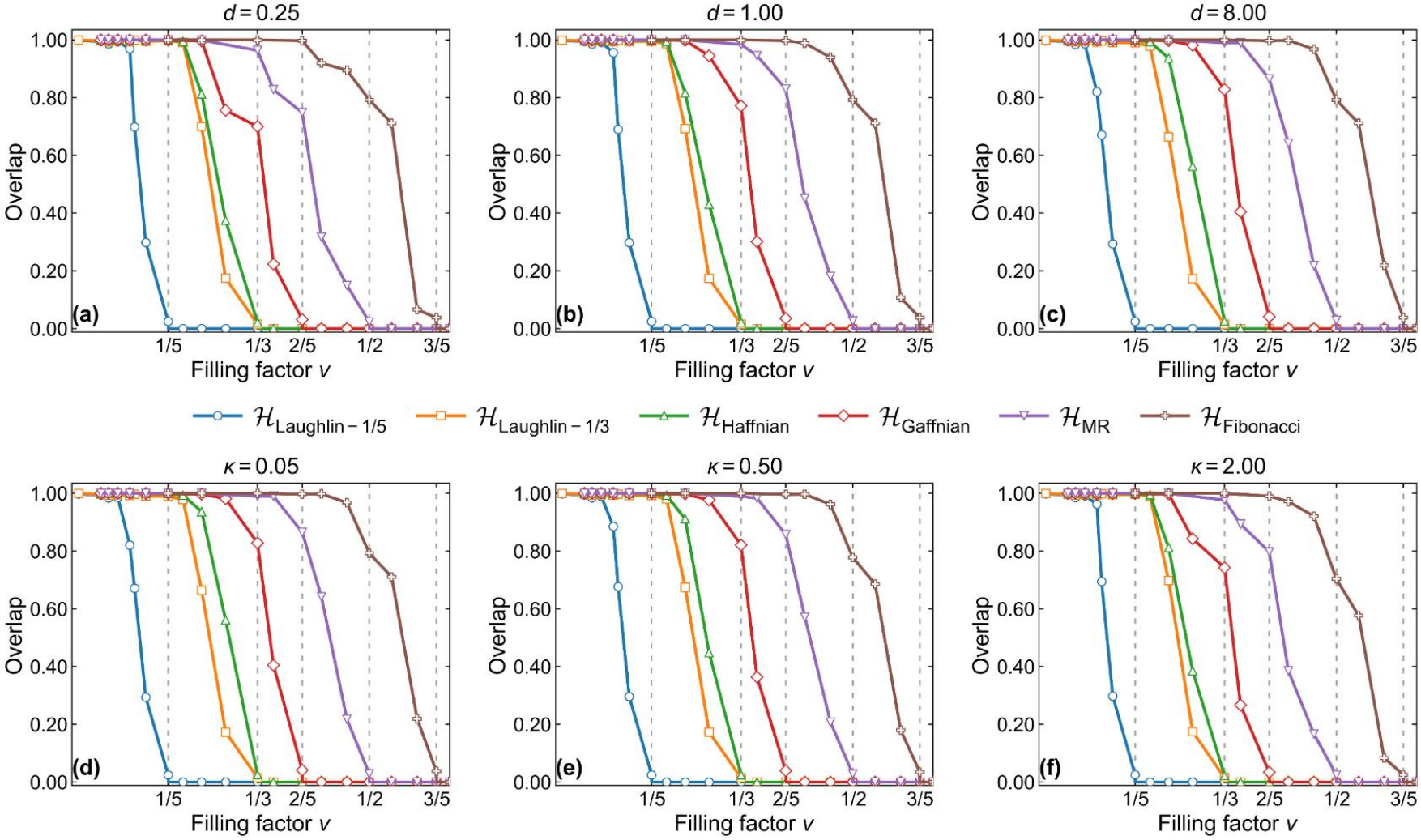}
    \caption{
Total overlap of the lowest 200 eigenstates of the screened Coulomb
interaction (\eref{eq:screen}), projected into the $\mathcal{C}=-1$
band of TBG, with representative CHS as a function of filling factor
for fermions.
(a)--(c) Results for $d=0.25$, $1$, and $8$, respectively.
(d)--(f) Results for $\kappa=0.05$, $0.5$, and $2$, respectively.
Different curves correspond to the indicated CHS.
}
    \label{fig:S610}
\end{figure}

\begin{figure}[!ht]
    \centering
    \includegraphics[width=.9\textwidth]{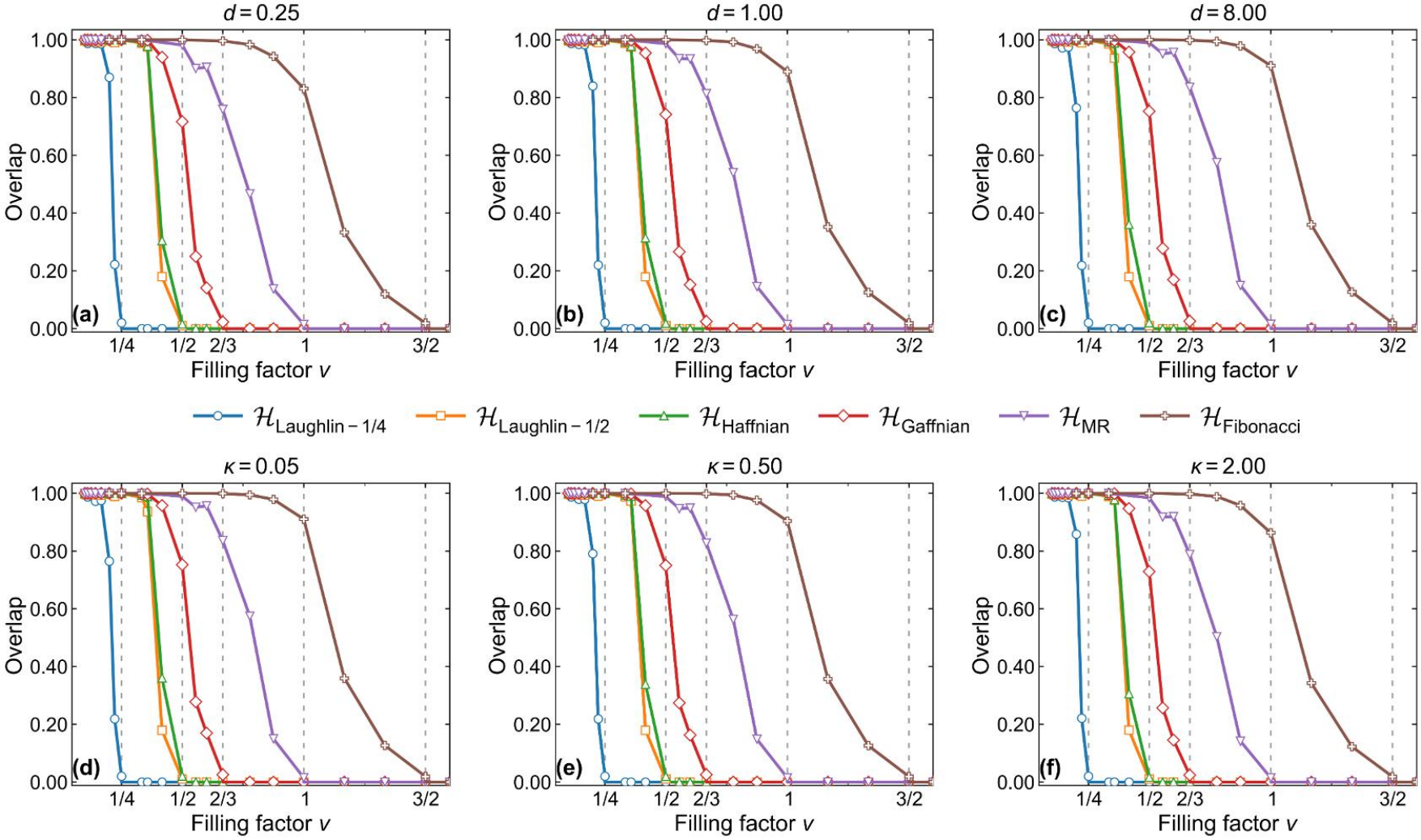}
    \caption{
Total overlap of the lowest 200 eigenstates of the screened Coulomb
interaction (\eref{eq:screen}), projected into the $\mathcal{C}=-1$
band of TBG, with representative CHS as a function of filling factor
for bosons.
(a)--(c) Results for $d=0.25$, $1$, and $8$, respectively.
(d)--(f) Results for $\kappa=0.05$, $0.5$, and $2$, respectively.
Different curves correspond to the indicated CHS.
}
    \label{fig:S611}
\end{figure}

\begin{figure}[!ht]
    \centering
    \includegraphics[width=.9\textwidth]{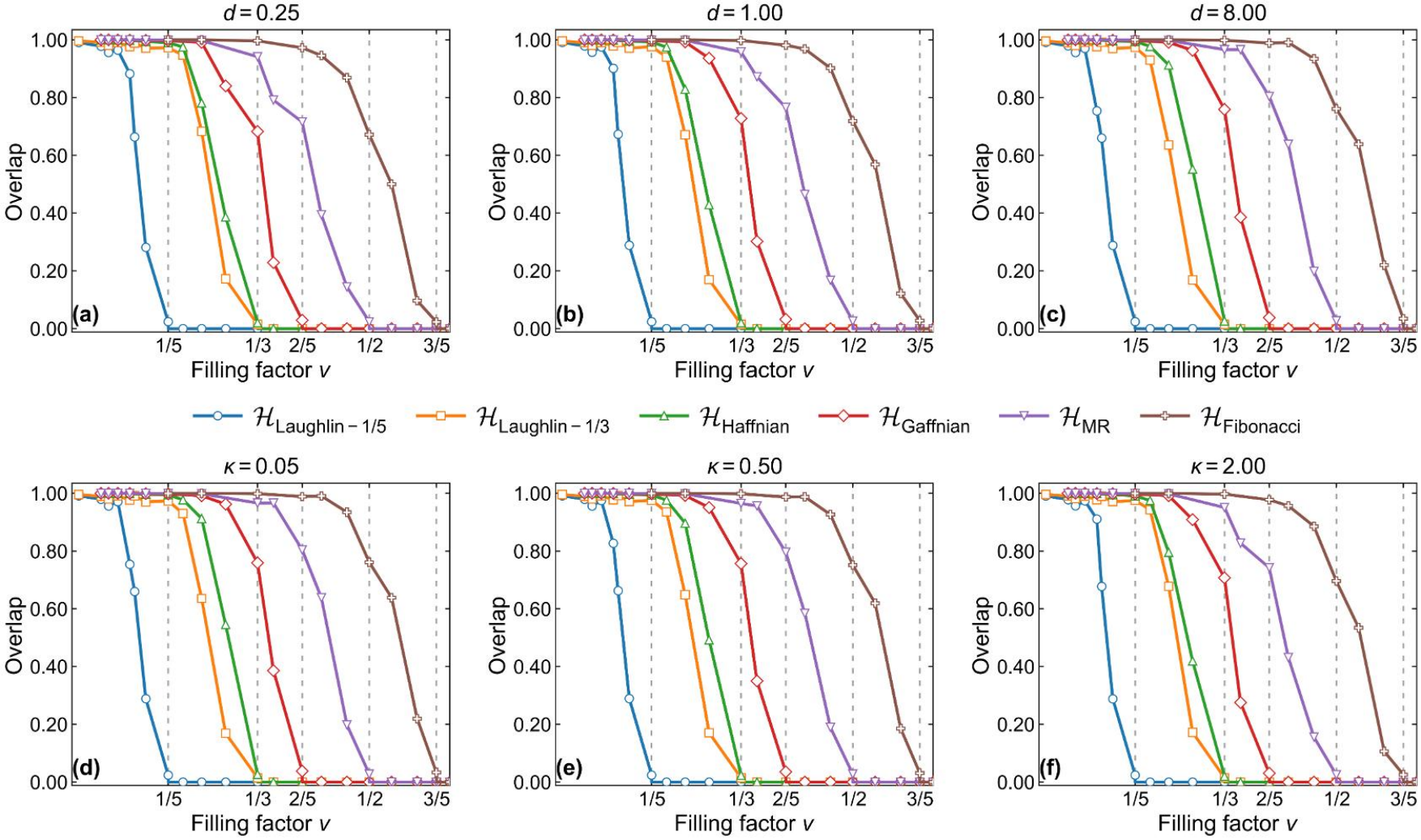}
    \caption{
Total overlap of the lowest 200 eigenstates of the screened Coulomb
interaction (\eref{eq:screen}), projected into the
$\vartheta=1.2^\circ$, $\chi^{-1}=1.14$ band of  tMoTe$_2$, with
representative CHS as a function of filling factor for fermions.
(a)--(c) Results for $d=0.25$, $1$, and $8$, respectively.
(d)--(f) Results for $\kappa=0.05$, $0.5$, and $2$, respectively.
Different curves correspond to the indicated CHS.
}
    \label{fig:S612}
\end{figure}

\begin{figure}[!ht]
    \centering
    \includegraphics[width=.9\textwidth]{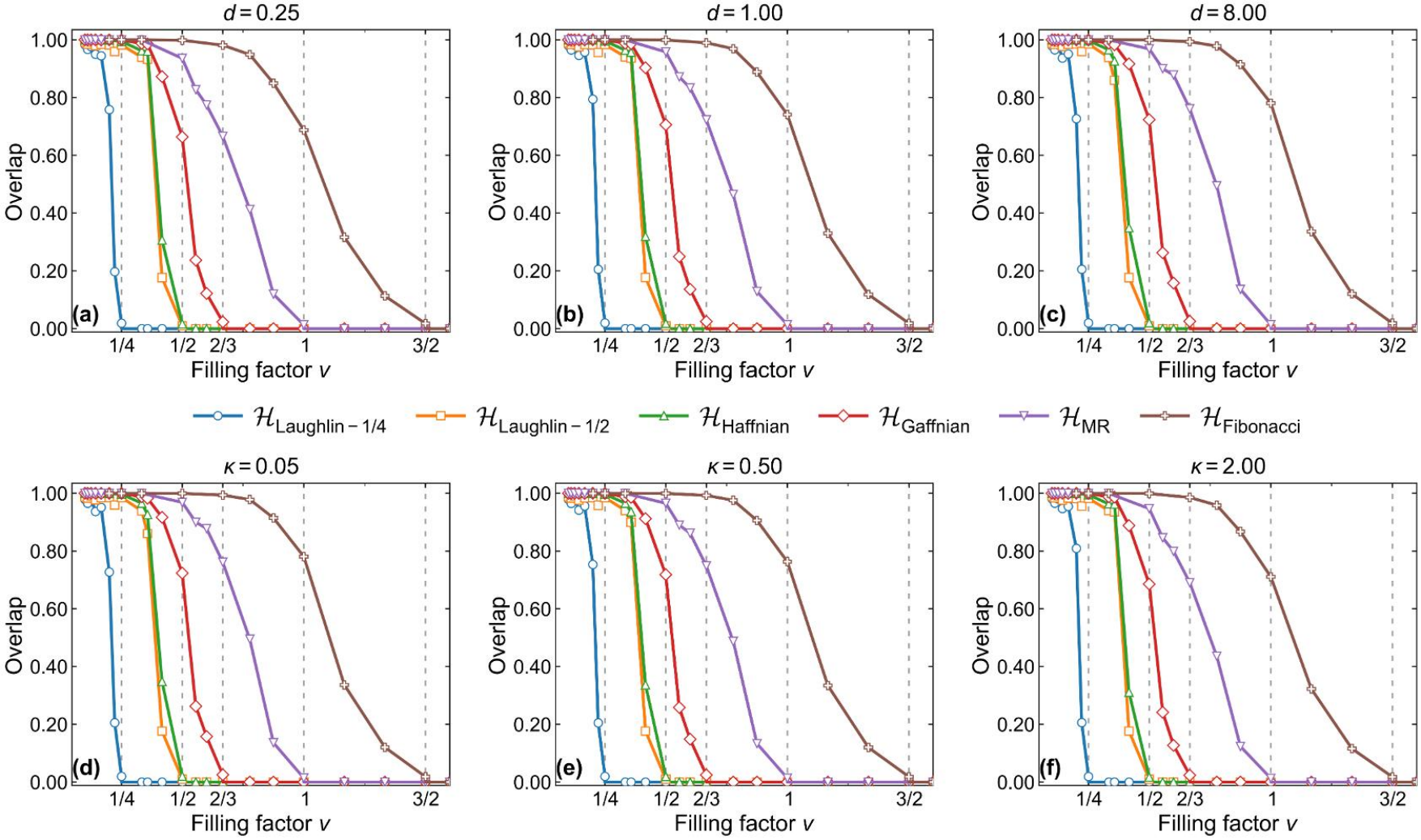}
    \caption{
Total overlap of the lowest 200 eigenstates of the screened Coulomb
interaction (\eref{eq:screen}), projected into the
$\vartheta=1.2^\circ$, $\chi^{-1}=1.14$ band of  tMoTe$_2$, with
representative CHS as a function of filling factor for bosons.
(a)--(c) Results for $d=0.25$, $1$, and $8$, respectively.
(d)--(f) Results for $\kappa=0.05$, $0.5$, and $2$, respectively.
Different curves correspond to the indicated CHS.
}
    \label{fig:S613}
\end{figure}

\begin{figure}[!ht]
    \centering
    \includegraphics[width=.9\textwidth]{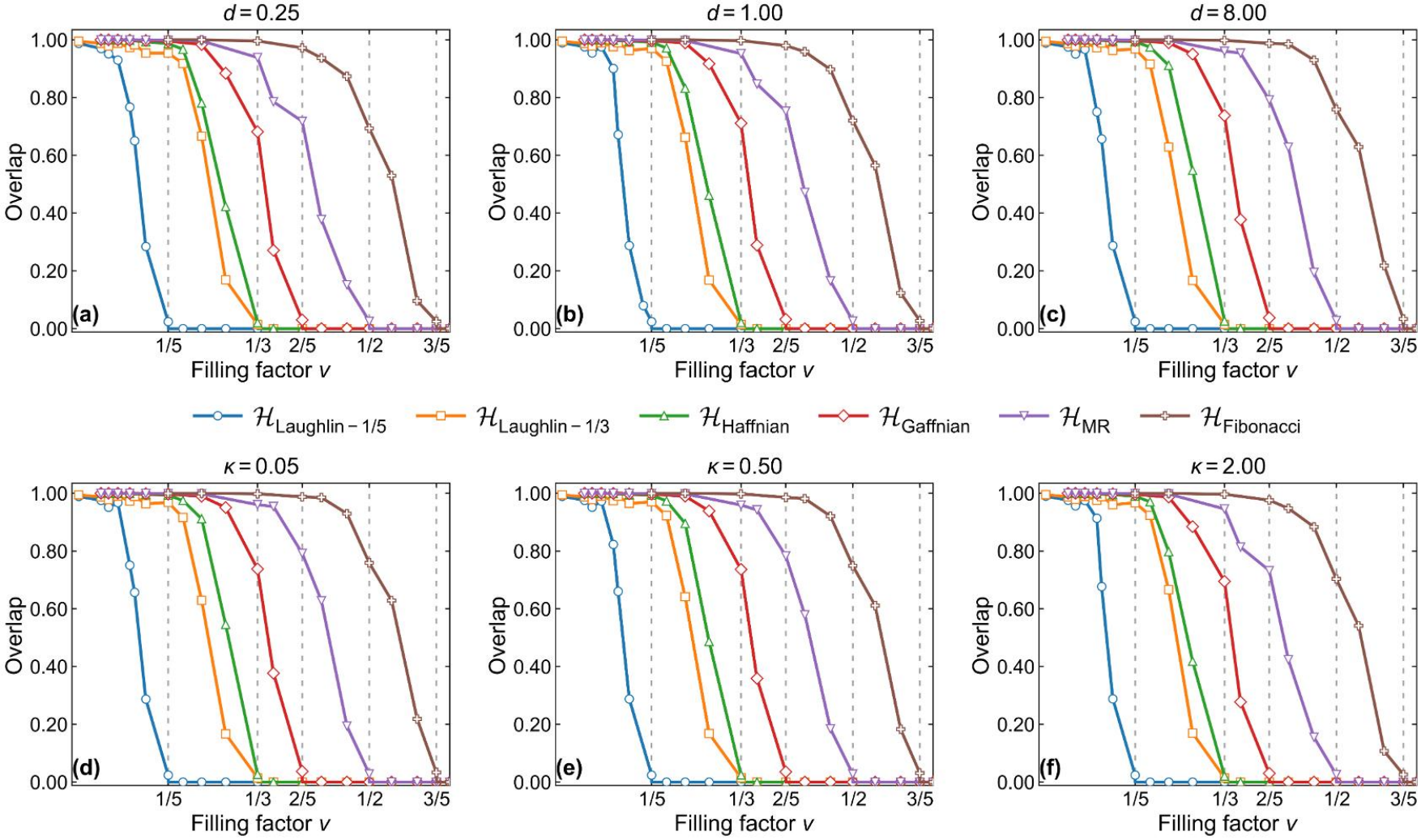}
    \caption{
Total overlap of the lowest 200 eigenstates of the screened Coulomb
interaction (\eref{eq:screen}), projected into the
$\vartheta=1.2^\circ$, $\chi^{-1}=1.30$ band of  tMoTe$_2$, with
representative CHS as a function of filling factor for fermions.
(a)--(c) Results for $d=0.25$, $1$, and $8$, respectively.
(d)--(f) Results for $\kappa=0.05$, $0.5$, and $2$, respectively.
Different curves correspond to the indicated CHS.
}
    \label{fig:S614}
\end{figure}

\begin{figure}[!ht]
    \centering
    \includegraphics[width=.9\textwidth]{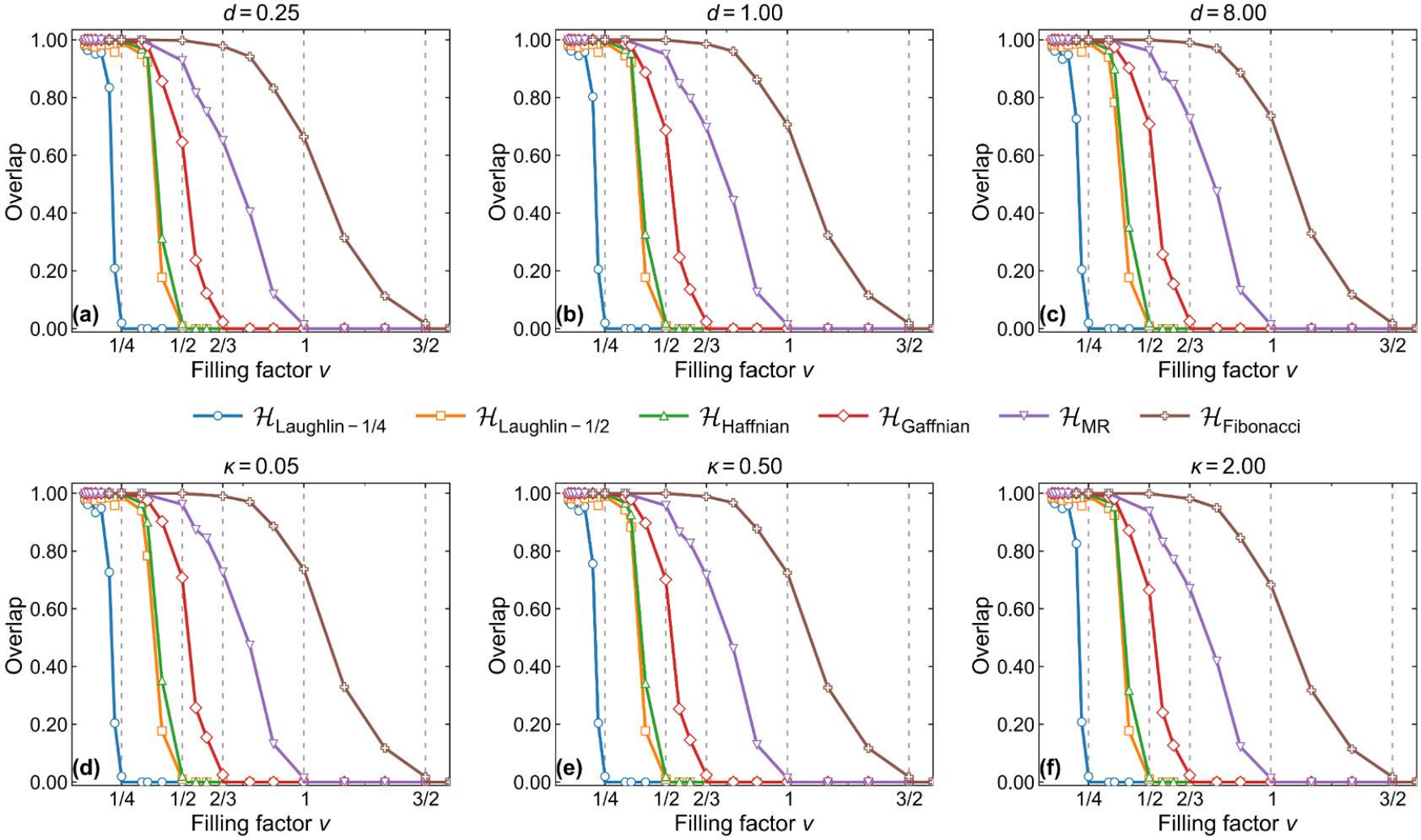}
    \caption{
Total overlap of the lowest 200 eigenstates of the screened Coulomb
interaction (\eref{eq:screen}), projected into the
$\vartheta=1.2^\circ$, $\chi^{-1}=1.30$ band of tMoTe$_2$, with
representative CHS as a function of filling factor for bosons.
(a)--(c) Results for $d=0.25$, $1$, and $8$, respectively.
(d)--(f) Results for $\kappa=0.05$, $0.5$, and $2$, respectively.
Different curves correspond to the indicated CHS.
}
    \label{fig:S615}
\end{figure}

\begin{figure}[!ht]
    \centering
    \includegraphics[width=.9\textwidth]{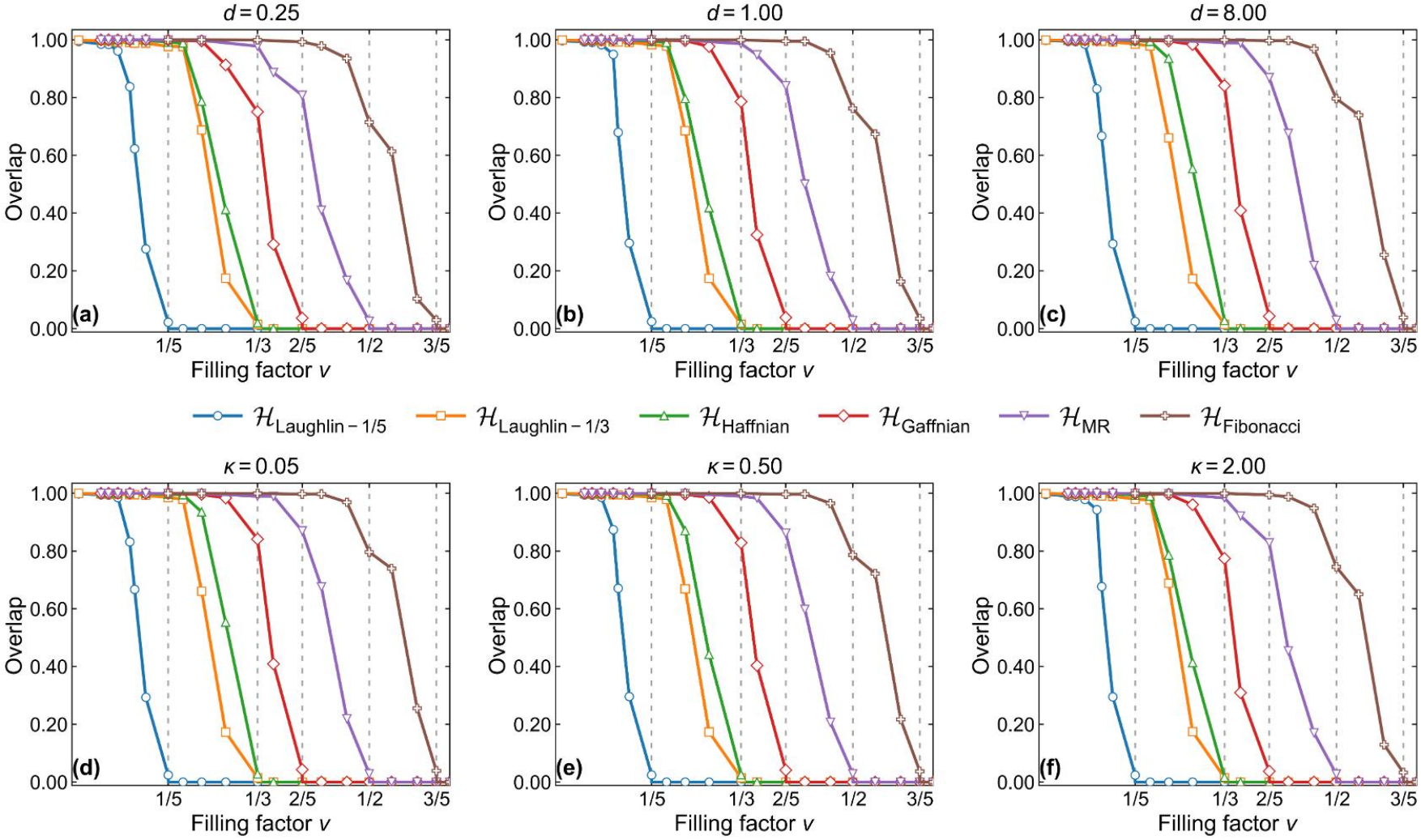}
    \caption{
Total overlap of the lowest 200 eigenstates of the screened Coulomb
interaction (\eref{eq:screen}), projected into the
$\vartheta=2^\circ$, $\chi^{-1}=1.15$ band of  tMoTe$_2$, with
representative CHS as a function of filling factor for fermions.
(a)--(c) Results for $d=0.25$, $1$, and $8$, respectively.
(d)--(f) Results for $\kappa=0.05$, $0.5$, and $2$, respectively.
Different curves correspond to the indicated CHS.
}
    \label{fig:S616}
\end{figure}

\begin{figure}[!ht]
    \centering
    \includegraphics[width=.9\textwidth]{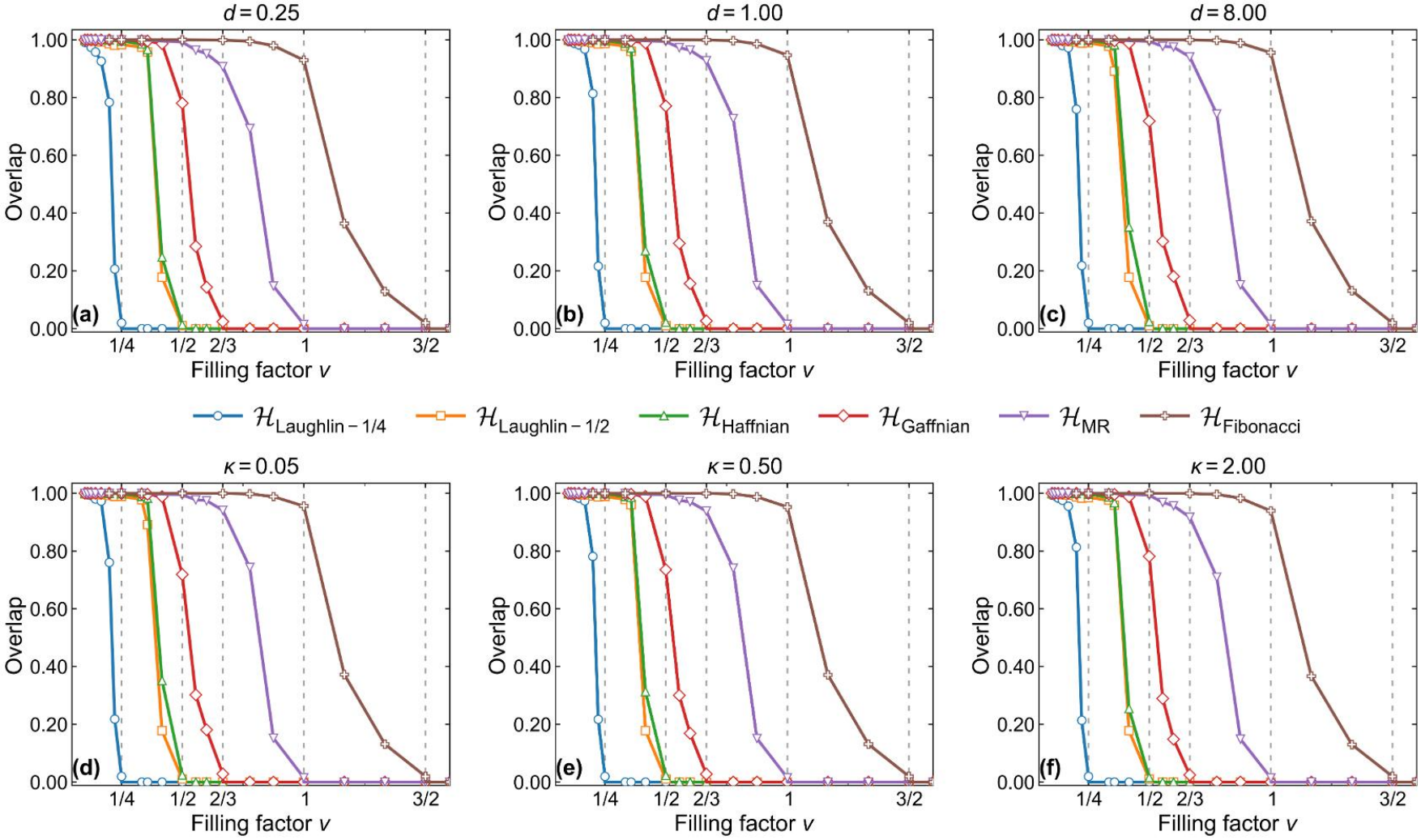}
    \caption{
Total overlap of the lowest 200 eigenstates of the screened Coulomb
interaction (\eref{eq:screen}), projected into the
$\vartheta=2^\circ$, $\chi^{-1}=1.15$ band of tMoTe$_2$, with
representative CHS as a function of filling factor for bosons.
(a)--(c) Results for $d=0.25$, $1$, and $8$, respectively.
(d)--(f) Results for $\kappa=0.05$, $0.5$, and $2$, respectively.
Different curves correspond to the indicated CHS.
}
    \label{fig:S617}
\end{figure}

\begin{figure}[!ht]
    \centering
    \includegraphics[width=.9\textwidth]{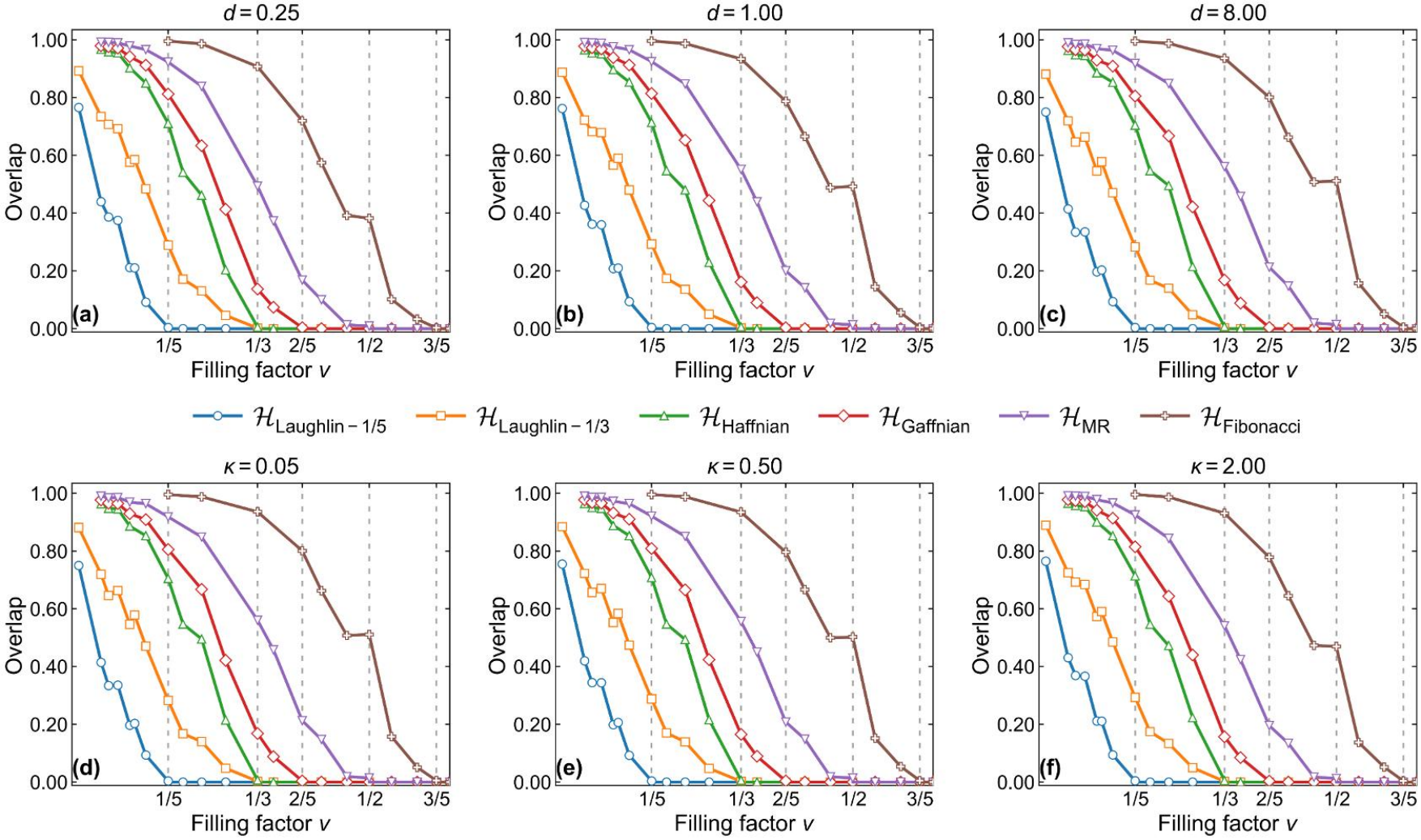}
    \caption{
Total overlap of the lowest 200 eigenstates of the screened Coulomb
interaction (\eref{eq:screen}), projected into the
$\vartheta=2^\circ$, $\chi^{-1}=3.74$ band of tMoTe$_2$, with
representative CHS as a function of filling factor for fermions.
(a)--(c) Results for $d=0.25$, $1$, and $8$, respectively.
(d)--(f) Results for $\kappa=0.05$, $0.5$, and $2$, respectively.
Different curves correspond to the indicated CHS.
}
    \label{fig:S618}
\end{figure}

\begin{figure}[!ht]
    \centering
    \includegraphics[width=.9\textwidth]{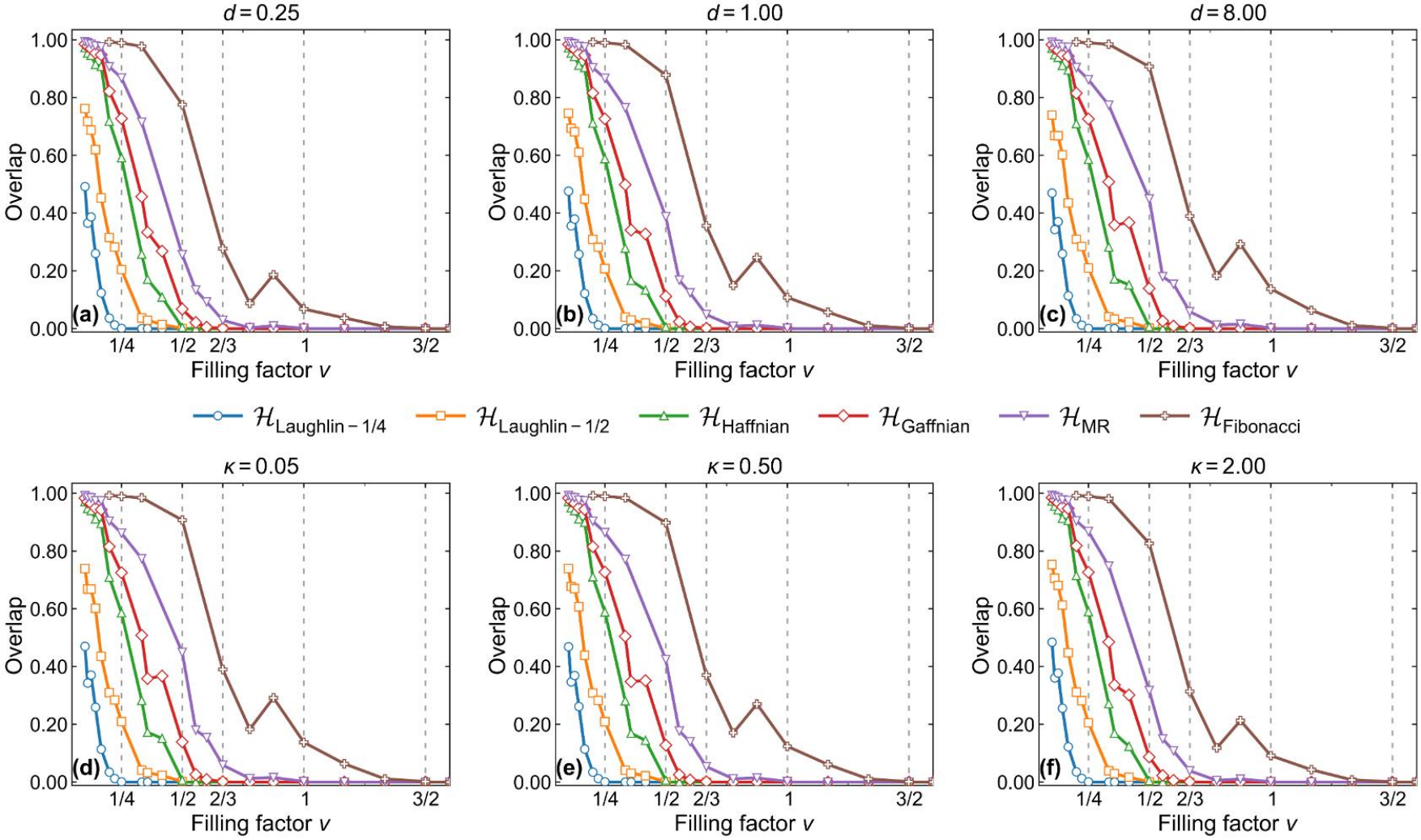}
    \caption{
Total overlap of the lowest 200 eigenstates of the screened Coulomb
interaction (\eref{eq:screen}), projected into the
$\vartheta=2^\circ$, $\chi^{-1}=3.74$ band of tMoTe$_2$, with
representative CHS as a function of filling factor for bosons.
(a)--(c) Results for $d=0.25$, $1$, and $8$, respectively.
(d)--(f) Results for $\kappa=0.05$, $0.5$, and $2$, respectively.
Different curves correspond to the indicated CHS.
}
    \label{fig:S619}
\end{figure}

Beyond the results presented in the main text, we further evaluate the
overlap between the lowest 200 eigenstates of the Coulomb interaction and
different CHS for the 1LL (\fref{fig:S601}), the TBG bands
(\fref{fig:S602}), and the tMoTe$_2$ bands
(\fref{fig:S603}) over a range of filling factors. In all cases, the
overlap exhibits the same hierarchical structure among the CHS.
It is particularly noteworthy that the tMoTe$_2$ band at
$\vartheta=2^\circ$ with $\chi^{-1}=3.74$ deviates substantially from the
ideal-band condition compared with the other bands considered here. As a
result, the projection from the LLL CHS into this band is subject to much
stronger deformation. Nevertheless, the same CHS hierarchy remains clearly
visible, indicating that the hierarchical structure is robust even away
from the ideal-band limit.
In addition, we examine screened Coulomb interactions with different
screening parameters across the various Chern bands considered above
(\fref{fig:S604}--\fref{fig:S619}). The same hierarchical structure is
observed throughout, indicating that the CHS hierarchy is robust against
both changes in the underlying band geometry and variations in the
interaction range, from long-ranged to strongly screened short-ranged
interactions.

\section{Finite-size scaling of band mixing}
\begin{figure}[!ht]
    \centering
    \includegraphics[width=.7\textwidth]{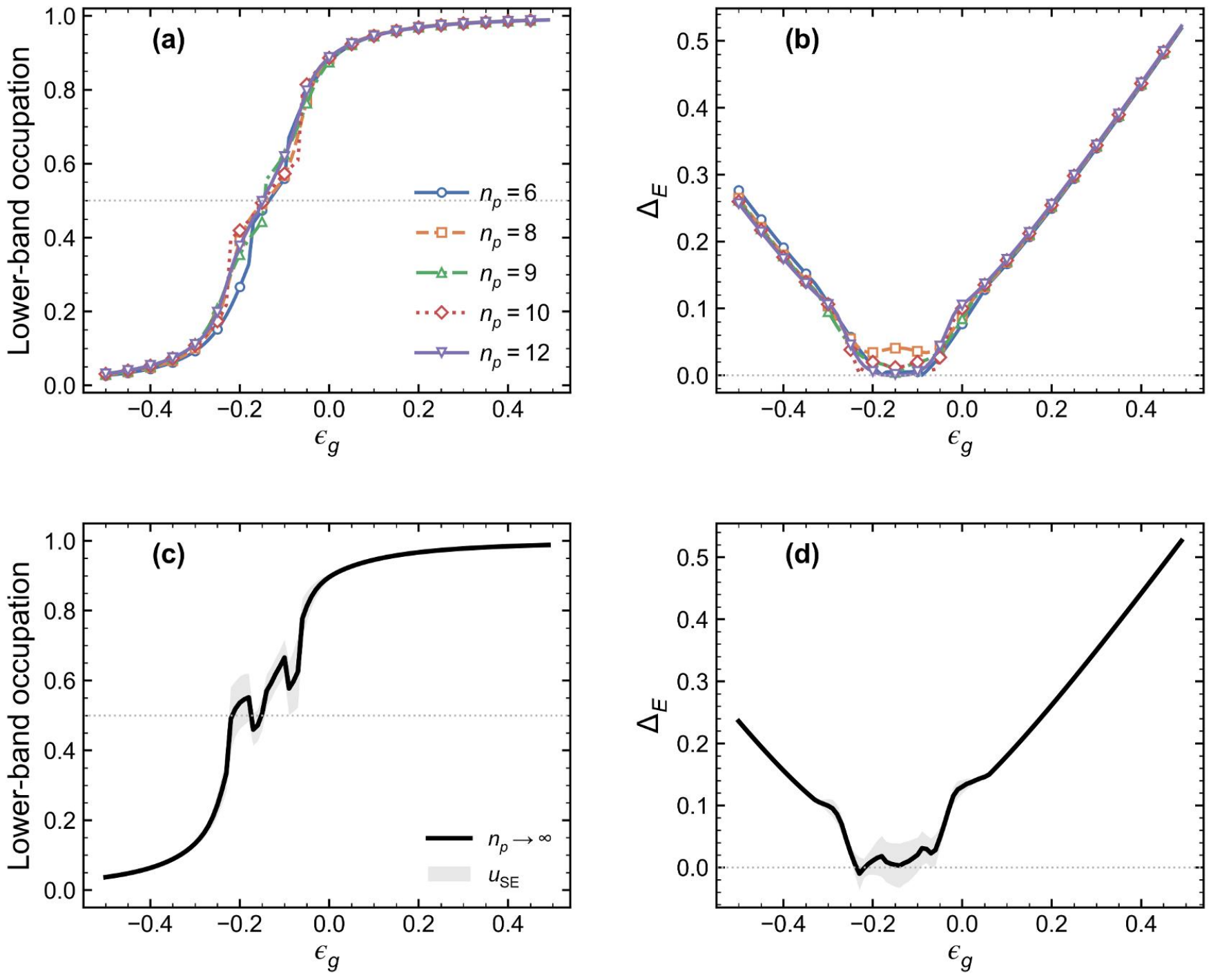}
    \caption{
Finite-size scaling of the LLL occupation and excitation gap
$\Delta E$ for the LLL--1LL system as functions of $\epsilon_g$.
(a) LLL occupation and (b) excitation gap for
$n_p=6$, $8$, $9$, $10$, and $12$.
(c) and (d) Corresponding thermodynamic-limit values obtained by
independently fitting the finite-size data at each $\epsilon_g$ using
unweighted ordinary least squares. The black curves show the extrapolated
intercepts, and the shaded regions indicate one standard error
$u_{\rm SE}$ of the intercepts.
}
    \label{fig:S701}
\end{figure}

\begin{figure}[!ht]
    \centering
    \includegraphics[width=.97\textwidth]{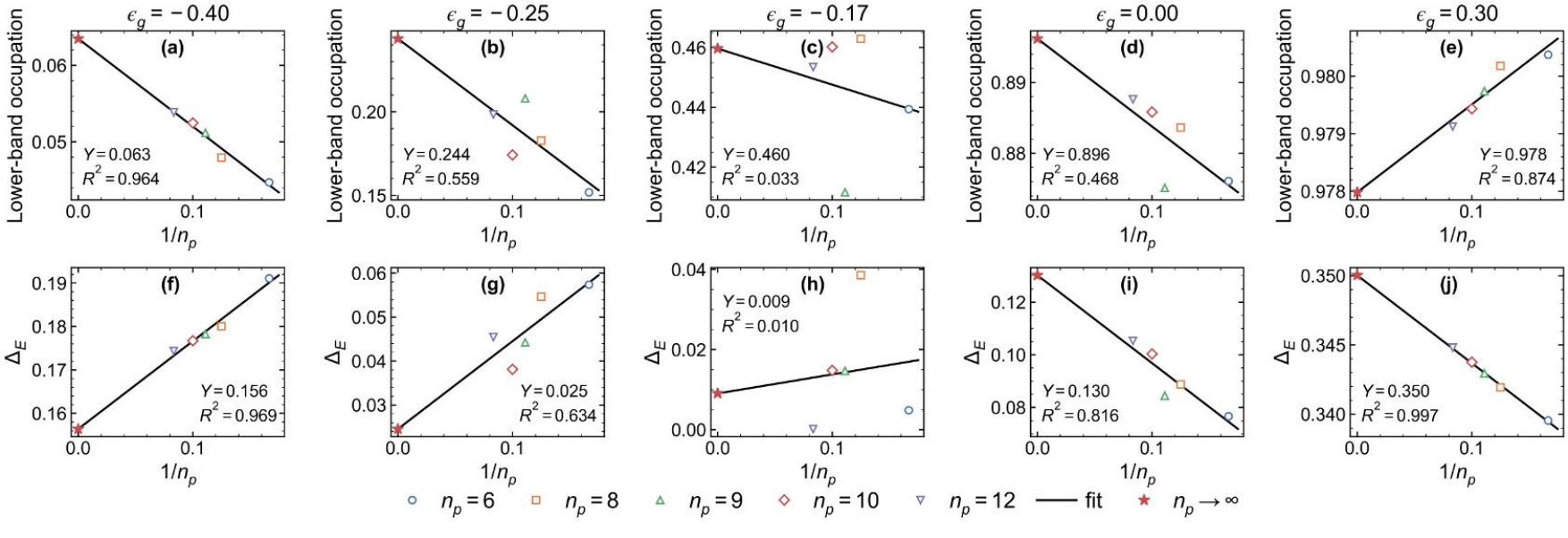}
    \caption{
Representative finite-size extrapolations for the LLL--1LL system.
From left to right, the columns correspond to
$\epsilon_g=-0.40$, $-0.25$, $-0.17$, $0$, and $0.30$.
(a)--(e) LLL occupation and (f)--(j) excitation gap $\Delta E$.
Symbols denote the finite-size data, and the black lines show the
corresponding unweighted linear fits.
Red stars at $1/n_p=0$ mark the extrapolated thermodynamic-limit values
$Y$. The fitted intercept $Y$ and coefficient of determination $R^2$
are indicated in each panel.
}
    \label{fig:S702}
\end{figure}

\begin{figure}[!ht]
    \centering
    \includegraphics[width=.7\textwidth]{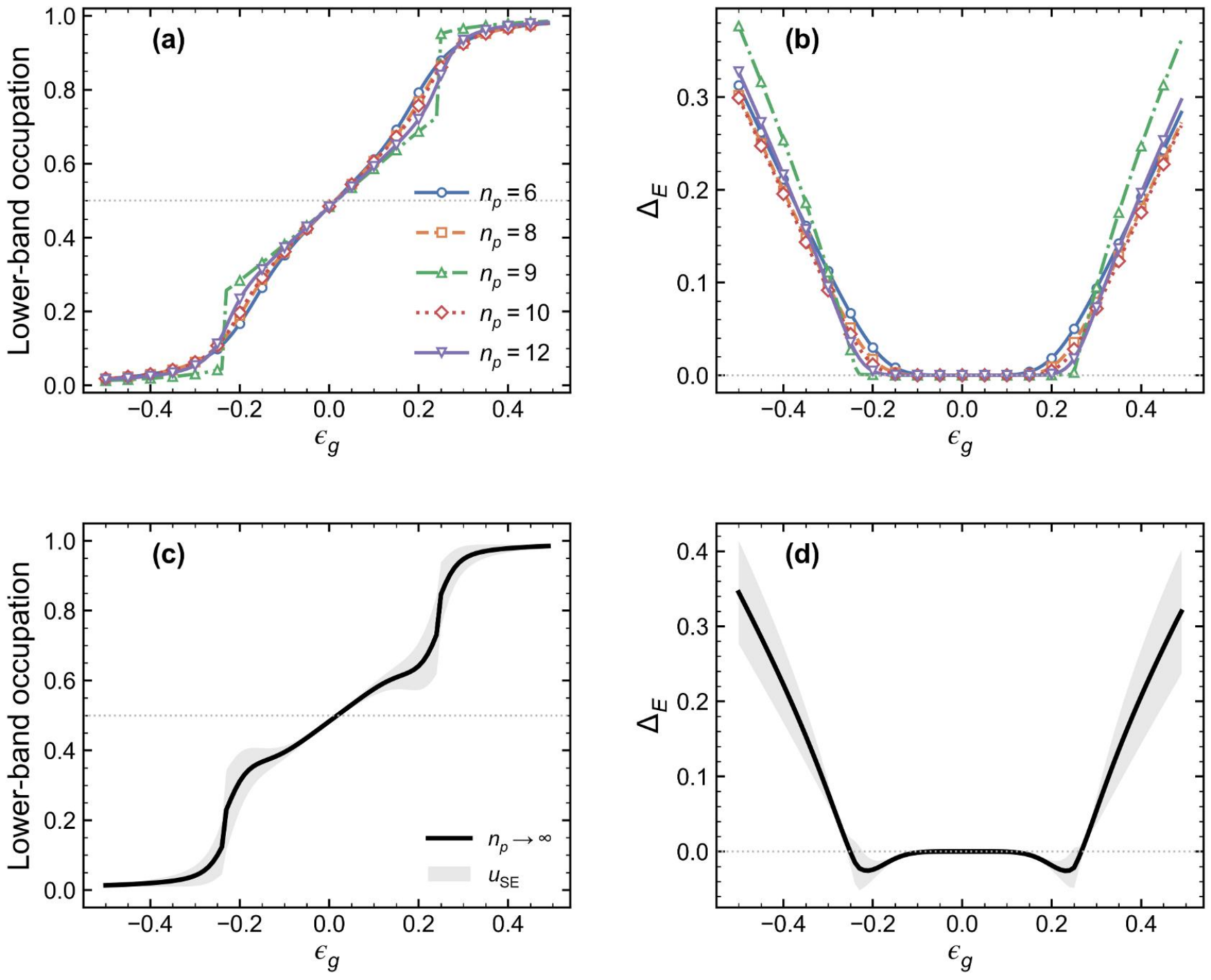}
    \caption{
Finite-size scaling of the occupation of the $\chi^{-1}=1.30$ band and
the excitation gap $\Delta E$ for tMoTe$_2$ at
$\vartheta=1.2^\circ$ as functions of $\epsilon_g$.
(a) Occupation of the $\chi^{-1}=1.30$ band and (b) excitation gap for
$n_p=6$, $8$, $9$, $10$, and $12$.
(c) and (d) Corresponding thermodynamic-limit values obtained by
independently fitting the finite-size data at each $\epsilon_g$ using
unweighted ordinary least squares. The black curves show the extrapolated
intercepts, and the shaded regions indicate one standard error
$u_{\rm SE}$ of the intercepts.
}
    \label{fig:S703}
\end{figure}

\begin{figure}[!ht]
    \centering
    \includegraphics[width=.97\textwidth]{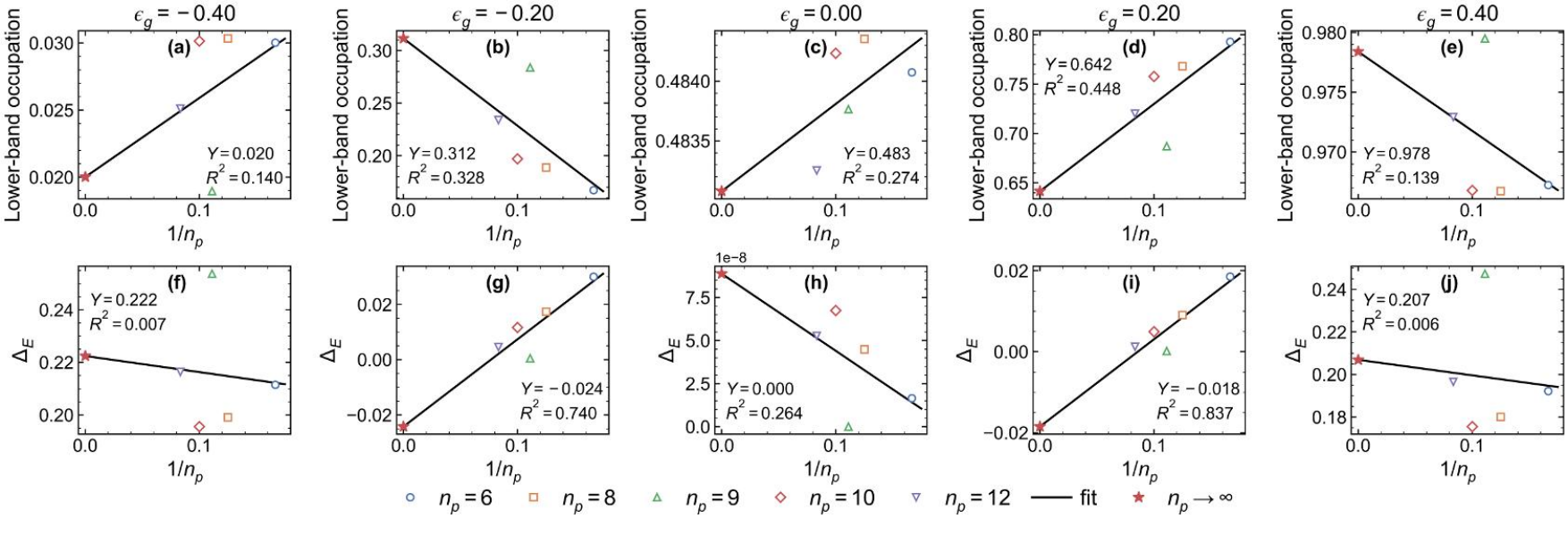}
    \caption{
Representative finite-size extrapolations for tMoTe$_2$ at
$\vartheta=1.2^\circ$.
From left to right, the columns correspond to
$\epsilon_g=-0.40$, $-0.20$, $0$, $0.20$, and $0.40$.
(a)--(e) Occupation of the $\chi^{-1}=1.30$ band and
(f)--(j) excitation gap $\Delta E$.
Symbols denote the finite-size data, and the black lines show the
corresponding unweighted linear fits.
Red stars at $1/n_p=0$ mark the extrapolated thermodynamic-limit values
$Y$. The fitted intercept $Y$ and coefficient of determination $R^2$
are indicated in each panel.
}
    \label{fig:S704}
\end{figure}

\begin{figure}[!ht]
    \centering
    \includegraphics[width=.7\textwidth]{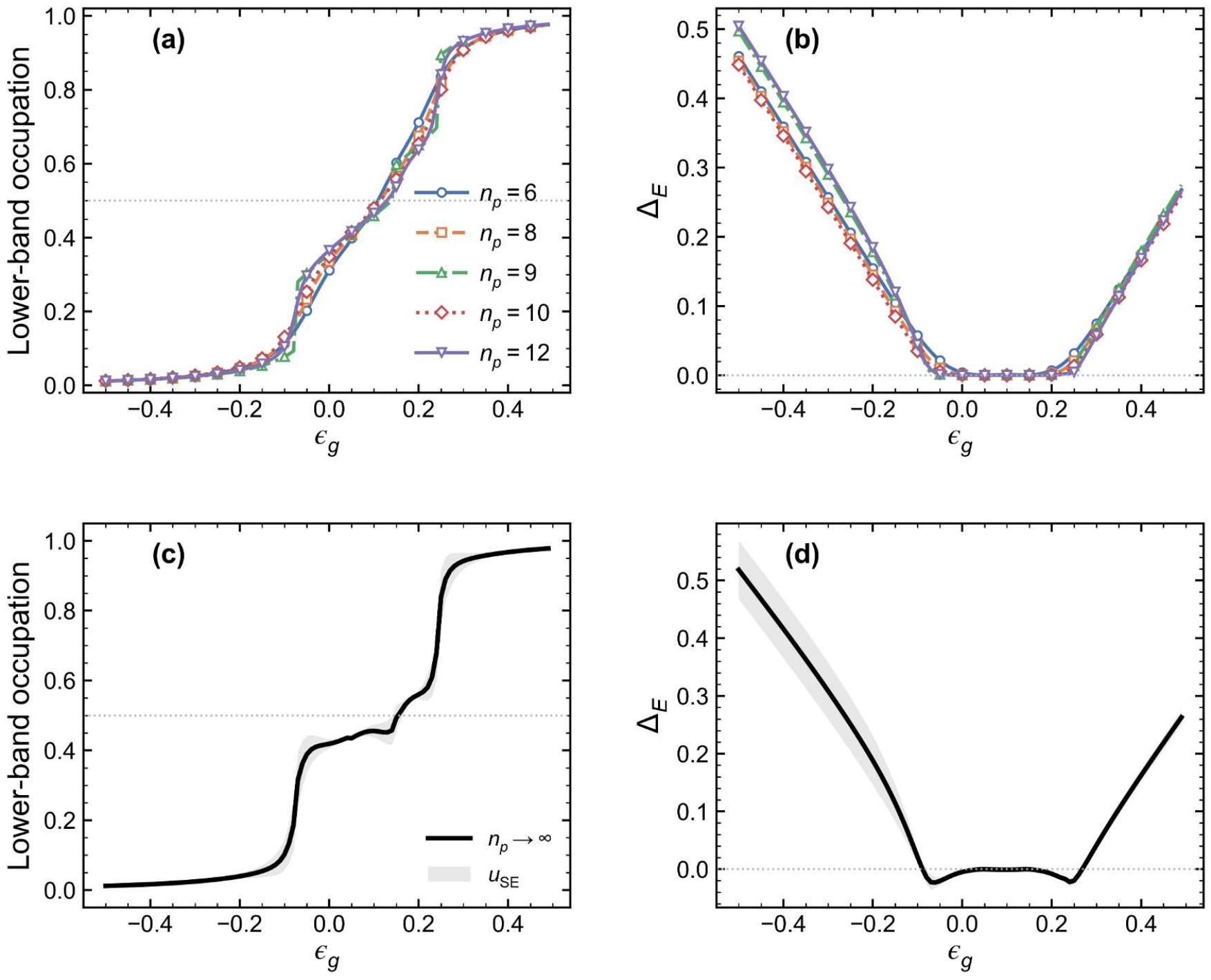}
    \caption{
Finite-size scaling of the occupation of the $\chi^{-1}=3.74$ band and
the excitation gap $\Delta E$ for tMoTe$_2$ at
$\vartheta=2^\circ$ as functions of $\epsilon_g$.
(a) Occupation of the $\chi^{-1}=3.74$ band and (b) excitation gap for
$n_p=6$, $8$, $9$, $10$, and $12$.
(c) and (d) Corresponding thermodynamic-limit values obtained by
independently fitting the finite-size data at each $\epsilon_g$ using
unweighted ordinary least squares. The black curves show the extrapolated
intercepts, and the shaded regions indicate one standard error
$u_{\rm SE}$ of the intercepts.
}
    \label{fig:S705}
\end{figure}

\begin{figure}[!ht]
    \centering
    \includegraphics[width=.97\textwidth]{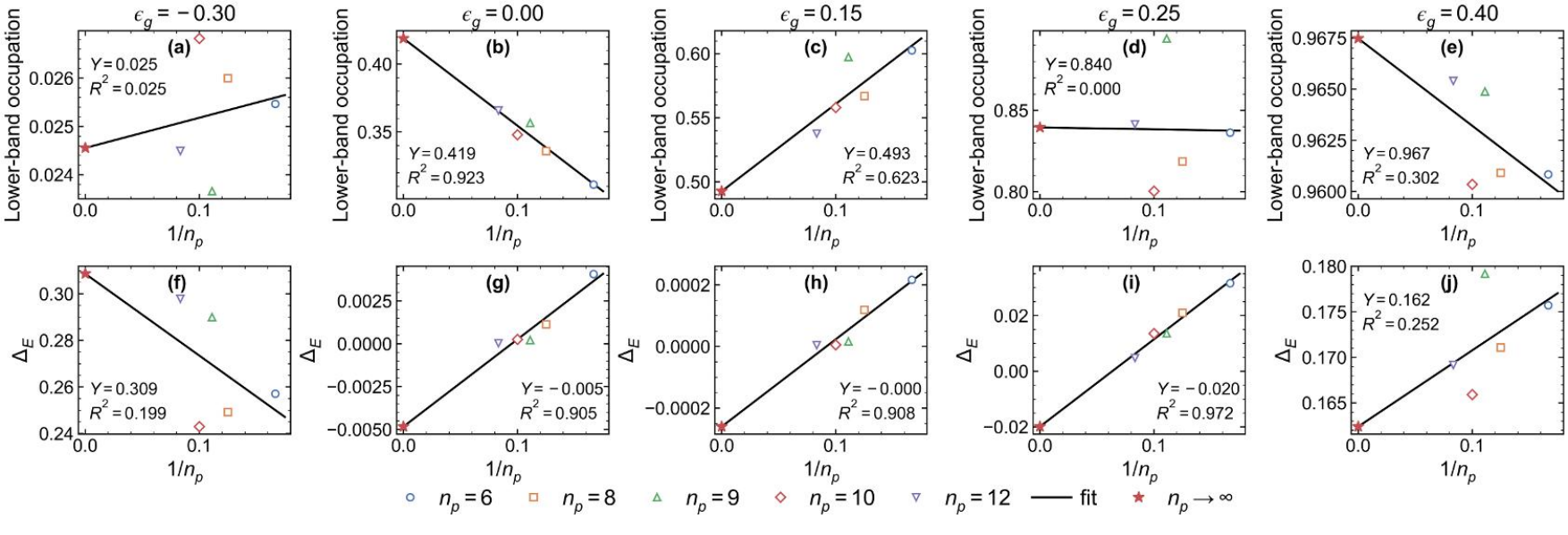}
    \caption{
Representative finite-size extrapolations for tMoTe$_2$ at
$\vartheta=2^\circ$.
From left to right, the columns correspond to
$\epsilon_g=-0.40$, $-0.20$, $0$, $0.20$, and $0.40$.
(a)--(e) Occupation of the $\chi^{-1}=3.74$ band and
(f)--(j) excitation gap $\Delta E$.
Symbols denote the finite-size data, and the black lines show the
corresponding unweighted linear fits.
Red stars at $1/n_p=0$ mark the extrapolated thermodynamic-limit values
$Y$. The fitted intercept $Y$ and coefficient of determination $R^2$
are indicated in each panel.
}
    \label{fig:S706}
\end{figure}

\begin{figure}[!ht]
    \centering
    \includegraphics[width=.7\textwidth]{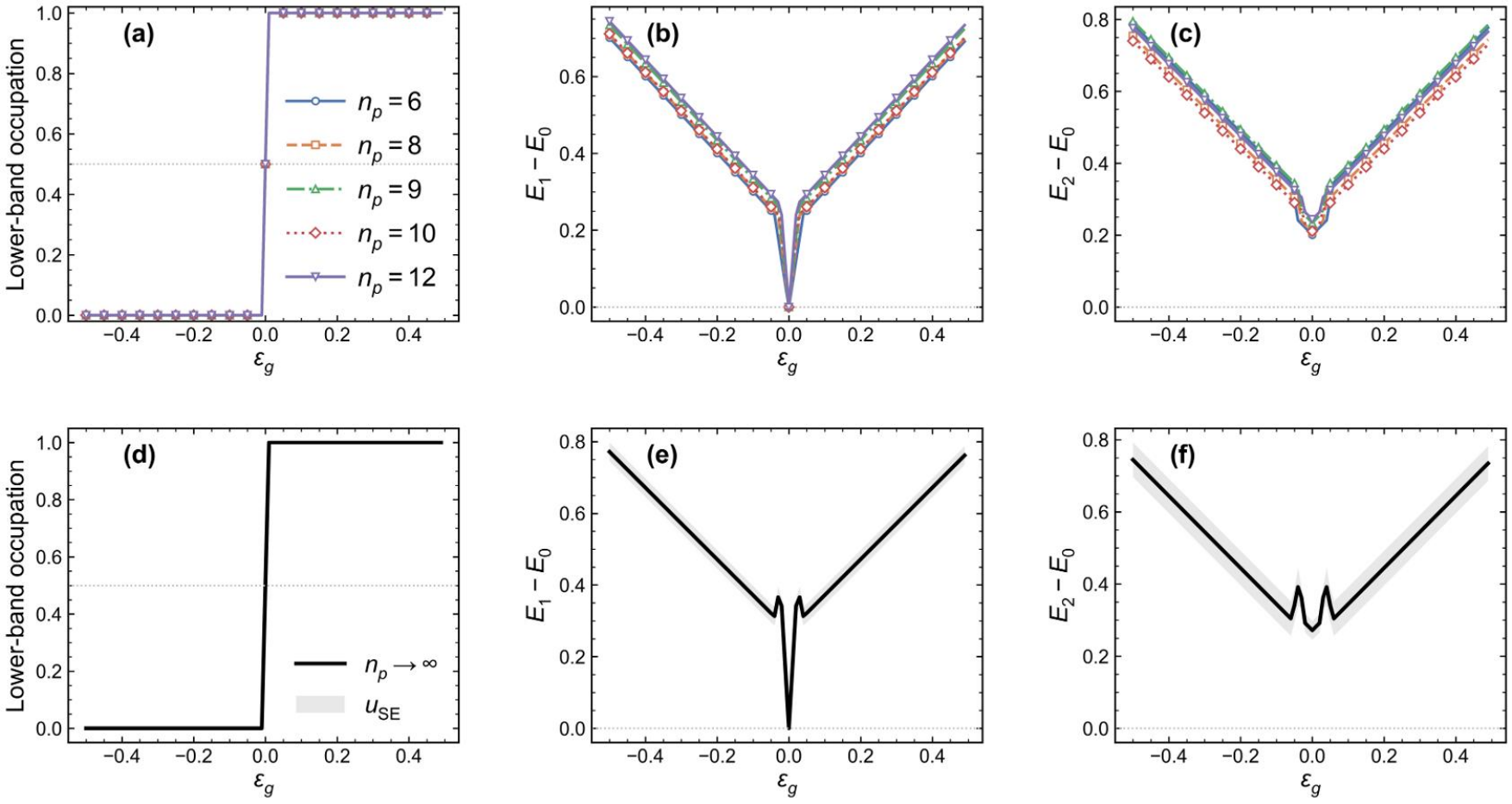}
    \caption{
Finite-size scaling of the occupation of the $\mathcal C=-1$ band and the
excitation gaps in TBG as functions of $\epsilon_g$.
(a)--(c) Occupation of the $\mathcal C=-1$ band, $E_1-E_0$, and
$E_2-E_0$, respectively, for $n_p=6$, $8$, $9$, $10$, and $12$.
(d)--(f) Corresponding thermodynamic-limit values obtained by independently
fitting the finite-size data at each $\epsilon_g$ using unweighted ordinary
least squares. The black curves show the extrapolated intercepts, and the
shaded regions indicate one standard error $u_{\rm SE}$ of the intercepts.
}
    \label{fig:S707}
\end{figure}

\begin{figure}[!ht]
    \centering
    \includegraphics[width=.97\textwidth]{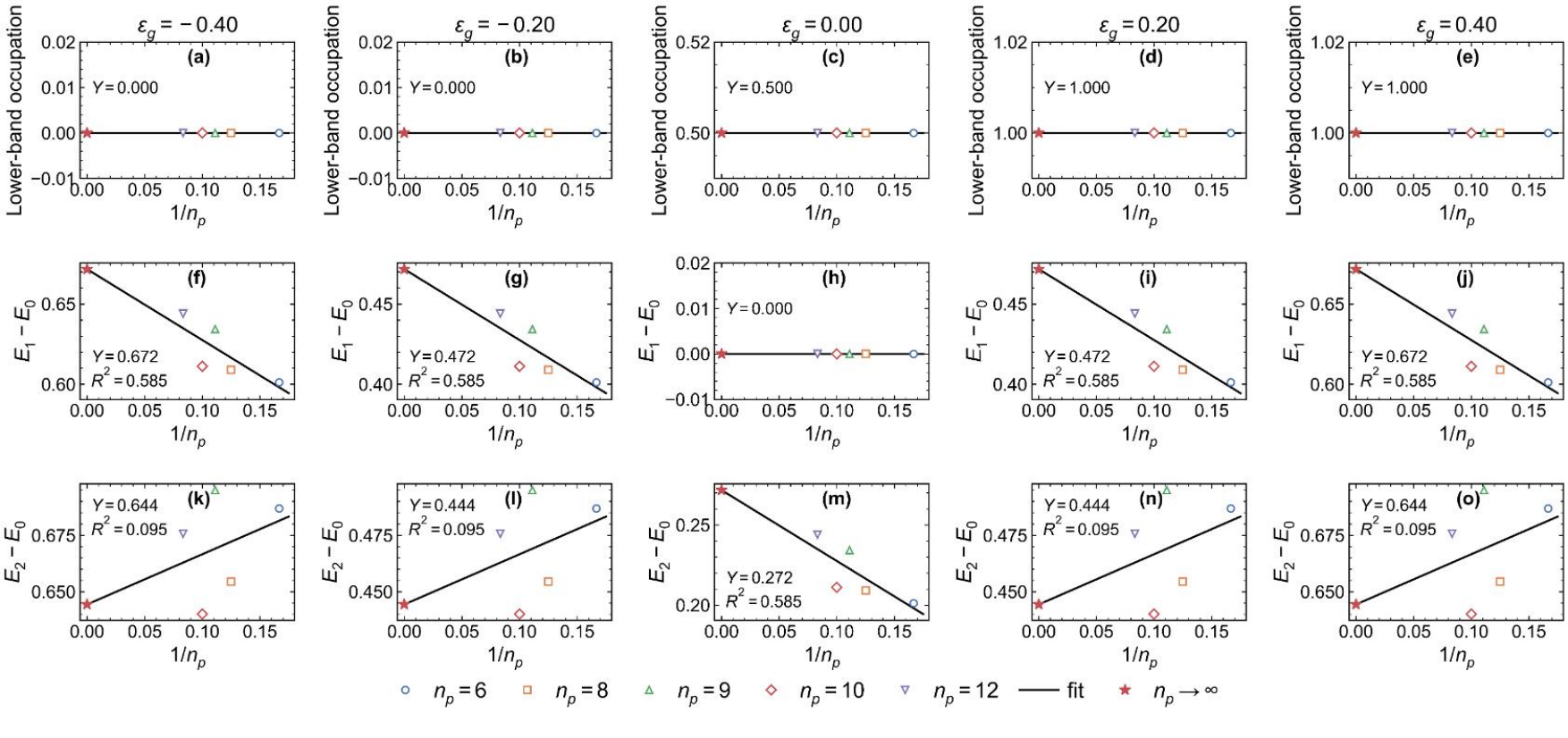}
    \caption{
Representative finite-size extrapolations for TBG.
From left to right, the columns correspond to
$\epsilon_g=-0.40$, $-0.20$, $0$, $0.20$, and $0.40$.
(a)--(e) Occupation of the $\mathcal C=-1$ band,
(f)--(j) $E_1-E_0$, and (k)--(o) $E_2-E_0$.
Symbols denote the finite-size data, and the black lines show the
corresponding unweighted linear fits.
Red stars at $1/n_p=0$ mark the extrapolated thermodynamic-limit values
$Y$. The fitted intercept $Y$ is indicated in each panel, whereas
$R^2$ is omitted in panels (a)--(e) and (h), where the finite-size data
are constant within numerical precision.
}
    \label{fig:S708}
\end{figure}

In this section, we examine the finite-size behavior of band mixing in the
LLL+1LL, TBG, and tMoTe$_2$ systems. We characterize the band-mixing
behavior using both the low-energy spectrum and the band occupation. As shown in \fref{fig:S701}--\fref{fig:S708}, for different band
splittings $\epsilon_g$, we calculate the energy gaps between the ground
state and the lowest excited states for several system sizes and
extrapolate them to the thermodynamic limit. We also perform finite-size
scaling of the individual band occupations, which provides a direct
measure of the degree of band mixing. At each $\epsilon_g$, the finite-size
data are extrapolated using an unweighted ordinary least-squares fit. The
uncertainty $u_{\rm SE}$ shown in the figures corresponds to the standard
error of the fitted intercept obtained from the least-squares
analysis~\cite{PhysRevB.88.035133}. To illustrate the quality of the
extrapolation, we additionally show the fits explicitly for several
representative values of $\epsilon_g$.

We first note that the quality of the finite-size extrapolation is not
uniform throughout the parameter range. One source of the scatter is the
sensitivity of finite-size spectra to the geometry and aspect ratio of the
torus~\cite{PhysRevX.1.021014}, which can lead to appreciable variations
among finite-size clusters with different shapes. In addition, the
extrapolation naturally becomes less reliable in regions where the
excitation gap is very small or tends to close, since finite-size effects
become increasingly important near a gapless regime. In such regions, an
unconstrained linear extrapolation may even yield a negative intercept.
Thus, the simple linear finite-size extrapolation is not quantitatively
reliable in the vicinity of a gapless regime.

Our main focus, however, is not to determine the thermodynamic-limit gap
with high quantitative precision, but rather to establish the qualitative
behavior at $\epsilon_g=0$, where the two bands are degenerate. For the
LLL--1LL system, the extrapolation indicates that more than $85\%$ of the
electrons remain in the LLL at $\epsilon_g=0$ (\fref{fig:S702}(d)),
while a finite excitation gap persists (\fref{fig:S702}(i)). As
$\epsilon_g$ is continuously reduced from the large-splitting limit to
zero, we find no indication of a closing of the excitation gap, showing
that the ground state remains in the quantum Hall ferromagnetic phase.
The precise thermodynamic-limit value of the gap is therefore not essential
for the conclusion drawn here. At $\epsilon_g=0$, the excitation gap shows an increasing trend with
increasing system size. This provides
clear qualitative evidence for an adiabatic connection between the
large-$\epsilon_g$ polarized state and the interaction-driven state at
$\epsilon_g=0$.

The behavior in tMoTe$_2$ is qualitatively different. We note that
\fref{fig:S704}--\fref{fig:S707} show the extrapolated occupation of
the band with the larger $\chi^{-1}$, while Fig.~1(c) of the main text
shows that of the band with the smaller $\chi^{-1}$, obtained as one minus
the former. As $\epsilon_g$ is
reduced, both twist angles exhibit substantially stronger band mixing than
the LLL--1LL system. At $\vartheta=1.2^\circ$, the occupations of the two bands are found to be close at $\epsilon_g=0$, indicating strong band
mixing. At $\vartheta=2^\circ$, despite the large difference in
$\chi^{-1}$ between the two bands, the band mixing remains pronounced
compared with the LLL--1LL system. At the same time, the many-body
excitation gap can collapse even before the two bands become degenerate.
As in the LLL--1LL case, our conclusion does not rely on a highly precise
determination of the thermodynamic-limit gap. Rather, the qualitative
features of the finite-size data---the substantially enhanced band mixing
and the strong suppression or closing of the excitation gap---are already
sufficient to distinguish tMoTe$_2$ from the LLL--1LL system. These
finite-size extrapolations therefore support the conclusion of the main
text that tMoTe$_2$ does not retain a robust band-polarized gapped state in
the small-$\epsilon_g$ regime.

We further consider the strong-interaction limit
$\epsilon_{\rm int}/\epsilon_g\rightarrow\infty$.
At low fractional fillings in lattice models, electrons have been found
to remain predominantly in the lower-energy band even in the
strong-interaction regime~\cite{PhysRevLett.112.126806}. From the
perspective of the CHS, this behavior admits a more general interpretation.
We therefore focus on the low-filling regime, where the suppression of band
mixing can be examined most clearly.
Finite-size scaling becomes increasingly demanding at larger fillings in
a two-band system because of the rapid growth of the many-body Hilbert
space. We thus perform finite-size extrapolations at
$\nu=1/7$, $1/6$, $1/5$, $1/4$, $1/3$, and $1/2$.

\begin{figure}[!ht]
    \centering
    \includegraphics[width=.7\textwidth]{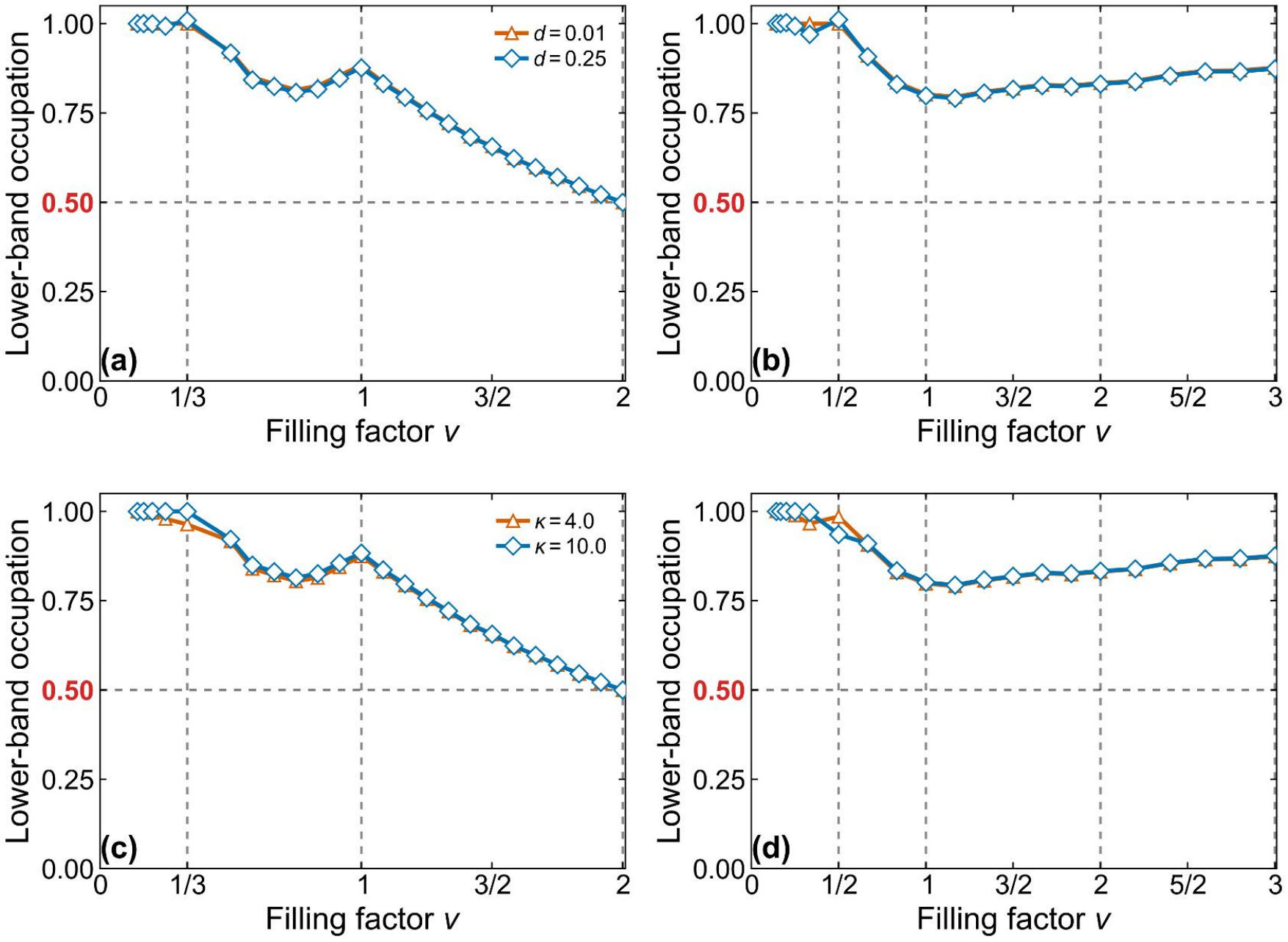}
    \caption{
(a) and (b) LLL occupation fraction in the LLL+1LL system for fermions
and bosons, respectively, with the dual-gate-screened Coulomb interaction
$U_{\mathrm{gate}}(q)$ (\eref{eq:screen}) at $d=0.01$ and $0.25$
in the strong-interaction limit
$\epsilon_{\rm int}/\epsilon_g\rightarrow\infty$.
(c) and (d) LLL occupation fraction in the LLL+1LL system for fermions
and bosons, respectively, with the Yukawa interaction
$U_{\mathrm{Y}}(q)$ (\eref{eq:screen}) at $\kappa=4$ and $10$
in the same strong-interaction limit.
}
    \label{fig:S709}
\end{figure}

As shown in \fref{fig:S711} and \fref{fig:S712}, together with
Fig.~2(a,b) of the main text, band mixing is strongly suppressed at low
filling in ideal bands, as exemplified by the Landau-level systems. This
behavior becomes particularly transparent for the shortest-ranged TK
interaction. For fermions at $\nu\leq 1/3$, the many-body ground states
lie entirely within the CHS $\mathcal H_{\rm Laughlin-1/3}$. Since this
subspace is an exact null space of the TK interaction, the particles can
remain completely confined to $\mathcal H_{\rm Laughlin-1/3}$, and no band
mixing is induced in this limit. As shown in \fref{fig:S709}, replacing the TK interaction by screened
Coulomb interactions with increasing range, and ultimately by the Coulomb
interaction (Fig.~2(a,b) of the main text), progressively enhances band mixing, reflecting the increasing
departure from the short-range TK limit.

\begin{figure}[!ht]
    \centering
    \includegraphics[width=.7\textwidth]{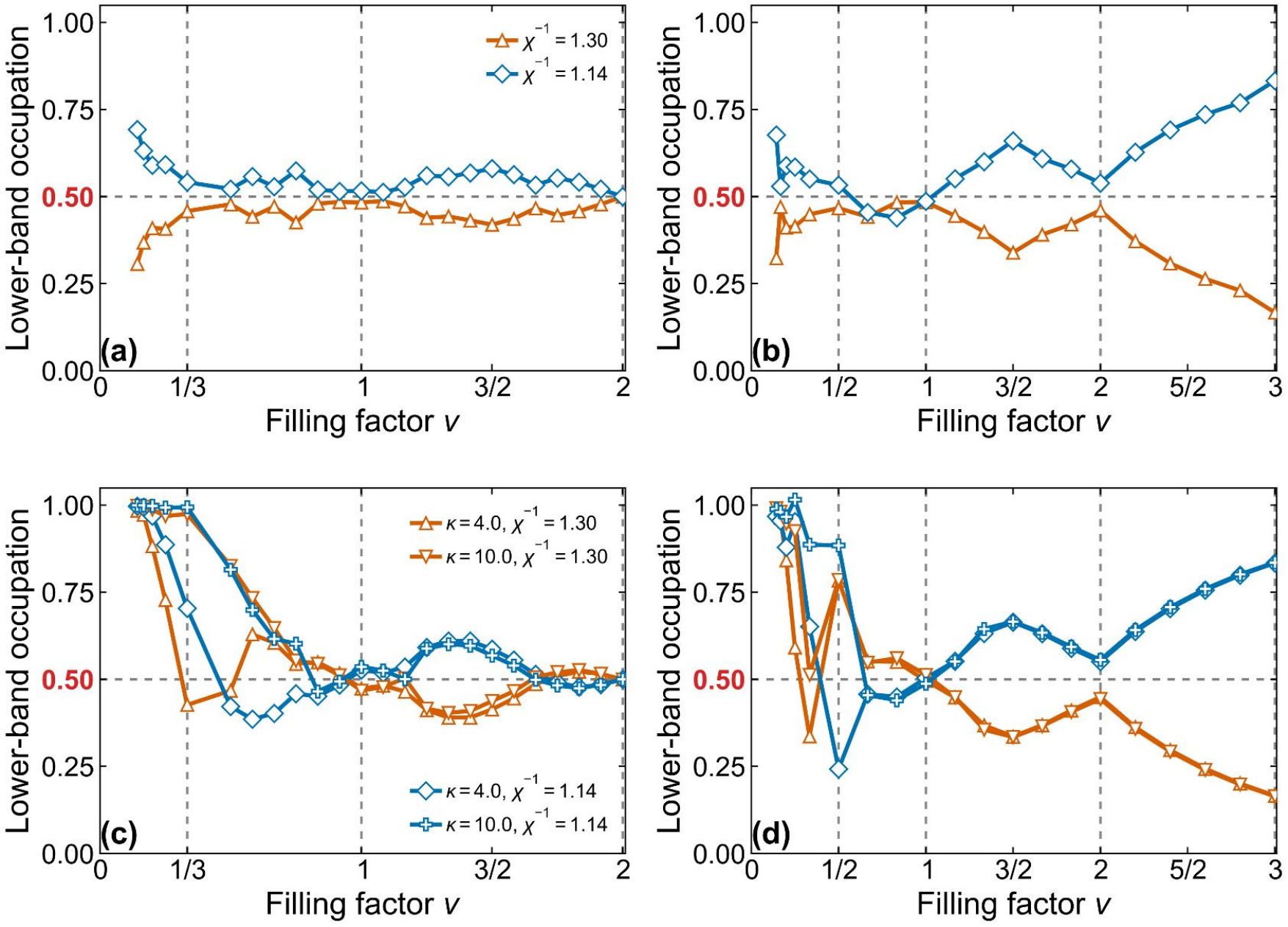}
\caption{
(a) and (b) Occupation fraction of the lower-energy band in tMoTe$_2$
at $\vartheta=1.2^\circ$ for fermions and bosons, respectively, with the
Coulomb interaction in the strong-interaction limit
$\epsilon_{\rm int}/|\epsilon_g|\rightarrow\infty$.
(c) and (d) Corresponding occupation fractions for fermions and bosons,
respectively, with the Yukawa interaction
$U_{\mathrm{Y}}(q)$ (\eref{eq:screen}) at $\kappa=4$ and $10$
in the same strong-interaction limit.
}
    \label{fig:S710}
\end{figure}

\begin{figure}[!ht]
    \centering
    \includegraphics[width=.97\textwidth]{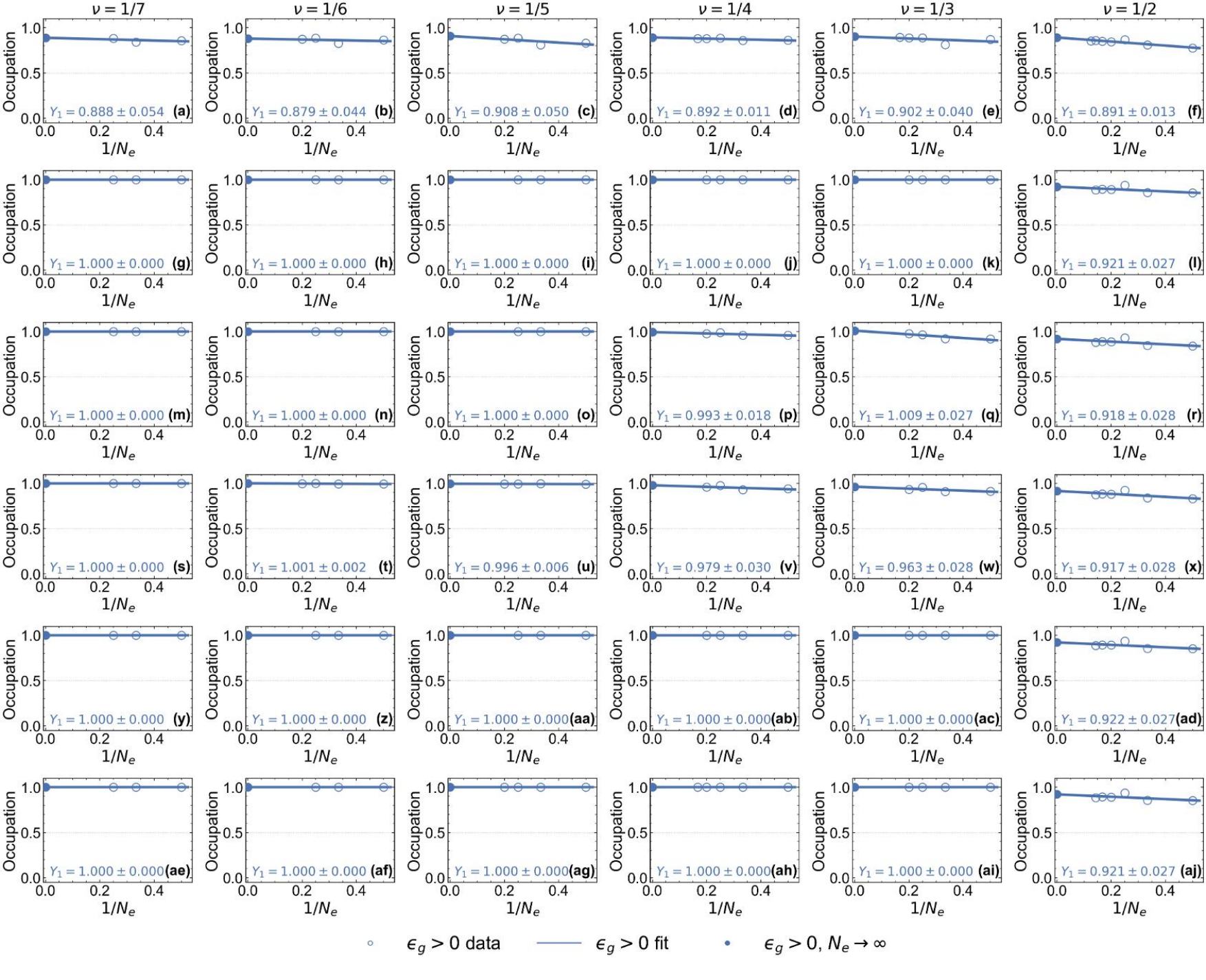}
    \caption{
Finite-size scaling of the LLL occupation fraction in the LLL+1LL system
for fermions in the strong-interaction limit
$\epsilon_{\rm int}/\epsilon_g\rightarrow\infty$.
Panels (a)--(f), (g)--(l), (m)--(r), (s)--(x), (y)--(ad), and
(ae)--(aj) correspond to the Coulomb interaction, screened Coulomb
interactions with $d=0.01$, $d=0.25$, $\kappa=4$, and $\kappa=10$,
and the TK interaction, respectively.
Within each group of six panels, the filling factors from left to right
are $\nu=1/7$, $1/6$, $1/5$, $1/4$, $1/3$, and $1/2$.
Open symbols denote the finite-size results, and the lines show linear
fits in $1/N_e$.
Filled symbols at $1/N_e=0$ indicate the extrapolated thermodynamic-limit
occupations $Y_1$.
The quoted uncertainties are the standard errors of the fitted intercepts.
}
    \label{fig:S711}
\end{figure}

\begin{figure}[!ht]
    \centering
    \includegraphics[width=.97\textwidth]{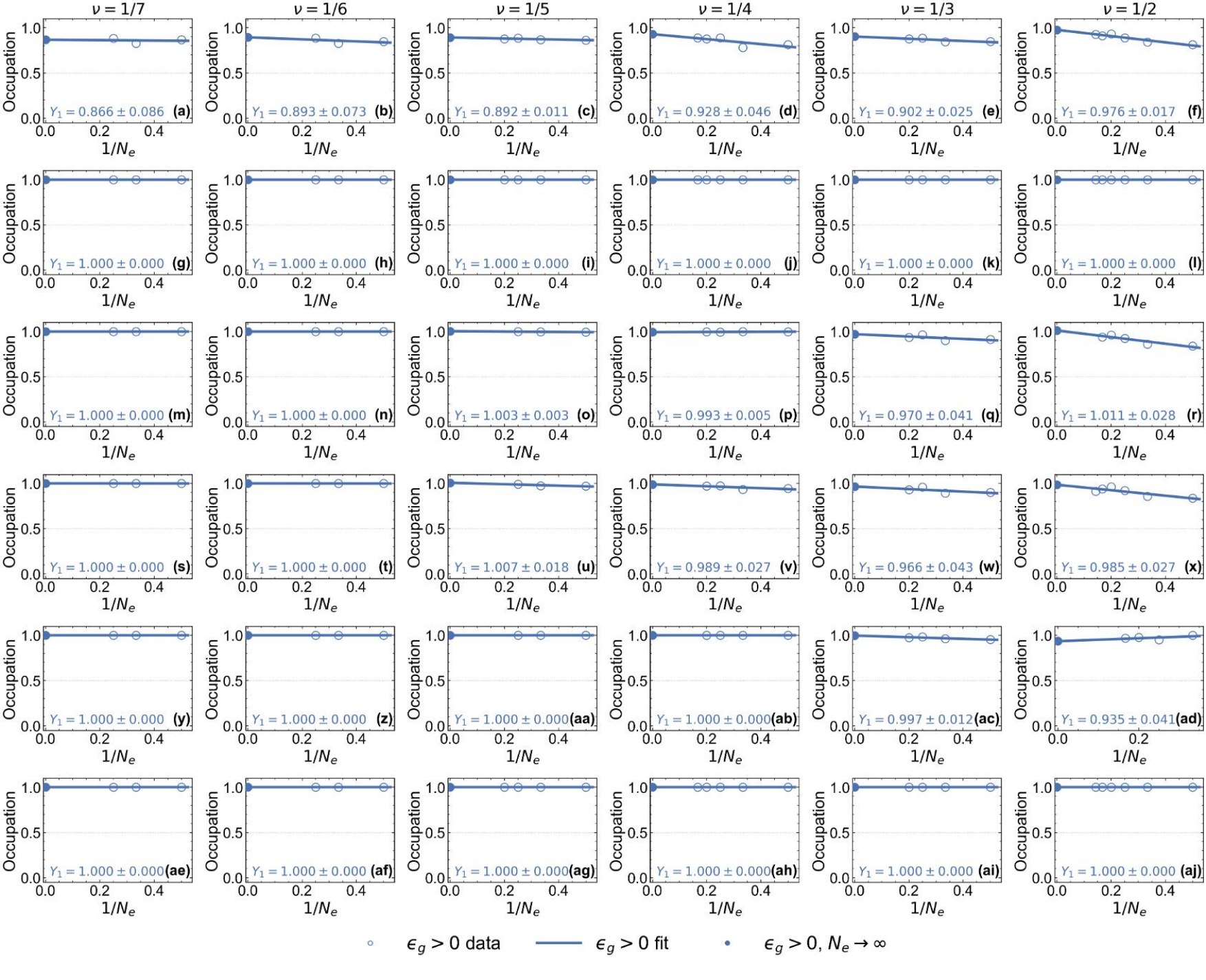}
    \caption{
Finite-size scaling of the LLL occupation fraction in the LLL+1LL system
for bosons in the strong-interaction limit
$\epsilon_{\rm int}/\epsilon_g\rightarrow\infty$.
Panels (a)--(f), (g)--(l), (m)--(r), (s)--(x), (y)--(ad), and
(ae)--(aj) correspond to the Coulomb interaction, screened Coulomb
interactions with $d=0.01$, $d=0.25$, $\kappa=4$, and $\kappa=10$,
and the TK interaction, respectively.
Within each group of six panels, the filling factors from left to right
are $\nu=1/7$, $1/6$, $1/5$, $1/4$, $1/3$, and $1/2$.
Open symbols denote the finite-size results, and the lines show linear
fits in $1/N_e$.
Filled symbols at $1/N_e=0$ indicate the extrapolated thermodynamic-limit
occupations $Y_1$.
The quoted uncertainties are the standard errors of the fitted intercepts.
}
    \label{fig:S712}
\end{figure}

\begin{figure}[!ht]
    \centering
    \includegraphics[width=.97\textwidth]{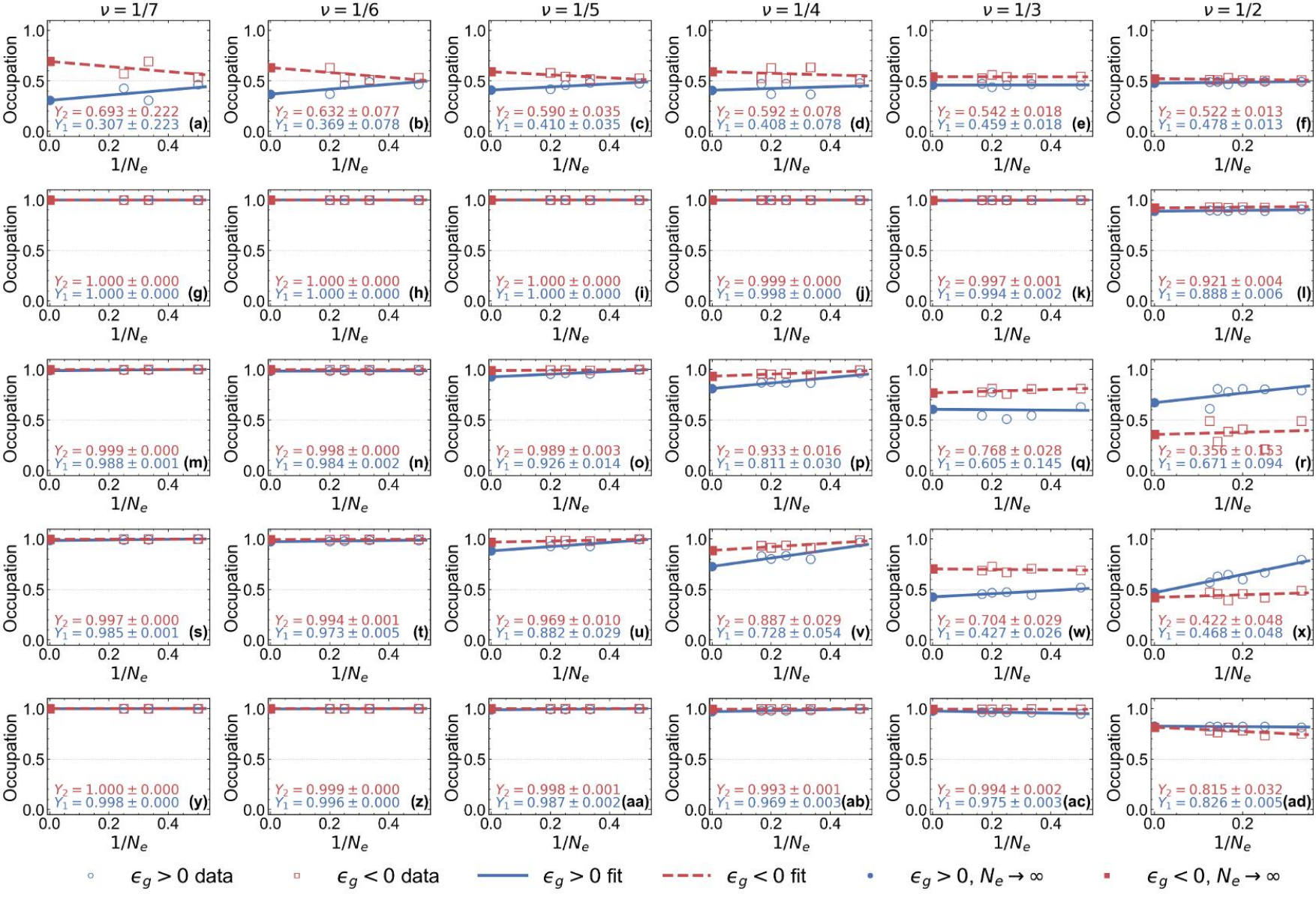}
    \caption{
Finite-size scaling of the lower-band occupation for fermions in
tMoTe$_2$ at $\vartheta=1.2^\circ$ in the strong-interaction limit
$\epsilon_{\rm int}/|\epsilon_g|\rightarrow\infty$.
Panels (a)--(f), (g)--(l), (m)--(r), (s)--(x), and (y)--(ad)
correspond to the Coulomb interaction and screened Coulomb interactions
with $d=0.01$, $d=0.25$, $\kappa=4$, and $\kappa=10$, respectively.
Within each group of six panels, the filling factors from left to right
are $\nu=1/7$, $1/6$, $1/5$, $1/4$, $1/3$, and $1/2$.
Open symbols denote the finite-size results for $\epsilon_g>0$ and
$\epsilon_g<0$, and the corresponding lines show linear fits in $1/N_e$.
Filled symbols at $1/N_e=0$ indicate the extrapolated thermodynamic-limit
occupations. Here, $Y_1$ and $Y_2$ denote the extrapolated occupations for
$\epsilon_g>0$ and $\epsilon_g<0$, respectively.
The quoted uncertainties are the standard errors of the fitted intercepts.
}
    \label{fig:S713}
\end{figure}

\begin{figure}[!ht]
    \centering
    \includegraphics[width=.97\textwidth]{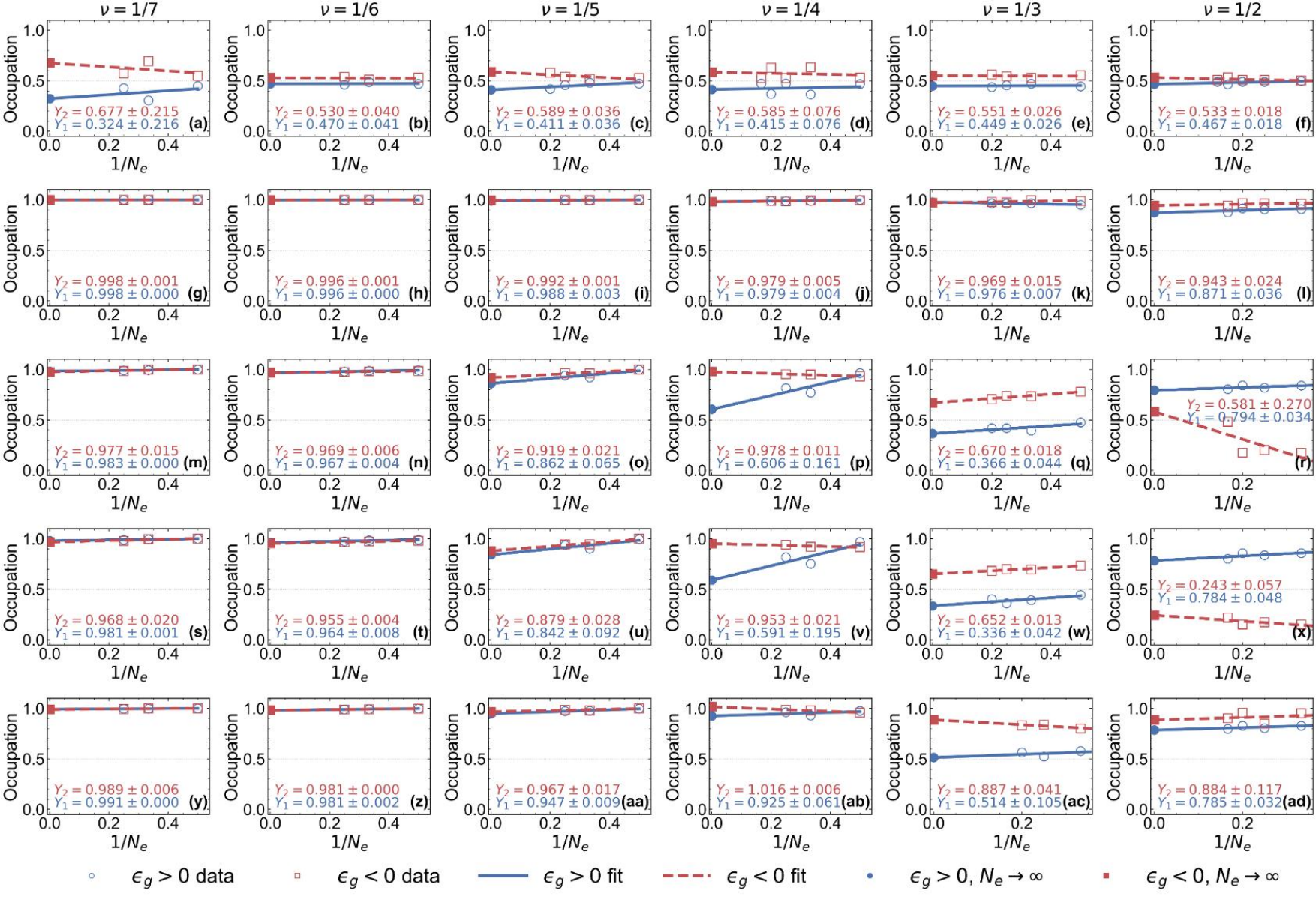}
    \caption{
Finite-size scaling of the lower-band occupation for bosons in
tMoTe$_2$ at $\vartheta=1.2^\circ$ in the strong-interaction limit
$\epsilon_{\rm int}/|\epsilon_g|\rightarrow\infty$.
Panels (a)--(f), (g)--(l), (m)--(r), (s)--(x), and (y)--(ad)
correspond to the Coulomb interaction and screened Coulomb interactions
with $d=0.01$, $d=0.25$, $\kappa=4$, and $\kappa=10$, respectively.
Within each group of six panels, the filling factors from left to right
are $\nu=1/7$, $1/6$, $1/5$, $1/4$, $1/3$, and $1/2$.
Open symbols denote the finite-size results for $\epsilon_g>0$ and
$\epsilon_g<0$, and the corresponding lines show linear fits in $1/N_e$.
Filled symbols at $1/N_e=0$ indicate the extrapolated thermodynamic-limit
occupations. Here, $Y_1$ and $Y_2$ denote the extrapolated occupations for
$\epsilon_g>0$ and $\epsilon_g<0$, respectively.
The quoted uncertainties are the standard errors of the fitted intercepts.
}
    \label{fig:S714}
\end{figure}

We next turn to generic Chern bands and consider the two bands of
tMoTe$_2$ at $\vartheta=1.2^\circ$, as shown in
\fref{fig:S710}, \fref{fig:S713}, and \fref{fig:S714}, together with
Fig.~4(a,b) of the main text. We examine two
representative situations. In the first, the lower-energy band has the
smaller quantum-metric trace $\chi^{-1}$, so that the kinetic-energy
preference and the interaction-induced preference both favor occupation of
the lower band. Band mixing is therefore suppressed. In the second, the
lower-energy band has the larger $\chi^{-1}$, so that the kinetic and
interaction preferences compete, leading to substantially stronger band
mixing. We further examine screened Coulomb interactions with different screening
parameters in both cases. The same trend persists:
shorter-ranged interactions suppress band mixing more effectively. This also provides a natural interpretation of the behavior observed in lattice Chern-band systems. Since these models typically involve short-ranged interactions, such as nearest-neighbor repulsion \cite{PhysRevLett.112.126806}. Band mixing is therefore strongly suppressed at low filling.

\oldsection*{Supplementary References}
\bibliography{ref.bib}

\newpage
\oldsection*{Supplementary Figures}

\clearpage

